\documentclass[aps,reprint,superscriptaddress,longbibliography,showpacs,superscriptaddress,notitlepage,nofootinbib]{revtex4-1}
\usepackage[dvips]{graphicx}
\usepackage{amsmath,amssymb,amsthm,mathrsfs,amsfonts,dsfont,amscd,keyval}
\usepackage{mathtools,mathrsfs}
\usepackage{yfonts} 
\usepackage{subfigure, epsfig}
\usepackage{titletoc} 
\usepackage[subfigure]{tocloft}
\usepackage{titlesec}

\usepackage{ytableau} \ytableausetup{boxsize = 2pt}
\usepackage{braket}
\usepackage{tensor}
\usepackage{bm}
\usepackage{bbm}
\usepackage{enumerate,enumitem}
\usepackage{color}
\usepackage{hyperref}
\hypersetup{breaklinks=true}
\hypersetup{colorlinks=true, linkcolor=blue, citecolor=magenta, urlcolor=blue, linktocpage=true}
\usepackage[hyphenbreaks]{breakurl}
\usepackage{tabularx}
\usepackage{times}
\usepackage{multirow}
\usepackage{float}
\usepackage{tcolorbox}

\renewcommand\vec{\mathbf}

\newtheorem*{theorem*}{Theorem}
\newtheorem*{definition*}{Definition}

\theoremstyle{definition}
\newtheorem{definition}{Definition}[section]

\newtheorem{proposition}[definition]{Proposition}

\theoremstyle{plain}
\newtheorem{theorem}[definition]{Theorem}
\newtheorem{lemma}[definition]{Lemma}

\newtheorem*{fact}{Fact}

\theoremstyle{remark}
\newtheorem*{remark}{Remark}

\newcommand{\comments}[1]{}

\renewcommand\vec{\mathbf}

\begin{document}

\title{SU$(d)$-Symmetric Random Unitaries:\\ Quantum Scrambling, Error Correction, and Machine Learning}% Force line breaks with \\

\author{Zimu Li$^{**}$}
\email{lizm@mail.sustech.edu.cn}
\affiliation{Yau Mathematical Sciences Center, Tsinghua University, Beijing 100084, China}

\author{Han Zheng$^{**}$}
\email{hanz98@uchicago.edu}
\affiliation{Pritzker School of Molecular Engineering, The University of Chicago, Chicago, IL 60637, USA}
\affiliation{Department of Computer Science, The University of Chicago, Chicago, IL 60637, USA}

\author{Yunfei Wang}
\affiliation{Martin A. Fisher School of Physics, Brandeis University, Waltham, MA, 02435, USA}

\author{Liang Jiang}
\email{liangjiang@uchicago.edu}
\affiliation{Pritzker School of Molecular Engineering, The University of Chicago, Chicago, IL 60637, USA}

\author{Zi-Wen Liu}
\email{zwliu0@tsinghua.edu.cn}
\affiliation{Yau Mathematical Sciences Center, Tsinghua University, Beijing 100084, China}
\affiliation{Perimeter Institute for Theoretical Physics, Waterloo, Ontario N2L 2Y5, Canada}

\author{Junyu Liu}
\email{junyuliu@uchicago.edu}
\affiliation{Pritzker School of Molecular Engineering, The University of Chicago, Chicago, IL 60637, USA}
\affiliation{Department of Computer Science, The University of Chicago, Chicago, IL 60637, USA}
\affiliation{Kadanoff Center for Theoretical Physics, The University of Chicago, Chicago, IL 60637, USA}
\affiliation{Department of Computer Science, The University of Pittsburgh, Pittsburgh, PA 15260, USA}

\begin{abstract}
    Quantum information processing in the presence of continuous symmetry is of wide importance and exhibits many novel physical and mathematical phenomena. SU$(d)$ is a continuous symmetry group of particular interest since it represents a fundamental type of non-Abelian symmetry  and also plays a vital role in quantum computation. Here, we explicate three particularly interesting applications of SU$(d)$-symmetric random unitaries in  diverse contexts ranging from physics to quantum computing: information scrambling with non-Abelian conserved quantities,  covariant quantum error correcting random codes, and geometric quantum machine learning. First, we show that, in the presence of SU$(d)$ symmetry, the local conserved quantities would exhibit residual values even at $t \rightarrow \infty$ which decays as $\Omega(1/n^{3/2})$ under local Pauli basis for qubits and $\Omega(1/n^{(d+2)^2/2})$ under local symmetric basis for general qudits with respect to the system size, in contrast to $O(1/n)$ decay for U(1) case and the exponential decay for no-symmetry case in the sense of out-of-time ordered correlator (OTOC). Second, we show that SU$(d)$-symmetric unitaries can be used to construct asymptotically optimal (in the sense of saturating the fundamental limits on the code error, or the approximate Eastin--Knill theorems) SU$(d)$-covariant codes (codes with universal transversal gates) that encode any constant number of logical qudits, extending [Kong \& Liu; PRXQ 3, 020314 (2022)]. Finally, we derive an overpartameterization threshold via the quantum neural tangent kernel (QNTK) required for exponential convergence guarantee of generic ansatz for geometric quantum machine learning, which reveals that the number of parameters required scales only with the dimension of desired subspaces rather than that of the entire Hilbert space.  Our work invites further research on quantum information with continuous symmetries, where the mathematical tools developed in this work are expected to be useful. 
\end{abstract}

\maketitle

%---------------------------------------------------------------------------------------------

\section{Introduction}

Symmetries serve as not only fundamental principles in theoretical physics, but also vital roles in modern quantum technologies. Specifically, within the vast array of quantum circuit models, the adoption of common symmetries such as rotations or permutations under group actions unveils innovative features across various facets of contemporary quantum information science and quantum computing. A principal emergence of symmetry in quantum information is to study random circuits with conservation laws, which has recently drawn significant attraction due to its fundamental relevance in theoretical physics and rich phenomena displayed therein, including quantum information scrambling \cite{Hayden-Preskill2007,hayden2016holographic, yuan2022quantum, landsman2019verified, von2018operator, choi2021emergent, nahum2018operator,yoshida:soft,junyu2020chargescrambler,Nakata2023}, quantum error correction \cite{Yoshida-Kitaev2017, gullans2021quantum, brown2013short,Zhou2021newperspectives,kong2022near}, quantum machine learning \cite{Huang_2021,nguyen2022theory,Zheng2021SpeedingUL,PRXQuantum.3.030341, Liu2022QNTK, Liu2023QNTK, you2023analyzing}, and random benchmarking protocols \cite{knill2008randomized, helsen2022general, kong2021framework}.

The famous Noether's theorem posits that in a closed quantum system, continuous symmetries are associated with conserved charges, namely, conservation laws. The entire Hilbert space is subsequently decomposed into charge sectors whose charges are reflected through the eigenvalues of the underlying conserved charge or the Casimir operators. 
Quantum systems governed by conservation laws play a key role in quantum information theory, exemplified by foundational results such as the Eastin--Knill theorem for quantum error-correcting codes \cite{eastin2009restrictions,faist20,Kubica21}. Recent work has uncovered further constraints imposed by conservation laws, including limitations on information recovery \cite{Yoshida-Kitaev2017,Tajima2021,Tajima2022,Nakata2023,Majidy2023review} and no-go theorems to achieve universal quantum computation via local interactions \cite{MarvianNature,Zheng2021SpeedingUL,MarvianSU2,SUd-k-Design2023}. Among these, random quantum circuits with conservation law produce many novel physical insights such as in operator spreading \cite{rakovszky2018diffusive, majidy2023non, agarwal2023charge, khemani2018operator, chang2024deep, Liu2024Mpemba}, covariant quantum error correction \cite{Zhou2021newperspectives, kong2022near}, and monitored circuit dynamics \cite{majidy2023critical, agrawal2022entanglement}. 

In this paper, we are specifically interested in the SU$(d)$ symmetry, a canonical model of non-Abelian symmetry governed by the special unitary group SU$(d)$ acting transversally on each qudit with local dimension $d$ over the system and tightly correlated with the permutation symmetry of the system, which widely exists in quantum many-body systems, quantum chemistry, and graph-based data structures, due to a mathematical formalism called \emph{Schur--Weyl duality} \cite{Goodman2009,Tolli2009}. The theory of random quantum circuits with SU$(d)$ symmetry has recently produced fruitful implications e.g., \cite{Zheng2021SpeedingUL, MarvianSUd, kong2022near, majidy2023critical} and drawn interest due to non-Abelian conserved charges. Here, we primarily focus on three different applications: quantum scrambling, quantum error correction, and quantum machine learning. 

Conserved quantities from SU$(d)$-symmetric quantum circuits, due to Schur--Weyl duality, are generated by the symmetric group $S_n$ which permutes the sites of qudits. Except for the aforementioned significance, SU$(d)$-conserved quantities serve as a natural and important generalization to that of U($1$) symmetry which has been systematically studied in terms of hydrodynamics. To be specific, it has been observed that the diffusion of local U($1$) conserved charges scale in power law in time and in late time post scrambling, the existence of hydrodynamic tails or slow-mode relaxation \cite{rakovszky2018diffusive, agarwal2023charge}. Moreover, in a generic chaotic system with energy conservation satisfying the eigenstate thermalization hypothesis (ETH), a similar late-time slow-mode relaxation exists which scales as power-law decay with system size \cite{Huang2019OTOC} as opposed to the exponentially fast decay proved in the case without any conserved quantity \cite{Hayden-Preskill2007,Hayden2013FastScrambe,Yoshida-Kitaev2017,Majidy2023review}. It is, hence, of interest to study the case with SU$(d)$-conserved quantities in the late time regime and check if the non-Abelian nature could further contribute to the slow-mode relaxation by probing \emph{out-of time-ordered correlator} (OTOC) \cite{OTOC1969,roberts2016chaos,RobertsChaos2017} which is widely used in qualifying quantum scrambling dynamics by measuring the spread of local charges under the dynamics in Heisenberg picture. As the first result of this paper, we mathematically derive a power inverse lower bound $\Omega (1/n^{3/2})$ for the finite-size residual value of the OTOC due to the spreading of local conserved quantities with respect to SU($2$) charge density (which is equivalent to using the local Pauli basis to probe the quantity). We further show that the finite-size residual value between two local conserved quantities via OTOC is  $\Omega(1/n^{(d+2)^2/2})$ for SU($d$)-symmetry on general qudits. 

In the case where the symmetry acts transversally on local degrees of freedom, the symmetry actions can be understood as transversal implementations of logical gates in the context of quantum error correction. Codes that respect such symmetries are called \emph{covariant codes}. For continuous symmetries like SU$(d)$ and U($1$), the Eastin--Knill theorem prevents the existence of covariant codes admitting perfect recovery without errors \cite{eastin2009restrictions}. Nevertheless, it is still possible to find approximate recovery schemes with approximate Eastin--Knill theorem \cite{faist20,Woods2020continuousgroupsof,Kubica21,Yang22,Zhou2021newperspectives}. It claims a fundamental limit in terms of infidelity which asymptotically scales as $O(1/n)$ where $n$ is the number of physical qudits. One question of great interest is the existence of optimal quantum error correction codes that would saturate this fundamental limit with the presence of symmetry. We improve upon the previous construction of optimal SU$(d)$-covariant random codes \cite{kong2022near} to encode arbitrary $k$ logical qudits. We prove that for $k$ which does not scale with the number $n$ of physical qudits, the code is asymptotically optimal to the fundamental limit against one-qudit erasure error. However, we provide numerical evidence that the code ceases to be optimal if we assume an overlarge coding rate against one-qudit erasure error.

Moreover, random circuits are currently applied to the study of the convergence rate of the near-term variational algorithms through \emph{quantum neural tangent kernel} (QNTK) \cite{Liu2022QNTK,Liu2023QNTK}. In variational quantum algorithms and quantum machine learning, QNTK indicates the theoretical predictions of the convergence efficiency during the gradient descent process for given variational circuits. For a wide array of problems, especially for applications in physics, designing a variational ansatz that respects the underlying symmetry is crucial because it not only reflects the physical properties but is also observed to significantly improve the convergence speed and require much fewer parameters \cite{sauvage2022building, Zheng2021SpeedingUL}. It is speculated that the superior performance of symmetric-adaptive variational algorithms likely stemmed from the reduced effect of the barren plateau from its reduced effective dimension of the dynamical Lie algebra. For this reason, a recent subfield called geometric quantum machine learning has emerged \cite{ragone2022representation, nguyen2022theory}, and it is of practical interest to study the convergence theory of near-time quantum variational algorithms that respect the underlying symmetry. To this end, we derive an overparametrization threshold for the existence of exponential convergence whose parameters scale with the dimension of a particular charge sector decomposed from the entire Hilbert space in which the information is encoded under SU$(d)$ symmetry and, more generally, under any continuous symmetry generated by Lie algebra or discrete cases like permutation symmetry. This result generalizes many previous studies on the theoretical convergence on permutation invariant ansatz to general symmetries.  

In this paper, we systematically investigate the above aspects in the presence of SU$(d)$ symmetry. In Section \ref{sec: backgrounds} we provide necessary backgrounds regarding random unitary circuits and SU$(d)$ symmetry. In Section \ref{sec: late-time-tail} we explain the power inverse lower bound of the OTOC under SU$(d)$ symmetry. In Section \ref{sec: codes} we present the optimal SU$(d)$-covariant code encoding $k$ qudits provided that $k$ does not scale up with $n$. In Section \ref{sec: qntk}, we calculate the overparametrization threshold for QNTK under general compact group symmetry and specify its applications to SU$(d)$ and $S_n$ symmetry. Our results may invite future research in several directions. On a fundamental level in quantum information, they motivate the study of the information scrambling from local interactions such as the local random circuits under non-Abelian conserved quantities. We anticipate that further work on entanglement generation and operator spreading with the non-Abelian conserved quantities would shed light on this phenomenon especially compared with the results in \cite{majidy2023non}, and in the monitored random circuits under SU$(d)$ symmetry. In addition, our work might connect with recent works on the fundamental limitation of recovering fidelity from the random circuits with a conservation law \cite{Tajima2021,Tajima2022, faist20,Nakata2023}, especially through the lens of studying the mutual information (see in Eq.\eqref{eq: main-fidelity} in Section \ref{sec: backgrounds}). Furthermore, we expect that the results would find towards the ongoing effort in geometric quantum machine learning by providing insights on its convergence criterion and potential robustness to noise and barren plateau. All calculations and proof details as well as necessary mathematical backgrounds are pathologically presented in the Appendix. 

%---------------------------------------------------------------------------------------------

\section{Entanglement, decoupling, and OTOC under SU$(d)$ Symmetry} \label{sec: backgrounds}

The above questions can be mathematically formulated in the same line through the concepts of OTOC and entanglement (mutual information). This connection is, perhaps, not surprising since all applications concern the questions of information detecting and recovery post scrambling. We borrow tools from Refs.~\cite{Zheng2021SpeedingUL,SUd-k-Design2023} developed to study these concepts under SU($d$) symmetry. In the case of SU$(d)$ symmetry, the Hilbert space $\mathcal{H}$ is decomposed according to the irreducible representation (irrep) of the symmetric group as:
\begin{align}\label{eq:SchurWeyl}
	\mathcal{H} = \bigoplus_{\lambda \vdash n } \operatorname{1}_{ m_{\lambda}} \otimes S^\lambda,
\end{align}
where $S^\lambda$ stands for an irrep with $\lambda \vdash n$ recording the irrep as a partition of the number $n$ of qudits in the system \cite{Sagan01,Tolli2009}. The number $m_\lambda$ denotes the multiplicity of $S^\lambda$ and $\dim S_\lambda \equiv d_\lambda$ is its dimension. A Haar SU$(d)$-symmetric random unitary $U \equiv \bigoplus_{\lambda \vdash n} I_{m_\lambda} \otimes U_\lambda$ with $U_\lambda \in \text{U}(S^\lambda)$ is drawn from the compact group $\mathcal{H}_{\times}$ of SU$(d)$-symmetric unitaries under Haar measure, whose balance-$k$th-order expander ($k$-design) is: 
\begin{align} \label{eq: k-design}
	T_k^{\text{Haar}} = \int_{\mathcal{U}_\times} dU U^{\otimes k} \otimes \bar{U}^{ \otimes k}.
\end{align}
Explicit expression of the above integral (presented in Appendix \ref{sec:Setting}) is at the center stage of most technical calculations in this paper for it connects well with OTOC and entanglement which we pay significant attention to. The relationship between OTOC and entanglement is explained gradually in the following context. As a reminder, even though they are tightly related physical quantities to probe the scrambling of quantum information, it is shown that separation of scales between these two quantities exists \cite{harrow2021separation}.

\begin{figure}[!ht]
	\centering
	\includegraphics[width=\linewidth]{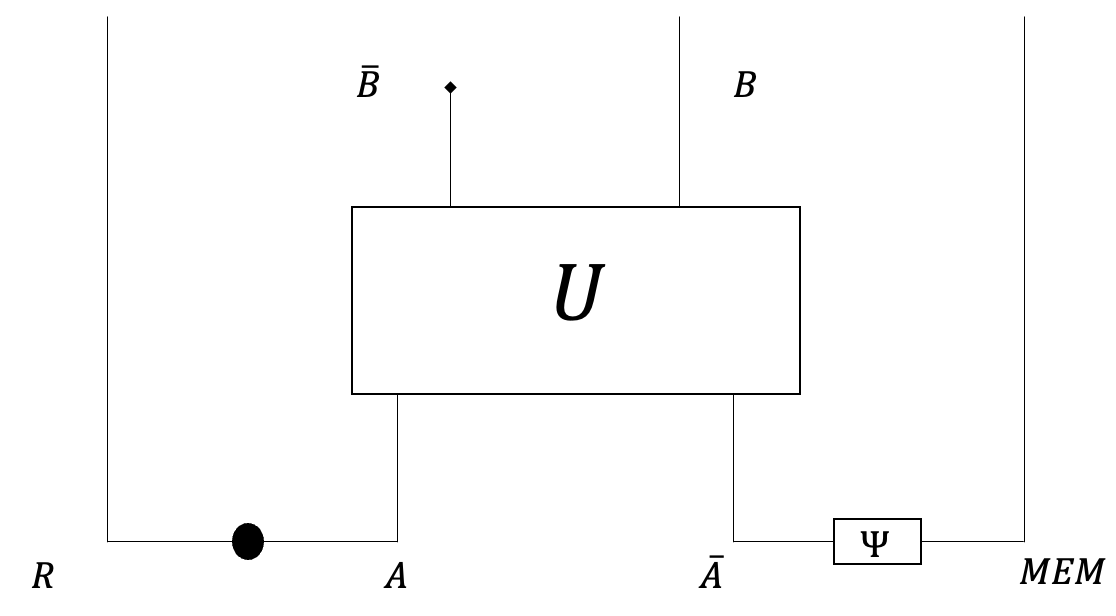}
	\caption{Hayden-Preskill decoupling as a general set-up where $R$ and $A$ form a maximally entangled pair and $\Psi$ is a generic density matrix in $\bar{A}$ which is purified with the memory register $\operatorname{MEM}$. In the case that $\Psi$ itself is a pure state, we omit the MEM register for the case of SU$(d)$-covariant codes.}
	\label{fig:hayden-preskill}
\end{figure}

To begin with, suppose that Alice prepares a Bell pair $\ket{\phi^{RA}} = \frac{1}{\sqrt{d_A}}\sum_i \ket{i}_A \ket{i}_R$ between the register $R$ and $A$ each endowed with $k$ qudits of the system with $d_A = d^k$ the dimension of $A$. Alice then transmits her information via the channel: 
\begin{align}
	\Phi^{RA} \mapsto U (\Phi^{RA} \otimes \Psi) U^\dag
\end{align}
which is a CPTP map given a proper density matrix $\Psi$ (which could be purified by adding memory qudits). At some time Bob may measure part of the system of $t$ qudits denoted by $B$ (as well as the memory) so that Bob might in principle extract information from the state: 
\begin{align}
	\rho_{B \cup \text{MEM}} = \operatorname{Tr}_{R \cup \bar{B}}U (\Phi^{RA} \otimes \Psi) U^\dag
\end{align}
A simple qualification of Bob's detection on Alice's information is to calculate the mutual information $I(R : B \cup \text{MEM}) = S(R) + S(B \cup \text{MEM}) - S(R \cup B \cup \text{MEM})$, and we say that Bob successfully detects all Alice's information if $I(R \cup B \cup MEM)$ saturates to its maximum value $2k$. This scenario depicts the so-called Hayden-Preskill protocol \cite{Hayden-Preskill2007}, which has many implications in black hole physics and quantum gravity. With recent developments in black hole information, we can see the correspondence between a black hole and the setup in Fig.\ref{fig:hayden-preskill}. To obtain the Page curve for the process of black hole radiation (working in the back ground of AdS/CFT), people isolated the radiation of the black hole from the CFT, which were usually called the reservoir \cite{penington2020entanglement} and then adding a bulk entanglement term to the Ryu-Takayanagi proposal \cite{engelhardt2015quantum, chen2022quantum}. It should be clear that this corresponds to the state $\Psi$ and MEM part in Fig.\ref{fig:hayden-preskill}. Then a physical reason that we can choose $\Psi$ to be pure is that we consider a black hole that were formed from a collapse of a pure state of matter consisting billions of atoms. The reservoir is entangled with the black hole interior which, at last, turns into the island. In Fig.\ref{fig:hayden-preskill}, $\bar{B}$ can be considered as the island, since after the action of a random unitary, we should be looking at an old black hole. As a result, the information recovery discussion here has many physical implications in AdS/CFT correspondence \cite{penington2020entanglement, almheiri2019entropy}, black hole information paradox \cite{hayden2016holographic, Hayden-Preskill2007, engelhardt2015quantum, almheiri2019islands}, and quantum many-body teleportation \cite{schuster2022many}. 

To lower-bound the mutual information, we may replace von Neumann entropy by Rényi entropy $S^{(2)}(\rho) = -\log \text{Tr}(\rho^2)$ as a lower bound to appraise the information recovery scheme, which also enjoys a more convenient computation integrating over the concerned group. Assume $\Phi^{RA},\Psi = \Psi^{\bar{A} \ \text{MEM}}$ are all given by Bell pairs, then
\begin{align}
	\rho_{\bar{B}} = \operatorname{Tr}_{R \cup B \cup \text{MEM}} U (\Phi^{RA} \otimes \Psi) U^\dag = \frac{I_{\bar{B}}}{d^{n-t}}
\end{align}
and one can check by Schmidt decomposition that $S^{(2)}(\bar{B}) = S^{(2)}(R \cup B \cup \text{MEM})$. Since von Neumann entropy $S(B \cup \text{MEM})$, in general, dominates $S^{(2)}(B \cup \text{MEM})$, $I(R: B \cup \text{MEM}) \geq I^{(2)}(R: B \cup \text{MEM})$ and we define the information recover fidelity under Haar average as (see computational details in Appendix \ref{sec:Setting})
\begin{align}\label{eq: main-fidelity}
	& \mathbb{E} \big[S^{(2)}(R) + S^{(2)}(B \cup MEM) - S^{(2)}(R \cup B \cup MEM) \big] \notag \\
	\geq & 2(n+k) - \log \Big(\sum_{P_A P_B} \int dU \operatorname{Tr}(\widetilde{P}_A^\dag P_B^\dag \widetilde{P}_A P_B ) \Big)
\end{align}
where $\widetilde{P}_A \equiv U P_A U^\dag$ for a given Haar random unitary and Pauli strings on $A$. We explicate in Eq.\eqref{eq: late-time-otoc} in the following that this lower bound can be obtained computing a sum of OTOC with respect to the Pauli basis elements on the respective subsystems.

Alternatively, since $I(R : B \cup \text{MEM}) + I(R : \bar{B}) = 2k$ due to Schmidt decomposition, we could observe whether the mutual information between the register and Bob's inaccessible region $\bar{B}$ is small. Then it is equivalent to lower bound the mutual information $I(R: \bar{B}) = D(\rho_{\operatorname{R \cup \bar{B}}} \| \sigma_{R \cup \bar{B}})$ equal to the quantum relative entropy with respect to the product state\begin{align}\label{eq:complementary}
	\sigma_{R \cup \bar{B}} = \rho_R \otimes \rho_{\bar{B}},
\end{align}
where $\rho_R$ and $\rho_{\bar{B}}$ are the respective reduced density matrices on the subsystem $R$ and $\bar{B}$, whose mutual information is zero in either von Neumann or Rényi sense. Therefore, mutual information is useful in probing any residual entanglements between $R$ and $\bar{B}$: the smallness of $I(R: \bar{B})$ can be seen as a necessary condition for the decoupling given by the trace norm distance or purified distance between $\rho_{R \cup \bar{B}}$ and $\rho_R \otimes \rho_{\bar{B}}$, as a consequence of the continuity relation of the \emph{Fannes–Audenaert inequality} (see Appendix \ref{sec:entropy}). This leads to the decoupling equation in Hayden-Preskill protocol \cite{Hayden-Preskill2007,junyu2020chargescrambler}, Page theorem \cite{page1993average} and the use of complimentary channel formalism in quantum error correction \cite{Beny2010}. Depending on the choice of norms such as the Schatten-$p$ norm or purified distance, the computation would resort to applications of $2$-designs and $1$-designs. In particular, as we show in Section \ref{sec: codes}, we need to compute the following quantities: 
\begin{align}\label{eq: codes-2-design}
	\sum_{P_A, P_B} \int dU \operatorname{Tr} \big( (P_A^\dag \otimes \Psi) \widetilde{P}_B^\dag (P_A \otimes \Psi) \widetilde{P}_B \big).
\end{align}
Simplification can be made if $\Psi$ is further selected as a pure state (so there is no need for MEM states to purify) and then we obtain the following expression during the calculation process: 
\begin{align}
	\frac{1}{d_{\bar{B}}^2 - 1} \frac{1}{d^n} \sum_{P_{\bar{B}} \neq I} \int dU \operatorname{Tr}(\widetilde{\Psi} P_{\bar{B}}^\dag \widetilde{\Psi} P_{\bar{B}}). 
\end{align}

The pure state $\Psi \equiv \ket{\psi} \bra{\psi}$ defined on a sub-region $\bar{A}$ has no locality assumption. The above in mathematical form suggests similarity to another physical quantity: out-of-time-ordered commutator which might generally overestimate how fast the information propagates than that of entanglement \cite{Hayden2013FastScrambe,harrow2021separation, Swingle2022}. Given a chaotic Hamiltonian $H$ the operator growth of a local observable under the Heisenberg picture $W(t) \equiv e^{iHt}We^{-iHt}$ is detected by the non-commutativity at certain observables at different sites initially. For the random circuit model, it is common to discretize the notion of time such as in the Brickwork model where each layer represents one time step. The non-commutativity is measured by the out-of-time-ordered commutator (OTOC): 
\begin{align}\label{eq: oto-commutator}
	C^{WV}(t) & = \frac{1}{2} \operatorname{Tr}\big( [W(t), V_r]^{\dagger}[W(t), V_r] \big) \notag \\
	& = 1 - \operatorname{Tr}\big( W(t)^\dag V_r^\dag W(t) V_r \big) \\
	& =  1 - F^{WV}(r, t), \notag
\end{align}
where $V_r$ is a local traceless and normalized observable at site $r$ such that $\operatorname{Tr}(V^\dag_r V_r)$ equals one. At sufficiently large time post scrambling, we expect that the circuit would become Haar random or at least be unitary $k$-design (matching the $k$-th moment of the Haar randomness \cite{Gross2006,Dankert2006PRA}). Then the OTOC may be reformulated as $F^{WV}(r) = \lim_{t \rightarrow \infty } F^{WV}(r, t)$ and we call it the \emph{finite-size residual value} detected by the OTOC.

To obtain a more precise understanding, it is useful to think about the growth of the Heisenberg operator $W(t)$ under a complete basis of operators, e.g., from the (generalized) Pauli group $\mathcal{P}_n = \{\langle X^i Z^j; i,j = 0,...,d-1 \rangle \}^{\otimes n} / \langle \omega \mathbf{1}_{d^n} \rangle $ on qudits where $\omega = \exp(i 2 \pi k /d)$ for $k=0,...,d-1$, which obeys the following normalization and completeness properties for $\mathcal{S} \in \mathcal{P}_n$. 
\begin{align}
	\begin{aligned}
		&\frac{1}{\operatorname{tr} I} \operatorname{tr}\left(\mathcal{S}^{\dagger} \mathcal{S}^{\prime}\right)=\delta_{\mathcal{S} \mathcal{S}^{\prime}}, \\
		&\frac{1}{\operatorname{tr} I} \sum_{\mathcal{S}} \mathcal{S}^{\dagger} \otimes \mathcal{S} = \Pi,
	\end{aligned}
\end{align}
where $\Pi$ is the SWAP operator between the first and second system, i.e., $\Pi \ket{i, j} = \ket{j, i}$ for $i,j \in [d^n]$. There are in total $d^{2n}$ different Pauli strings for $n$ qudits. When $d = 2$ for qubits, this recovers the familiar Pauli matrices with identity $\sigma^x, \sigma^y, \sigma^z, I$. Then the Heisenberg operator $W(t)$ can be expanded under the basis as
\begin{align}
	W(t) = \sum_{\mathcal{S}} \operatorname{Tr}(\mathcal{S}^\dag W(t)) \mathcal{S}  = \sum_{\mathcal{S}} \alpha(\mathcal{S},t) \mathcal{S}.
\end{align}
Obviously $V_r = I$ provides no nontrivial dynamics, we hence define the OTOC at late time by 
\begin{align} \label{eq: late-time-otoc}
	F^{W}(r) = \frac{1}{d^n} \frac{1}{d^2 -1} \sum_{V_r \neq I} \int_{\mathcal{U}_\times} dU  \operatorname{Tr}(\widetilde{W}^\dag V^\dag_r \widetilde{W} V_r),
\end{align}
where $V_r$ are taken to be nontrivial Pauli matrices supported on the site $r$  with asymptotically Heisenberg operator $\lim_{t \rightarrow \infty}W(t)$ being replaced by its Haar random dynamics $\widetilde{W} = U W U^\dag$ based on the assumption that we enter the scrambled regime that satisfies at least $2$-design at late time. Up to coefficients, this definition uncovers the physical meaning of the lower bound of mutual information in our previous discussion. Without the presence of symmetry, the residual value $F^{WV}(r)$ given by the OTOC is exponentially small in the system size $n$ and independent of $r$ as a consequence of non-locality \cite{Hayden2013FastScrambe,Yoshida-Kitaev2017,roberts2016chaos,RobertsChaos2017,Swingle2022}. The intuition behind this exponential suppression of residual value in OTOC can be given in the following: $\widetilde{W}$ in the fully scrambled regime has support on all sites so that its expansion coefficients associated with the Pauli basis behave like Gaussian random variables from the law of large numbers. Hence, given $V_r$ to be a single Pauli group element, half part of the expansion in Eq.\eqref{eq: late-time-otoc} would anticommute and half commute so that they finally cancel. However, we present in Section \ref{sec: late-time-tail} that the late-time spreading of local SU$(d)$ conserved quantities scale in an inverse power law with respect to the number $n$ of sites in the system, which agrees with the generic observation of scrambling in the presence of conservation law \cite{rakovszky2018diffusive, Huang2019OTOC,yoshida:soft,junyu2020chargescrambler}. 

To be noted, analytical computation to be above equations in the presence of SU$(d)$ conservation law is highly non-trivial. Due to the generally super-polynomial scaling of the number of irreps $\lambda \vdash n$, the number of basis elements spanning even $2$-design commutant is intractable (see Appendix \ref{sec:SnCommutant} for more details). To tackle the issue, we borrow techniques from Ref.~\cite{Zheng2021SpeedingUL,SUd-k-Design2023} where we employ group representation-theoretical tools such as the Okounkov-Vershik approach \cite{Onishchik1990}. Additional information can be found in the self-contained Appendix provided in this paper.

%---------------------------------------------------------------------------------------------

\section{Results}

\subsection{Late-time Saturation of SU$(d)$ Conserved Quantities} \label{sec: late-time-tail}

The late-time hydrodynamics of other classes of chaotic systems with conservation laws has drawn significant interest in physics lately. It was shown \cite{Huang2019OTOC} that for general energy-conserving quantum chaotic Hamiltonians assuming the eigenstate thermalization hypothesis (ETH), the OTOC for finite-size systems has a residual that scales as $O({1}/{\operatorname{poly}(n)})$ at late time. More precisely, the energy conservation in this setting is defined by imposing symmetry conditions on partitions of spectrum range by the energy difference $\Delta$. The operator growth or transport of local charges under e.g.~U($1$) symmetry has received significant attention. It has been shown that the conservation law slows relaxation in OTOC,  inducing a ``hydrodynamic tail'' at late times \cite{rakovszky2018diffusive}. In the post scrambling regime $t \rightarrow \infty$, the locality is lost due to the assumption of Haar randomness (or at least being $2$-design) under U($1$) conservation. In this regime, it has been shown that the finite-size residual values between the local charges $Z$ and the ``raising charges'' (defined by eigenmatrix of the adjoint action of the total charge $Z_{\operatorname{tot}}$) scale $O(1/n)$ (see also Appendix \ref{sec:U(1)charge} for more details).

The SU($d$) symmetry is generated by a set of non-commuting charges from the elements of the Lie algebra $\mathfrak{su}(d)$. The non-commuting nature implies that our charge sectors cannot exactly correspond to the spectrum of these conserved charges but rather its Casimir operators. The transversal action of the symmetry group SU$(d)$ then implies there exists a sequence of coupled Casimir operators. For instance, in the case of SU($2$) action, they are built by sequential coupling $\{\vec{S}_1 \cdot \vec{S}_2, \vec{S}_1\cdot \vec{S}_2 \cdot \vec{S}_3, \vec{S}_1 \cdot \vec{S}_2 \cdot \vec{S}_3 \cdots \vec{S}_n \}$ of spin operators \cite{pqc, sergii1}. Hence, the first sharp contrast between Abelian group symmetry and non-Abelian symmetry group is that the spectrum of conserved charges (or Lie algebra generators) in the latter case cannot explicitly correspond to charge sectors so that generally conserved quantities are \emph{not} the same as conserved charges. 

Here we define the conserved quantities to be elements generated by the exchange interactions $\vec{S}_i \cdot \vec{S}_j$ acting on the $(i, j)$-pair of qudits. We pay special attention to the dynamics of the $2$-local conserved quantities given by the exchange interactions. In particular, we consider the OTOC between the \emph{charge density} and exchange interactions under SU($2$) symmetry action where the charge density is given by the non-identity single-qubit Pauli elements. We also consider the OTOC between exchange interactions for general transversal SU($d$) symmetry. At the first and second qudits, it reads 
\begin{align}
	W = \frac{1}{\sqrt{d^2 - 1}} \vec{S}_1 \cdot \vec{S}_2 = \frac{1}{\sqrt{d^2 - 1}} \sum_{P \neq I} P \otimes P^\dagger
\end{align}
where $P$ are generalized Pauli matrices acting on one qudit. In the case of $d = 2$ we have $\vec{S}_i = (X_i, Y_i, Z_i)$. The first result of this paper is a rigorous mathematical verification that under SU$(d)$ symmetry the late-time OTOCs of these local conserved quantities scale as an inverse polynomial in contrast to exponential decay when there is no symmetry. At late time we work with the Heisenberg operator $\widetilde{W} = U W U^\dag$ where $U$ is randomly taken from SU$(d)$-symmetric unitaries. As mentioned at the end of Section \ref{sec: backgrounds}, we take a local basis of operators to probe the information. We show that when evolving with SU$(d)$-symmetric random unitaries obeying the Haar distribution, 
\begin{align}\label{eq: op-spreading-Pauli}
	\frac{1}{2^n} \frac{1}{3} \sum_{P_r \neq I} \int_{\mathcal{U}_\times} dU  \operatorname{Tr}(\widetilde{W}^\dag P^\dag_r \widetilde{W} P_r) = \Omega(\frac{1}{n^{3/2}}).
\end{align}
This contrasts the case with conservation law as first noted by \cite{Huang2019OTOC} with random unitaries with energy conservation. One would expect that under symmetry, the operator $\widetilde{W}$ cannot have support over entire systems. For instance, in the case of U($1$) symmetry, any conserved quantities would only have non-trivial support on local conserved quantities \cite{fisher2023random}. 

We investigate finite-size residual value via OTOC between two $2$-local conserved quantities (exchange interactions), which also constitute the smallest non-trivial $2$-local SU($d$)-symmetric basis elements. Denote $V_r \equiv (1/\sqrt{d^2 -1} )\vec{S}_r \cdot \vec{S}_{r+1}$ such that
\begin{align}
	\text{Tr}(V_r I) = 0, \ \text{Tr}(V_r^\dagger V_r) = d^n \text{ and } [V_r, U^{\otimes n}] = 0.
\end{align}
Then we prove the following lower bound for OTOC at late time: 
\begin{align}
	\frac{1}{d^n} \int_{\mathcal{U}_\times}  dU  \operatorname{Tr}(\widetilde{W}^\dag V^\dag_r \widetilde{W} V_r) = \Omega(\frac{1}{n^{(d+2)^2 / 2}} ),
\end{align}
by making explicit use of the expansion of Eq.\eqref{eq: k-design} for SU$(d)$-symmetric 2-deign as well as well-known facts like $S_n$ characters theory (see Appendix \ref{sec:Charge} for more details). In the special case of qubits ($d = 2$), the lower bound can be further improved to $\Omega (1/n^{3})$. 

In both cases, a nonvanishing finite-size residual OTOC is present which only exhibits a power law decay with respect to $n$. It is interesting to observe that OTOC with respect to the local SU($2$)-symmetric basis with local dimension $2$ has a faster mode of decay, which may indicate that there is further residual information that is leaked out to the charge sectors which is not probed by the locally symmetric basis. In summary, we lower bound the OTOC of SU$(d)$-symmetric conserved quantities given by exchanging interactions under both Pauli basis and SU$(d)$-symmetric local basis. The calculation with respect to the charge density (or equivalently single-quibt Pauli group element), although using a similar strategy such as that under the symmetric basis, is more complicated. It is conceivable that in order to integrate over the group $\mathcal{U}_\times$ of SU$(d)$-symmetric unitaries, a sensible way is looking at the charge sector decomposition in Eq.\eqref{eq:SchurWeyl} and seek to perform the integration on each inequivalent irrep block $S^\lambda$. However, since $V_r$ does not obey the SU$(d)$ symmetry, its matrix representation does not fit into the charge sector decomposition and poses a challenge for our computation. We employ mathematical methods used in quantum angular momentum (see Appendix \ref{sec:Charge} for more details) for the qubit ($d=2$) case and leave the computation of general qudits for future work. Note that these lower bounds are not necessarily tight. Further analysis of the quantum hydrodynamics with non-Abelian conserved quantities would be worthwhile. 

%--------------------------------------------------------------------------------------

\subsection{Near-optimal SU$(d)$-covariant Codes}\label{sec: codes}
 
Random unitaries exhibit good error-correction and decoupling properties. In the presence of continuous symmetries, U($1$) and SU$(d)$-symmetric random unitaries generate nearly optimal covariant error-correcting codes \cite{kong2022near} that they saturate the fundamental limits of the error correction imprecision with scaling $O(\frac{1}{n})$ as identified by the approximate Eastin--Knill theorems (see e.g.~ \cite{faist20,Woods2020continuousgroupsof,Kubica21,Yang22,Zhou2021newperspectives}. A key feature for non-perfect error correction is due to the fact that there is always logical information leaking into the environment even encoding only one logical qudit and against one-qudit erasure error. We slightly improve the SU$(d)$-covariant random codes by encoding $k$ logical qudits while still achieving $O(1/n)$ in averaged Choi error asymptotically against single qudit erasure. The actual computation only requires the SU$(d)$-covariant codes to satisfy a $2$-design condition and it is proved in Ref.~\cite{CQAconvergence2024} that there are symmetric local ensembles capable of converging to this SU$(d)$-symmetric 2-design in polynomials steps. This provides further motivation to study the SU$(d)$-covariant random codes, especially on certain natural physical platforms \cite{divincenzo2000universal, gao2018entangling} since our encoding of $\Psi$ on $\bar{A}$ can be taken as pure states for all local dimension $d$ (see \cite{Zheng2021SpeedingUL}). Concerning quantum information recovery especially in physical contexts, the decoding procedures through e.g.~Kitaev--Yoshida decoding protocol in the presence of charge conservation \cite{Yoshida-Kitaev2017,yoshida:soft,junyu2020chargescrambler,Nakata2023} remain to be studied. 

We adopt the complementary channel formalism \cite{Beny2010,kong2022near}. We partial trace out $B$ instead of $\bar{B}$ in the discussion around Eq.\eqref{eq:complementary} and obtain 
\begin{align}
	\rho_{R \cup \bar{B}} = \operatorname{Tr}_{B} (U (\Phi^{RA} \otimes \Psi^{\bar{A}}) U^\dag)
\end{align}
from the encoding protocol. We also assume that $\Psi^{\bar{A}}$ is a pure and SU$(d)$-symmetric state. It is straightforward to check that this defines an \emph{SU$(d)$-covariant} encoding map in the following sense:
\begin{align}
	\begin{aligned}
		\hat{U}^{\otimes n} & \big( U (\Phi^{RA} \otimes \Psi^{\bar{A}} ) U^\dagger \big) \hat{U}^{\dagger \otimes n} \\
		& = \big( U (\hat{U}^{\otimes k} \Phi^{RA} \hat{U}^{\dagger \otimes k} \otimes \Psi^{\bar{A}}) U^\dagger \big),
	\end{aligned}
\end{align}
where $\hat{U}^{\otimes k}$ is the transversal action of the group SU$(d)$ acting on $A$ of $k$ qudits. 

Then we compute Choi error against the decoupled states $\frac{I}{d_A} \otimes \zeta$ where $\zeta $ is some quantum state in the environment $\bar{B}$ that the erasure error occurs:
\begin{align}\label{eq: choi-error}
	\epsilon_{\operatorname{Choi}} = \min_{\zeta} P( \rho_{R \cup \bar{B}}, \frac{1}{d_A} I_R \otimes \zeta )
\end{align}
The quantity $P(\rho, \sigma) = \sqrt{1 - F(\rho, \sigma)^2}$ is called \emph{purified distance} with $F(\rho, \sigma) = \operatorname{Tr}\sqrt{\sqrt{\sigma} \rho \sqrt{\sigma}}$ being the fidelity between two density matrices. The following inequality with the 1-norm distance also holds:
\begin{align}
	\frac{1}{2}\|\rho-\sigma\|_1 \leq P(\rho, \sigma) \leq \sqrt{2\|\rho-\sigma\|_1}.
\end{align}
We can bound the Choi error Eq.\eqref{eq: choi-error} by triangle inequality:
\begin{align}\label{eq: choi-error-triangle}
	& \min_{\zeta} P( \rho_{R \cup \bar{B}}, \frac{1}{d_A} I_R \otimes \zeta ) \\
	\leq & P( \rho_{R \cup \bar{B}}, \rho_{R \cup \bar{B}, \text{avg}} ) + \min_{\zeta} P(\rho_{R \cup \bar{B}, \text{avg}}, \frac{1}{d_A} I_R \otimes \zeta) \notag \\
	\leq & \sqrt{2\Vert \rho_{R \cup \bar{B}} - \rho_{R \cup \bar{B}, \text{avg}} \Vert_1} + \min_{\zeta} P(\rho_{R \cup \bar{B}, \text{avg}}, \frac{1}{d_A} I_R \otimes \zeta), \notag
\end{align}
with $\rho_{R \cup\bar{B}, \text{avg}} = \int \rho_{R \cup \bar{B} } dU$ being averaged over the group of SU$(d)$-symmetric unitaries. To find the expectation of $\epsilon_{\operatorname{Choi}}$, we integrate the above inequality under SU$(d)$-symmetric Haar distribution and the integral only uses its first and second moments (1- and 2-designs). Instead of using the partial decoupling theorem \cite{Wakakuwa2021Decoupling}, we can analytically compute the first term from the triangle inequality Eq.~\eqref{eq: choi-error-triangle} thanks to the fact that $\Psi^{\bar{A}}$ is a pure state, which we show the techniques in Appendix \ref{sec:Covariant} when $k,t = o(n)$,
\begin{align}\label{eq: sud-page}
	& \mathbb{E} [\sqrt{\Vert \rho_{R \cup \bar{B}} - \rho_{R \cup \bar{B}, \text{avg}} \Vert_1} ]
	\leq \sqrt{\mathbb{E} [ \Vert \rho_{R \cup \bar{B}} - \rho_{R \cup \bar{B}, \text{avg}} ] \Vert_1} \notag \\
	\leq & \sqrt{ (d_R d_{\bar{B}})^{1/2} \big(\mathbb{E} [ \text{Tr} (\rho_{R \cup \bar{B}}^2)] - \text{Tr}(\rho_{R \cup \bar{B}, \text{avg}}^2) \big)^{1/2} } \\
	\leq & \frac{1}{e^{\Omega(n)}}. \notag
\end{align}
Similar to the familiar Page theorem \cite{page1993average}, the exponential rate of suppression is achieved, which is due to the selection of relevant charge sectors whose dimension scales exponentially with respect to the system size $n$ (see Appendix \ref{sec:SchurWeyl} for more details). 

If the erasure environment $\bar{B}$ contains a single qudit, we have
\begin{align}\label{eq: rho-avg}
	\begin{aligned}
		\rho_{\operatorname{avg}}(R \cup \bar{B}) = \frac{1}{n} &\sum_{a=1}^k \frac{1}{d^{k-1}} I_{R-a} \otimes\left|\Phi_{a, \bar{B}}\right\rangle\left\langle\Phi_{a, \bar{B}}\right| \\
		&+\frac{n-k}{n} \frac{1}{d^k} I_R \otimes \frac{1}{d} I_{\bar{B}},
	\end{aligned}
\end{align}
where the state $\left|\Phi_{a, \bar{B}}\right\rangle$ denotes the maximally entangled Bell pair between the $a$th qudit in the register $R$ and that erased qudit in $\bar{B}$. As expected, this is a slight generalization of the form given in \cite{kong2022near} with $k=1$ using a single Bell pair. By choosing $\zeta = \frac{I}{d_{\bar{B}}}$ we bound the second term in \eqref{eq: choi-error-triangle} by several matrix inequalities:
\begin{align}\label{eq:code-error}
	\mathcal{P} & (\rho_{R \cup \bar{B}, \text{avg}}, \frac{I}{d_A} \otimes \frac{I}{d_{\bar{B}}} )  \notag \\
	\leq & \sqrt{1 - \frac{1}{d^2} \Big( \sqrt{ \frac{k}{n} + \frac{n-k}{n} \frac{1}{d^2} } + \sqrt{ \frac{n-k}{n} \frac{1}{d^2}} (d^2 - 1) \Big)^2 } \notag \\
	\approx & \frac{k\sqrt{d^2 - 1} }{2n}.
\end{align}
\begin{figure}[!ht]
	\centering
	\includegraphics[width=\linewidth]{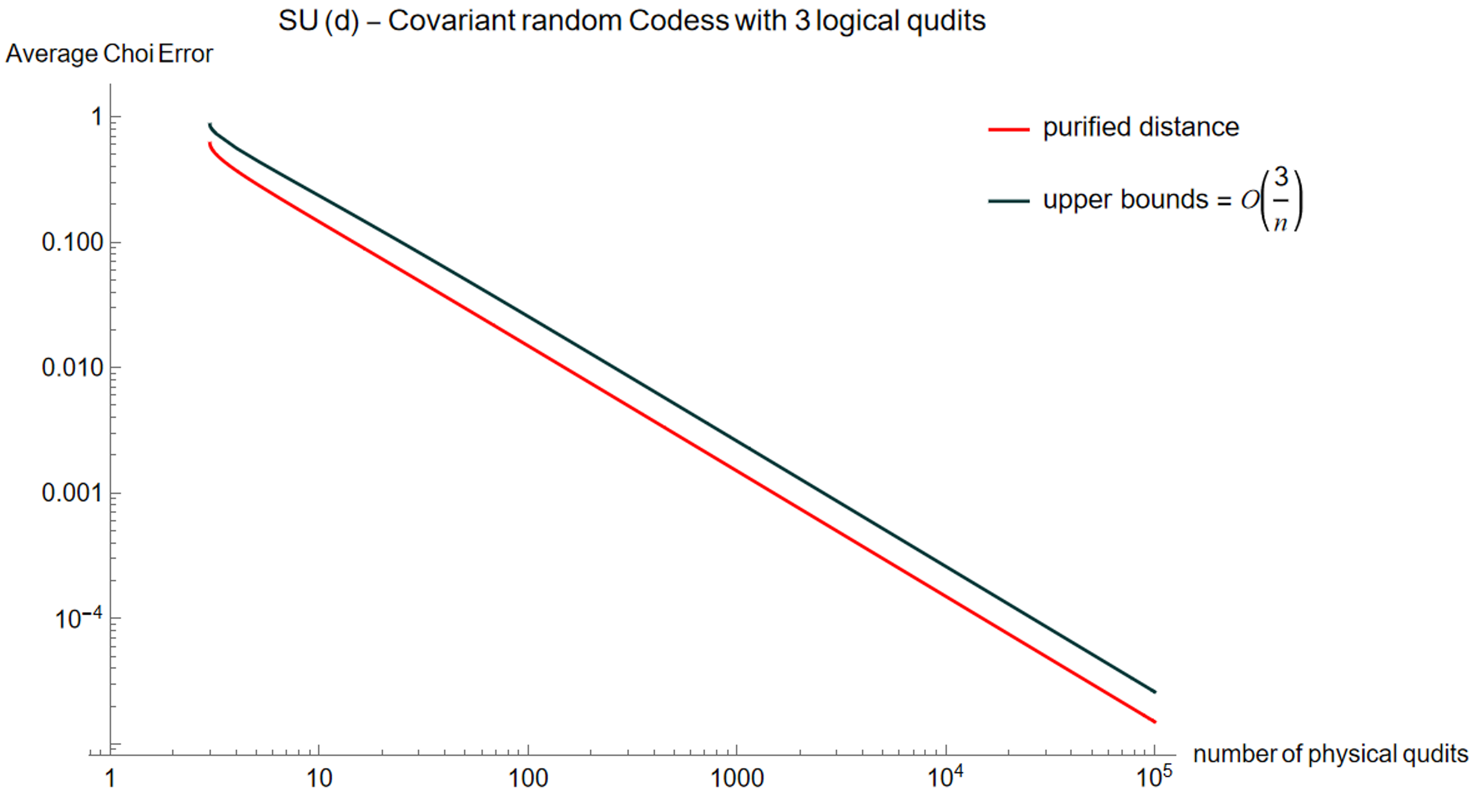}
	\caption{The log-log plot Purified distance between the decoupled states between $R$ and $\bar{B}$ and $\rho_{\operatorname{avg}}(R \cup \bar{B})$ with three logical qubits. The residual entanglement displayed is a consequence of the Eastin--Knill theorems and imperfect information recovery due to conservation law \cite{faist20,Tajima2021,Tajima2022, Nakata2023}}
	\label{fig:covariant codes}
\end{figure}
As expected, the inevitable leakage of information is due to the residual Bell pair coupled between the register $R$ and $\bar{B}$. For erasure beyond one qudit, the entanglement between register $R$ and $\bar{B}$ is likely to take a more complicated form and it would be interesting to investigate the optimality of SU$(d)$-covariant codes under multiple qudits erasure in future study. For now, we show that with pure state encoding $\Psi^{\bar{A}}$, the SU$(d)$-covariant codes which encode $k$ logical qudits against one qudit erasure are asymptotically optimal covariant codes saturating the scaling limits given by the approximate Eastin--Knill theorems \cite{faist20,Woods2020continuousgroupsof,Kubica21,Yang22,Zhou2021newperspectives}. We note that this upper-bound is tight and it might indicate that for a non-constant coding rate $k = O(f(n))$, the asymptotic optimality might lose. Indeed, if $k = O(f(n))$ for a non-constant $f(n)$, we have that 
\begin{align}
	\mathcal{P} (\rho_{\operatorname{avg}}(R \cup \bar{B})), \frac{I}{d^k} \otimes \frac{I}{d_{\bar{B}}} )   
    \geq \alpha \frac{f(n)}{n}
\end{align}
for some constant $\alpha$. Hence, as shown in Figure \ref{fig:covariant codes}, the SU$(d)$ random covariant codes fail to be close to the fundamental limit if the coding rate is non-constant. This stands a sharp contrast to random codes without symmetry, where the error is suppressed exponentially with $n_{\bar{A}} - n_{A} = n - 2k$.

%-------------------------------------------------------------------------------------------------

\subsection{Overparametrization Regime with Geometric Quantum Machine Learning} \label{sec: qntk}

The third application concerns a general class of geometric quantum machine learning \cite{ragone2022representation,PRXQuantum.3.030341,nguyen2022theory} models where the ansatze respect the underlying symmetry of the problem. With the success of the classical geometric and equivariant machine learning models, there has been a surge of interest in adapting symmetry to quantum machine learning ansatze \cite{Roth_2021,sauvage2022building, Zheng2021SpeedingUL, nguyen2022theory, meyer2022exploiting}. These symmetry-respecting or equivariant quantum machine learning ansatze can significantly outperform ones without symmetry in many tasks such as learning ground states of the frustrated antiferromagnetic Heisenberg model \cite{NetKet,Seki_2020,VieijraRBM,Vieijra_2021,Zheng2021SpeedingUL,sauvage2022building} and weighted graphs \cite{skolik2023equivariant}. Despite the superior performance and parameter efficiency from empirical observations, it is imperative to ask if there exists a theoretical guarantee for exponential convergence in the number of gradient descent steps or queries. This exponential convergence is highly desirable in the near-term application of QML and variational quantum eigensolver (VQE) in learning complex quantum many-body physics and beyond. For this, we generalize the quantum neural tangent kernel (QNTK) which states that, if the ansatz mimics up to the second moment of the concerned Haar distribution, i.e., achieves the unitary 2-design, then the exponential convergence would arrive at the overparametrization regime. 

To begin with, let us consider the variational CQA ansatz $U(\vec{\theta})$ that respects SU$(d)$ symmetry \cite{Zheng2021SpeedingUL}:
\begin{align}
	\begin{aligned}
		&U(\vec{\theta}) = \exp(-i H_{\text{YJM}}(\beta) ) \exp(-i H_S(\gamma) ) \\& =
		\exp(-i \sum_{k,l} \beta_{kl} X_k X_l) \exp(-i\gamma \sum_{j=1}^{n-1} (j,j+1) ),
	\end{aligned}
\end{align}
where $(j,j+1)$ are adjacent SWAPs on qudits and $X_k = (1,k) + (2,k) + \cdots + (k-1, k)$ is the so-called \emph{Young-Jucys-Murphy element}, or YJM-element for short \cite{Young1977,Jucys1974,Murphy1981,Okounkov1996}, which is essential in the study of SU$(d)$-symmetric universality theorem as well as $k$-designs \cite{Zheng2021SpeedingUL,SUd-k-Design2023}. The reason to incorporate second-order products $X_k X_l$ of YJM-elements is also explained in detail in these papers. Note that YJM-elements commute with each other, so $\exp(-i \sum_{k,l} \beta_{kl} X_k X_l) = \prod_{k,l} \exp(-i \beta_{kl} X_k X_l)$. Decomposing the second exponential of adjacent SWAPs, however, introduces Trotter errors. A initial state $\ket{\psi^\lambda} = \ket{\alpha_T^\lambda,m}$ is then taken from one $S_n$ irrep $S^\lambda$. We would also consider the statistical ensemble $\rho_\lambda$ of these states later.

For a given observable $\hat{O}$ that is SU$(d)$-symmetric, e.g., the Heisenberg Hamiltonian with the form 
\begin{align}
	\hat{O} = \sum_{i=1}^N O_i,
\end{align}
where each $O_i$ contains a single term. In  quantum many-body theory with locality assumption, it is typically the case that the number $N$ of $O_i$ scales linearly or polynomially with the number of qubits. Let the loss be defined as 
\begin{align}
	\begin{aligned}
		& \frac{1}{2} \Big( \text{Tr} (U(\theta) \rho_\lambda U^\dag(\theta)  \hat{O}) - E_0 \Big)^2 \\
		= & \frac{1}{2}\Big( \text{Tr} ( \rho_\lambda U^\dag(\theta) \hat{O} U(\theta) ) - E_0 \Big)^2 \equiv \frac{1}{2} \varepsilon^2,
	\end{aligned}
\end{align}
where $E_0$ is the ground truth label (normally real-or integer-valued) during a supervised learning process for regression and classification purposes ($E_0$ might subtly relate to the frozen kernel claim and could trigger phase transitions in quantum machine learning dynamics \cite{toappearq}). QNTK concerns the question of the number of iterations needed in order for a hybrid classical-quantum variational algorithm to converge. In other words, we could define each interaction along with its classical computing resources as a query and minimizing the number of queries is a key aspect in observing any potential quantum advantage. It is shown in Refs.~\cite{Liu2022QNTK,Liu2023QNTK} that an exponential convergence guarantee can be achieved when
\begin{align}
	\epsilon(t) \approx (1 - \eta K)^t \epsilon(0), 
\end{align}
if $K$ does not fluctuate too much around its mean $\bar{K}$ and for learning rate $\eta$ sufficiently small. The detailed concentration conditions for which the QNTK needs to satisfy are given in Refs.~\cite{Liu2022QNTK,Liu2023QNTK}. In the case where $K$ is sufficiently close to its average case, we say that we have reached the \emph{overparametrization regime} where an exponential convergence guarantee is observed if the resulting $\eta \bar{K}$ is of order $O(1)$. The threshold of the overparametrization regime accounts for how many variational parameters---a quantity hiding in $\bar{K}$---are needed to achieve this. We show that, very generally, under the assumption that the Hilbert space is decomposed into a direct sum of invariant controlled charge sectors with multiplicities, the overparametrization regime still occurs at $O(d_\lambda)$ where $d_\lambda = \dim S^\lambda$ is the dimension of the charge sector for which the initial states encode. 

The following computation actually holds for symmetries governed by general compact groups, e.g., the permutation symmetry defined through permuting qudits by the group $S_n$ or U$(1)$ symmetry, so let us rewrite a generic ansatz as 
\begin{align}
	U(\theta) = U_{-, l} U_{+, l},
\end{align}
where $l = 1, \cdots, L$ is the index for the variational angles. Note that our ansatz can be further Trotterized into products of the time evolution of unitaries and each of which is only parameterized by one variational angle. Then the differentiation with chain rule easily reads:
\begin{align}
	\begin{aligned}
		&	\frac{\partial U(\theta)}{\partial \theta_l} = U_{-, l} (-i H_l ) U_{+,l}, \\
		& \frac{\partial U^\dag (\theta)}{\partial \theta_l } = U^\dag_{+, l} (i H_l ) U^\dag_{-, l},
	\end{aligned}
\end{align}
where $H_l$ denotes the Hamiltonian generator of the ansatz driven by the parameter $\theta_l$. The QNTK is given by: 
\begin{align}
	K &= \sum_l \frac{\partial \bar{\varepsilon}}{\partial \theta_l}\frac{\partial \varepsilon}{\partial \theta_l}
	\\&= \sum_l \Big\vert \bra{\psi^\lambda} U^\dag_{+, l} \left[ H_l, U^\dag_{-, l} \hat{O} U_{-, l}  \right]  U_{+, l} \ket{\psi^\lambda} \Big\vert^2 \notag \\
	& = - \Big( \bra{\psi^\lambda} U^\dag_{+, l} \left[ H_l, U^\dag_{-, l} \hat{O} U_{-, l}  \right]  U_{+, l} \ket{\psi^\lambda} \Big)^2 \notag \\
	&= - \Big( \sum_{i=1}^N \bra{\psi^\lambda} U^\dag_{+, l} \left[ H_l, U^\dag_{-, l} O_i U_{-, l}  \right]  U_{+, l} \ket{\psi^\lambda} \Big)^2, \notag
\end{align}
where we assume that the observable $\hat{O}$ respects a certain symmetry whose matrix representation decomposes in a way resembling those in Eq.\eqref{eq:SchurWeyl} and $\ket{\psi^\lambda}$ (or $\rho_\lambda$) is taken from one charge sector, still labeled by $\lambda$ for brevity. Due to this choice, we only need to concern the unitary $U^\lambda$ restricted to that irrep and
\begin{align}
	\begin{aligned}
		K & = - \sum_l \sum_{i,j=1}^N \text{Tr}( \rho_\lambda U^{\lambda \dag}_{+, l} \left[  H_l^{\lambda}, O^{\lambda,I} _i \right]  \\
		& U^\lambda_{+, l} \rho_\lambda U^{\lambda \dag}_{+, l} \left[ H_l^{\lambda}, O^{\lambda,I}_j  \right]  U^\lambda_{+, l} ),
	\end{aligned}
\end{align}
where we define the interaction picture observable $\hat{O}^I = U^{\lambda \dag}_{-, l} \hat{O} U^\lambda_{-, l}$ (and similarly for $O^{\lambda, I}_i$) with general statistical ensemble $\rho_\lambda$. The assumption of respecting the symmetry is practical in the experiment which also provides an accessible way to average the above identity by Haar randomness. As explained in the previous context, twirling a general operator $M$ with no assumption on symmetry leads to the most intricate issues when we study OTOC and covariant codes. 

%--------------------------------------------------------------------------------------------------------

We put the detailed computation to obtain the average $\bar{K}$ in Appendix \ref{sec:QNTK}. In conclusion,
\begin{align}
	\begin{aligned}
		\bar{K} = & - \sum_l \int \Big( \int \text{Tr}( \rho_\lambda U^{\lambda \dag}_{+, l} \left[  H_l^{\lambda}, O^{\lambda,I}  \right]  \\
		& U^\lambda_{+, l} \rho_\lambda U^{\lambda \dag}_{+, l} \left[  H_l^{\lambda}, O^{\lambda,I}  \right]  U^\lambda_{+, l} ) dU_{+,l} \Big) dU_{-,l} \\
		= & \frac{2}{(d_\lambda + 1)(d_\lambda^2 - 1)} \Big( \text{Tr}[(O^{\lambda})^2] - \frac{[\text{Tr}(O^\lambda)]^2 }{d_\lambda} \Big) \\
		& \sum_l \Big( \text{Tr}[(H_l^{\lambda})^2] - \frac{[\text{Tr}(H_l^{\lambda})]^2 }{d_\lambda} \Big).
	\end{aligned}
\end{align}

In the special case of SU$(d)$ symmetry, we expand $O$ and $H_l$ by SWAPs and their products in YJM elements. Then their trace restricted to one irrep is just given by $S_n$ group characters \cite{Ingram1950,Roichman1996,Lassalle2008}. Unfortunately, the character formula becomes intractable for general qudits and arbitrary $S_n$ irreps, but when $d = 2$ for qubits, we can explicitly check that they scale as $\Theta(d_\lambda)$ for large $n$. Therefore,
\begin{align}\label{eq: qntk}
	\bar{K} \approx \sum_{i, j=1}^N \frac{L}{d_\lambda^2}\left(\operatorname{Tr}\left(O_i^\lambda O_j^\lambda\right)-\frac{\operatorname{Tr}\left(O_i^\lambda\right) \operatorname{Tr}\left(O_j^\lambda\right)}{d_\lambda}\right) \approx \frac{N^2 L}{d_\lambda},
\end{align}
where the dimension $d_\lambda$ of the irrep $S^\lambda$ can be evaluated by the so-called hook length formula in $S_n$ representation theory \cite{Sagan01,Tolli2009}. Some of them scale exponentially with respect to the system size $n$ and we also provide a detailed introduction at the end of Appendix \ref{sec:SnTheory}. Moreover, if one concerns with the permutation symmetry given by permuting qudits through the group $S_n$, Schur--Weyl duality asserts that the decomposed charge sectors are given by SU$(d)$ irreps whose dimensions scales as $O(n^d)$. Especially when $d = 2$ for qubits, it is familiar that the largest charge sector, spin-$n/2$ irrep, preserving permutation invariance is of dimension $n+1$.

%------------------------------------------------------------------------------------------------------------------------------------------------

\section{Discussion}

In this work, we explore the physical applications of random quantum circuits with SU$(d)$ symmetry. We first study the late-time residual of local SU$(d)$ non-Abelian quantities through OTOC, where we show that the residual part decays only inverse polynomials with respect to the total sites $n$. These asymptotic results stand in agreement with general observation for chaotic dynamics with conservation laws \cite{rakovszky2018diffusive,Huang2019OTOC}. The late time (post-scrambling) analysis in our results could motivate a further study of quantum hydrodynamics of quantum circuits with SU($d$) conservation laws, where in Ref.~\cite{SUd-k-Design2023} a local random circuit ensemble is provided for which to converge to unitary $k$-designs for general qudits. Furthermore, our results may also invite further research in line with studying the entanglement generation through Page curve or short-range chaotic models with non-Abelian quantities of which SU$(d)$-symmetric quantum circuits could serve as a principal model. It is also worth investigating various SU$(d)$-symmetric monitored circuit dynamics of local charges such as \cite{majidy2023critical} where novel physical phenomena may appear due to the nature of non-commuting charges. In quantum error correction, as stated in \cite{kong2022near}, it is of interest to investigate the optimality of random SU$(d)$-covariant codes against multiple qudit erasure error. Furthermore, it is of interest to find covariant codes that could be capable of a non-constant coding rate. It is also interesting to implement such SU$(d)$-covariant codes in certain platforms where exchange interaction is naturally implemented \cite{gao2018entangling, divincenzo2000universal} and logical information is restricted to a certain charge sector. In geometric quantum machine learning our results might give an indication of exponential convergence in the presence of symmetry. For instance, in the presence of permutation symmetry on qubit systems where the charge sectors scale at most $O(n)$, exponential convergence can be quickly achieved by requiring variational parameters linearly in $n$. It is worth mentioning that many physical interesting problems still bear exponentially scaling charge sectors such as the frustrated Heisenberg models \cite{Zheng2021SpeedingUL}. Hence, our work might motivate further study on the performance of these symmetry-respecting variational ansatz in geometric quantum machine learning. 

%------------------------------------------------------------------------------------------------------------------------------------------------

\section*{Data Availability}
The relevant data used for simulation can be retrieved by contacting \url{hanz98@uchicago.edu}.

\section*{Author Contributions}
Z.L and H.Z contributed equally in this work. Z.L and H.Z conceived the idea and developed the theoretical proofs. Y.W contributed to the analysis of Hayden-Preskill protocol. L.J, ZW.L and J.L proposed the research topic and supervised the project. All authors contributed to valuable discussions. All authors drafted and revised the manuscript.

\section*{Competing interests}
The Authors declare no Competing Financial or Non-Financial Interests.

\section*{Acknowledgment}

 We acknowledge helpful discussions with Gregory S. Bentsen, Bill Fefferman, Christopher Kang, Laimei Nie, Brian Swingle, Shengqi Sang, Sergii Strelchuk, Pei Zeng, among others. L.J. acknowledges support from the ARO(W911NF-23-1-0077), ARO MURI (W911NF-21-1-0325), AFOSR MURI (FA9550-19-1-0399, FA9550-21-1-0209), NSF (OMA-1936118, ERC-1941583, OMA-2137642), NTT Research, Packard Foundation (2020-71479), and the Marshall and Arlene Bennett Family Research Program. This material is based upon work supported by the U.S. Department of Energy, Office of Science, National Quantum Information Science Research Centers. This research used resources of the Oak Ridge Leadership Computing Facility, which is a DOE Office of Science User Facility supported under Contract DE-AC05-00OR22725. J.L. is supported in part by the University of Pittsburgh, School of Computing and Information, Department of Computer Science, Pitt Cyber, and the PQI Community Collaboration Awards, and by International Business Machines (IBM) Quantum through the Chicago Quantum Exchange, and the Pritzker School of Molecular Engineering at the University of Chicago through AFOSR MURI (FA9550-21-1-0209).

\bibliography{CQA_OTOC.bib}

\widetext
\appendix

\pagebreak
\begingroup
\titleformat{\section}[block]{\large\bfseries\filcenter}{}{0pt}{}
\section*{Appendix}
\endgroup

\setcounter{tocdepth}{2}

\startcontents[sections]
\titlecontents{section}[0pt]{\vspace{5mm}}{\thecontentslabel\hspace{1em}}{}{\titlerule*[1pc]{.}\contentspage}
\titlecontents{subsection}[1.5em]{}{\thecontentslabel\hspace{1em}}{}{\titlerule*[1pc]{.}\contentspage}
\printcontents[sections]{}{1}{\section*{}\vspace{-10mm}}

%------------------------------------------------------------------------------------------------------------------------------------------------

\section{Preliminaries}

We introduce some basic notions and facts from $S_n$ representation theory as well as our CQA model to lay the foundation for later mathematical proofs. We also refer interested readers to Refs.~\cite{Fulton1997,Sagan01,Goodman2009,Tolli2009,Zheng2021SpeedingUL} for more systematic presentations on these topics.

\subsection{Miscellaneous facts about $S_n$ representation theory}\label{sec:SnTheory}

Irreducible representations (irreps) of the symmetric group $S_n$ permuting $n$ nodes are in one-to-one correspondence with the so-called \emph{Young diagrams}. For instance, for $S_6$, the following two Young diagrams stand for the \emph{trivial} and the \emph{standard} representation, respectively:
\begin{align*}
	\ytableausetup{boxsize=1.25em} \ydiagram{6}, \qquad \ydiagram{5,1},
\end{align*}
the direct sum of which is more familiar as the six-dimensional \emph{defining} representation under which each $\sigma \in S_6$ permutes the components of vectors from $\mathbb{R}^6$. 

\begin{definition}\label{def:YoungDiagram}
	Formally, let $\lambda = (\lambda_1,\ldots,\lambda_r)$ be a collection of positive integers such that $\lambda_i \geq \lambda_{i+1}$ and $\sum_i \lambda_i = n$. Then $\lambda$ is called a \emph{partition} of the integer $n$, denoted by $\lambda \vdash n$. Obviously, $\lambda$ defines a Young diagram abstractly and the $S_n$ irrep corresponding to this Young diagram is always denoted as $S^\lambda$. The dimension of this irrep is given by the \emph{hook-length formula}:
	\begin{align}
		\dim S^\lambda = \frac{n!}{\prod_{(x,y) \in \lambda} h_{x,y} },
	\end{align}
	where $(x,y)$ specifies a box from $\lambda$ by its \emph{row} and \emph{column numbers}, and the \emph{hook-length} $h(x,y)$ counts the number of all boxes to the right of or below $(x,y)$ plus itself.
\end{definition}

Given an arbitrary $S_n$ irrep $S^\lambda$, there is a canonical way to label a basis, called the \emph{Gelfand--Tsetlin (GZ) basis} or \emph{Young--Yamanouchi basis}, on the representation space, using \emph{standard Young tableau} $T$, which are defined by filling into each box of $\lambda$ a positive integer from $1,2,\ldots,n$ in an increasing order from left to right and top to bottom. For instance, the standard representation of $S_6$ mentioned above is five-dimensional with five basis vectors labeled as
\begin{align*}
	\ytableaushort{12345, 6}, \quad  \ytableaushort{12346, 5}, \quad \ytableaushort{12356, 4} \quad \ytableaushort{12456, 3}, \quad \ytableaushort{13456, 2}.
\end{align*}
When we study the group of $\text{SU}(d)$-symmetric unitaries in the main text, the Young basis $\{\ket{\alpha_T}\}$ labeled by standard tableaux is implicitly used and a detailed treatment can be found in Section \ref{sec:SnCommutant}. 

\begin{definition}\label{def:YJM}
	For $1 < k \leq n$, the \textit{Young--Jucys--Murphy element}, or YJM element for short, is defined as a (formal) sum of \emph{transpositions} or SWAPs 
	\begin{align}\label{eq:A-YJM}
		X_i = (1,i) + (2,i) + \cdots + (i-1,i).
	\end{align}
	We set $X_1 = 0$ as a convention. 
\end{definition} 

The YJM element is a central concept used in our work developed by Young \cite{Young1977}, Jucys \cite{Jucys1974} and Murphy \cite{Murphy1981} and later used by Okounkov and Vershik \cite{Okounkov1996}. Under any $S_n$ representation, it may be more comprehensible to treat $X_i = (1,i) + (2,i) + \cdots + (i-1,i)$ as the sum of matrix representations of these transpositions, or we can say that the representation is extended to the \emph{group algebra} 
\begin{align}
	\mathbb{C}[S_n] = \left\{ \sum_i c_i \sigma_i; \sigma_i \in S_n \right\}
\end{align}
consisting of formal finite linear combinations of $S_n$ group elements. By the Wedderburn theorem \cite{Sagan01,Goodman2009}, $\mathbb{C}[S_n]$ is isomorphic with the direct sum of all inequivalent $S_n$ irreps $\bigoplus_\lambda \mathrm{1}_{\dim S^\lambda} \otimes (S^\lambda)$ with multiplicities equal to their dimension. 

Let us consider coordinate differences $x-y$ of boxes from a Young diagram $\lambda$. Given any standard tableau $T$ of $\lambda$, its \emph{content vector} is defined by rearranging them with respect to the order of boxes determined by the tableau. For instance, the content vectors of the above five standard tableaux are listed as follows:
\begin{align*}
	(0,1,2,3,4,-1), \quad (0,1,2,3,-1,4), \quad (0,1,2,-1,3,4), \quad (0,1,-1,2,3,4), \quad (0,-1,1,2,3,4).
\end{align*} 
An important feature of YJM elements is their special actions under the Young basis as revealed by content vectors: 
\begin{enumerate}
	\item They are diagonal matrices under the Young basis (even each single transposition $(i,j)$ from Eq.~\eqref{eq:A-YJM} may not be diagonal).
	
	\item The diagonal entry of $X_i$ under the Young basis vector $\ket{\alpha_T}$ corresponding to standard tableau $T$ is just the $i$-th component of the content vector.
\end{enumerate}
On the irrep $S^{(5,1)}$, 
\begin{align}\label{eq:YJMExample}
	\begin{aligned}
		& X_1 = \begin{pmatrix} 0 & 0 & 0 & 0 & 0 \\ 0 & 0 & 0 & 0 & 0 \\ 0 & 0 & 0 & 0 & 0 \\  0 & 0 & 0 & 0 & 0 \\	0 & 0 & 0 & 0 & 0 \end{pmatrix}, \quad\ \ \ 
		X_2 = \begin{pmatrix} 1 & 0 & 0 & 0 & 0 \\ 0 & 1 & 0 & 0 & 0 \\ 0 & 0 & 1 & 0 & 0 \\  0 & 0 & 0 & 1 & 0 \\	0 & 0 & 0 & 0 & -1 \end{pmatrix}, \quad 
		X_3 = \begin{pmatrix} 2 & 0 & 0 & 0 & 0 \\ 0 & 2 & 0 & 0 & 0 \\ 0 & 0 & 2 & 0 & 0 \\  0 & 0 & 0 & -1 & 0 \\	0 & 0 & 0 & 0 & 1 \end{pmatrix}, \\
		& X_4 = \begin{pmatrix} 3 & 0 & 0 & 0 & 0 \\ 0 & 3 & 0 & 0 & 0 \\ 0 & 0 & -1 & 0 & 0 \\  0 & 0 & 0 & 2 & 0 \\	0 & 0 & 0 & 0 & 2 \end{pmatrix}, \quad
		X_5 = \begin{pmatrix} 4 & 0 & 0 & 0 & 0 \\ 0 & -1 & 0 & 0 & 0 \\ 0 & 0 & 3 & 0 & 0 \\  0 & 0 & 0 & 3 & 0 \\	0 & 0 & 0 & 0 & 3 \end{pmatrix}, \quad
		X_6 = \begin{pmatrix} -1 & 0 & 0 & 0 & 0 \\ 0 & 4 & 0 & 0 & 0 \\ 0 & 0 & 4 & 0 & 0 \\  0 & 0 & 0 & 4 & 0 \\	0 & 0 & 0 & 0 & 4 \end{pmatrix}.
	\end{aligned}
\end{align}
In summary, Young basis vectors $\ket{\alpha_T}$, standard tableaux $T$ and content vectors $\alpha_T$ are in one-to-one correspondence and uniquely determine the matrix representations of YJM elements. 

The matrix representation of each \emph{adjacent transposition} $(i,i+1)$ can be explicitly read off in the Young basis by the \emph{Young orthogonal form}: Let $r = \alpha_T(i+1) - \alpha_T(i)$ be the \emph{axial distance} and let $(i,i+1) \cdot T$ denote the tableau defined by exchanging integers $i, i +1$ from $T$. It is easy to check that as long as $r \neq \pm 1$, $(i,i+1) \cdot T$ is still a standard Young tableau. Then
\begin{align}\label{eq:YoungOrthogonal}
	(i,i+1) \ket{\alpha_T} = \frac{1}{r} \ket{\alpha_T} + \sqrt{1 - \frac{1}{r^2}} \ket{\alpha_{(i,i+1) \cdot T}}, \quad  (i,i+1) \ket{\alpha_{(i,i+1) \cdot T}} = \sqrt{1 - \frac{1}{r^2}} \ket{\alpha_T} - \frac{1}{r} \ket{\alpha_{(i,i+1) \cdot T}}.  
\end{align} 
On the irrep $S^{(5,1)}$, 
\begin{align}
	\renewcommand\arraystretch{1.25}
	\begin{aligned}
		& (1,2) = \begin{pmatrix} 1 & 0 & 0 & 0 & 0 \\ 0 & 1 & 0 & 0 & 0 \\ 0 & 0 & 1 & 0 & 0 \\  0 & 0 & 0 & 1 & 0 \\	0 & 0 & 0 & 0 & -1 \end{pmatrix}, \quad\quad 
		(2,3) = \begin{pmatrix} 1 & 0 & 0 & 0 & 0 \\ 0 & 1 & 0 & 0 & 0 \\ 0 & 0 & 1 & 0 & 0 \\  0 & 0 & 0 & -\frac{1}{2} & \frac{\sqrt{3}}{2} \\	0 & 0 & 0 & \frac{\sqrt{3}}{2} & \frac{1}{2} \end{pmatrix}, \quad 
		(3,4) = \begin{pmatrix} 1 & 0 & 0 & 0 & 0 \\ 0 & 1 & 0 & 0 & 0 \\ 0 & 0 & -\frac{1}{3} & \frac{2\sqrt{2}}{3} & 0 \\  0 & 0 & \frac{2\sqrt{2}}{3} & \frac{1}{3} & 0 \\	0 & 0 & 0 & 0 & 1 \end{pmatrix}, \\
		& (4,5) = \begin{pmatrix} 1 & 0 & 0 & 0 & 0 \\ 0 & -\frac{1}{4} & \frac{\sqrt{15}}{4} & 0 & 0 \\ 0 & \frac{\sqrt{15}}{4} & \frac{1}{4}  & 0 & 0 \\  0 & 0 & 0 & 1 & 0 \\ 0 & 0 & 0 & 0 & 1 \end{pmatrix}, \
		(5,6) = \begin{pmatrix}  -\frac{1}{5} & \frac{2\sqrt{6}}{5} & 0 & 0 & 0 \\  \frac{2\sqrt{6}}{5} & \frac{1}{5} & 0 & 0 & 0 \\ 0 & 0 & 1 & 0 & 0 \\  0 & 0 & 0 & 1 & 0 \\	0 & 0 & 0 & 0 & 1 \end{pmatrix}.
	\end{aligned}
\end{align}
Having access to the matrix representations of all adjacent transpositions allows us to numerically calculate the matrix representing any permutation $\sigma \in S_n$. There are also classical or quantum \emph{$S_n$-fast Fourier transform} methods designed for such tasks \cite{Clausen_1993,Maslen_1998}.

%----------------------------------------------------------------------------------------------------------

\begin{definition}\label{def:CycleType}
	We say that a permutation $\sigma \in S_n$ is of \emph{cycle type} $\lambda = (\lambda_1,\ldots,\lambda_r)$ where $\lambda \vdash n$ corresponds to a partition or Young diagram, if it is decomposed into cycles of lengths $\lambda_1,\ldots,\lambda_r$. 
\end{definition}

For instance, $\sigma = (134)(56) \in S_6$ is of cycle type $\lambda = (3,2,1)$. The trivial permutation $\operatorname{id} \in S_n$ is of type $(1,\ldots,1)$. Transpositions or SWAPs $(i,j)$ are just 2-cycles, so products of transpositions such as $(i,j)(k,l) \cdots (s,t)$ are of type $(2,2,\ldots,2,1,\ldots,1)$. 

\begin{definition}\label{def:PartitionFunction}
	Let $p(n)$ denote the {number of partitions} of $n$. It equals the number of all inequivalent $S_n$ irreps, as well as the number of possible cycle types in $S_n$. Analogously, we define $p(n,d)$ as the {number of partitions of $n$ with at most $d$ parts}, i.e., the number of all Young diagrams of $n$ boxes with at most $d$ rows. 
\end{definition}

We encounter $p(n)$ and $p(n,d)$ in the main text when discussing the difficulty of computing frame potentials in the presence of $\text{SU}(d)$ symmetry. They also appear when we study the locality required to achieve arbitrary $k$-designs (Section \ref{sec:SnCommutant}). Due to the celebrated work of Ramanujan and Hardy \cite{Ramanujan1918} and Uspensky, \cite{Uspensky1920}, 
\begin{align}\label{eq:Ramanujan}
	p(n) \sim \frac{e^{\sqrt{\pi^2 2n /3}}}{4n\sqrt{3}}, n \to \infty.
\end{align}
There has been further study on this \cite{Rademacher1938,Erdos1942}, and various useful bounds on $p(n)$ have been found later, such as \cite{Maroti2003, Wladimir2009}
\begin{align}
	\frac{e^{2\sqrt{n}}}{an} < p(n) < e^{b\sqrt{n}}.
\end{align}
If $d = 2$, $p(n,2) = \lfloor \frac{n}{2} \rfloor + 1$. However, there are no closed-form formulas for these partition functions in general. 

%----------------------------------------------------------------------------------------------------------

\begin{proposition}
	Let $c_\mu \in \mathbb{C}[S]$ be the sum of all $\sigma \in S_n$ with cycle type $\mu$. Considering all possible Young diagrams of size $n$, the collection $\{ c_\mu \}_{\mu \vdash n}$ forms a basis for the \emph{center} $Z(\mathbb{C}[S_n])$ consisting of all elements that commute with $\mathbb{C}[S_n]$. 
\end{proposition}  

By definition, $c_\mu$ commutes with any $\sigma \in S_n$. By the Wedderburn theorem \cite{Sagan01,Goodman2009}, its matrix representation, still denoted by $c_\mu$ for simplicity, under any $S_n$ irrep $S^\lambda$ is just a scalar. As a result, the representation of $Z(\mathbb{C}[S_n])$ consists of scalar matrices within any $S_n$ irrep. Being a basis of $Z(\mathbb{C}[S_n])$ means being a basis capable of spanning all scalar matrices, called \emph{relative phase factors} when we study $k$-design with symmetry, respecting the direct sum $\bigoplus S^\lambda$ of all inequivalent $S_n$ irreps. 

Besides the basis $\{ c_\mu \}$ defined above, we still have the following two kinds of center bases:

\begin{theorem}\label{thm:CenterBases}
	The following two collections also constitute bases for $Z(\mathbb{C}[S_n])$:
	\begin{enumerate}
		\item Consider the $S_n$ group character
		\begin{align}\label{eq:Character}
			\chi_\lambda (\sigma) = \operatorname{tr}_\lambda \sigma
		\end{align}
		defined by taking the trace of $\sigma \in S_n$ restricted to the irrep $S^\lambda$. Then
		\begin{align}\label{eq:OrthogonalProjection}
			\Pi_\mu \vcentcolon = \frac{\dim S^\mu}{n!} \sum_{\sigma \in S_n} \bar{\chi}_\mu(\sigma) \sigma
		\end{align}
		is a projection exclusively into the irrep $S^\mu$. The collection $\{\Pi_\mu \}$ is an orthonormal basis. 
		
		\item Consider the YJM elements $X_i$. For any $\mu = (\mu_1,\ldots,\mu_r) \vdash n$, we set
		\begin{align}
			X_\mu = \sum_{2 \leq i_1 \neq i_2 \neq \cdots \neq i_r \leq n} X_{i_1}^{\mu_1 - 1} X_{i_2}^{\mu_2 - 1} \cdots X_{i_r}^{\mu_r - 1}
		\end{align} 
		The collection $\{X_\mu\}$ is also a basis for $Z(\mathbb{C}[S_n])$ \cite{Jucys1974,Murphy1981,Tolli2009}.
	\end{enumerate}
\end{theorem}

A basis element $c_\mu$ is defined by summing over all permutations of a given cycle type $\mu$, and $\Pi_\mu$ even requires doing this over the whole symmetric group with $n!$ elements. A more constructive way to build basis elements with a desired locality is to employ the YJM elements, which turns out to enable $2k$-local CQA to form an exact $k$-design for any constant $k$ unconditionally in $n$ \cite{SUd-k-Design2023}.

%----------------------------------------------------------------------------------------------------------

In Section \ref{sec:Charge} and \ref{sec:QNTK}, we also need to compute the character $\chi_\lambda (\sigma)$ explicitly for some $\sigma \in S_n$, so we briefly introduce the method here. We first present the following proposition.

\begin{proposition}\label{prop:Integer}
	For any $\sigma \in S_n$, $\chi_\lambda(\sigma) \in \mathbb{Z}$. That is, $S_n$ characters are integer-valued.
\end{proposition}

One can prove a variety of similar facts using Galois theory for general finite groups. For our purpose, we simply note that there is the so-called \emph{Young's natural representation} which is a \emph{non-unitary} representation of $S^\lambda$ under which each $\sigma$ is expressed as matrices with integer entries \cite{Sagan01}. As trace is invariant under matrix similarity, $\chi_\lambda (\sigma) = \operatorname{tr}_\lambda \sigma \in \mathbb{Z}$ in general. 

It is also well known that permutations $\sigma$ and $\sigma'$ with the same cycle type are conjugate to each other, so $\chi_\lambda (\sigma) = \chi_\lambda (\sigma')$ and hence we only care about the value of $\chi_\lambda$ for a given cycle type $\mu$. The so-called \emph{Frobenius character formula} \cite{Goodman2009} expresses $S_n$ character values as coefficients of a power series. The coefficients can be formally computed by, e.g., contour integrals using the residue theorem. However, closed-form formulas only exist for very few simple cases \cite{Ingram1950,Roichman1996}. For instance, the characters for 2-cycles (transpositions/SWAPs) are
\begin{align}\label{eq:CharacterValue}
	& \frac{\chi_\lambda(i,j)}{\dim S^\lambda} = \frac{2}{n(n-1)} \sum_i \Big( \binom{\lambda_i}{2} - \binom{\lambda_i'}{2} \Big) = \frac{1}{n(n-1)} \sum_i \big[ (\lambda_i - 1)(\lambda_i - 1 + i) - i(i-1) \big], 
\end{align}
where $\lambda'$ denotes the conjugate of $\lambda$, e.g.,
\begin{align*}
	\lambda = \ytableausetup{boxsize=1.25em,aligntableaux=center} \ydiagram{4,2,1} \qquad \lambda' = \ydiagram{3,2,1,1}
\end{align*}
If $\lambda_i < 2$, the corresponding binomial coefficient is set to zero.

%----------------------------------------------------------------------------------------------------------

Let us relate the above character formula to some techniques involving YJM elements. Restricted to any irrep $S^\lambda$, we can associate the following invariants: 
\begin{align}
	P_l = \left(\sum_i X_i\right)^l, l = 1,2,\ldots
\end{align}
where $\sum_i X_i$ is the summation of all YJM elements. Let us check its matrix form under the Young basis $\{\ket{\alpha_T}\}$:
\begin{align}
	\left(\sum_i X_i\right) \ket{\alpha_T} = \sum_i \alpha_T(i) \ket{\alpha_T},
\end{align}
where the $\alpha_T$ are the content vectors. Obviously, for any fixed Young diagram $\lambda$, the sum of all components of any of its content vector $\alpha_T$ is simply equal to the sum of all coordinate differences, and we denote it as $\alpha_\lambda$. Then
\begin{align}\label{eq:P_l}
	P_l \ket{\alpha_T} = (\alpha_\lambda)^l \ket{\alpha_T} 
\end{align}
for all standard tableaux or Young basis vectors of the Young diagram $\lambda$.  

Let $\operatorname{tr}_\lambda$ denote the trace within $S^\lambda$ (it is just the $S_n$ character in Eq.~\eqref{eq:Character}). When $l = 1$, we note that
\begin{align}\label{eq:P_l2}
	\alpha_\lambda = \frac{\operatorname{tr}_\lambda (P_l) }{\dim S^\lambda} = \frac{\operatorname{tr}_\lambda \sum_i X_i}{\dim S^\lambda} = \frac{n(n-1)}{2} \frac{\operatorname{tr}_\lambda (i,j)}{\dim S^\lambda} = \frac{n(n-1)}{2} \frac{\chi_\lambda(i,j)}{\dim S^\lambda},
\end{align}
which gives another way to compute the character value of 2-cycles by summing all components from the content vector. The method using YJM elements and content vectors to express general $S_n$ characters can be found in Ref.~\cite{Lassalle2008}.

%----------------------------------------------------------------------------------------------------------

\begin{definition}\label{def:Dominance}
	Given two partitions $\lambda = (\lambda_i), \mu = (\mu_i) \vdash n$. We say that $\lambda$ \emph{dominates} $\mu$, denoted by $\lambda \unrhd \mu$, if for all $j > 0$, $\sum^j_i \lambda_i \geq \sum^j_i \mu_i$. 
\end{definition}

For instance, we have
\begin{align}
	(6) \unrhd (5,1) \unrhd (4,2) \unrhd (4,1^2) 
\end{align}
where $(4,1^2)$ is the abbreviation of $(4,1,1)$. The dominance relation is not \emph{totally ordered}, e.g., we cannot compare $(4,1^2)$ and $(3,3)$. However, in the case of qubits ($d = 2$), only two-row Young diagrams need to be considered (see Section \ref{sec:SchurWeyl}) and partitions $\lambda = (\lambda_1, \lambda_2)$ with $\lambda_1 \geq \lambda_2$ clearly give rise to a total ordering. 

\begin{lemma}\label{lemma: total-ordering2}
	For any two unequal partitions  $\lambda, \mu \vdash n$, if $\lambda \unrhd \mu$, then $\alpha_\lambda > \alpha_\mu$.
\end{lemma}
\begin{proof}
	We prove this lemma by induction. Suppose that the statement holds for $n-1$. Given unequal $\lambda, \mu \vdash n$ with $\lambda \unrhd \mu$, there should be some $i$ such that $\lambda_i > \mu_i$, where $\lambda_i$ and $\mu_i$ are the lengths of the $i$-th rows of $\lambda$ and $\mu$, respectively. If $\lambda_i > \lambda_{i+1}$ and $\mu_i > \mu_{i+1}$, then we discard the right-hand side boxes on the $i$-th rows of $\lambda$ and $\mu$. The resultant Young diagrams, denoted by $\lambda'$ and $\mu'$, still satisfy the relation $\lambda' \unrhd \mu'$. Then, by the induction hypothesis, $\alpha_{\lambda'} > \alpha_{\mu'}$. On the other hand, the content of the discarded box from $\lambda$ is larger than that from $\mu$ by definition, hence we conclude that $\alpha_\lambda > \alpha_\mu$.
	
	Suppose that $\lambda_i = \lambda_{i+1} = \cdots = \lambda_r$ or $\mu_i = \mu_{i+1} = \cdots = \mu_s$. Then we are only allowed to discard the right-hand side boxes of $\lambda_r$ and $\mu_s$ to ensure that $\lambda'$ and $\mu'$ are well-defined Young diagrams. Even when $r \neq s$, the dominance relation still holds because $\lambda_r > \mu_s$. By the same argument as above, we complete the proof. 
\end{proof}

By Eq.~\eqref{eq:P_l2}, the above lemma says that $\frac{\chi_{\lambda}(1,2)}{\dim S^\lambda}$ is strictly increasing with respect to the dominance order of $\lambda$. Lots of counterexamples occur when this order fails to hold: e.g., $(3,3)$ and $(4,1^2)$.

%----------------------------------------------------------------------------------------------------------

Let us end this subsection with some explicit analysis on the dimension of $S_n$ irreps of two-row Young diagrams. In this circumstance, the hook-length formula in.\ref{def:YoungDiagram} can be further simplified as  \cite{Ingram1950,Sagan01}
\begin{align}\label{eq:two-row-dim}
	\dim S^\lambda = d_\lambda = \binom{n}{r} - \binom{n}{r-1} = \frac{n - 2r +1}{n - r + 1}\binom{n}{r}.
\end{align}
For binomial coefficients, we have another two useful bounds (assume that $n = 2m$) \cite{Gallier2011}:
\begin{align}
	\frac{2^n}{\sqrt{\pi(\frac{n}{2} + \frac{1}{3})}} \leq \binom{n}{n/2} \leq \frac{2^n}{\sqrt{\pi(\frac{n}{2} + \frac{1}{4})}}, \quad e^{-(\frac{n}{2} - r)^2/(n-r+1)} \leq \binom{n}{r} \Big/ \binom{n}{\frac{n}{2}} \leq e^{-(\frac{n}{2} - r)^2/(n-r)}.
\end{align}
Therefore, the ratio of $\dim S^\lambda$ to the dimension of the entire Hilbert space is
\begin{align}\label{eq:dim-Comparision}
	\frac{1}{\sqrt{\pi(\frac{n}{2} + \frac{1}{3})}} \frac{n - 2r +1}{n - r + 1}\binom{n}{r} \Big/ \binom{n}{\frac{n}{2}} \leq \frac{\dim S^\lambda}{2^n} = \frac{1}{2^n} \frac{n - 2r +1}{n - r + 1}\binom{n}{r} \leq \frac{1}{\sqrt{\pi(\frac{n}{2} + \frac{1}{4})}} \frac{n - 2r +1}{n - r + 1}\binom{n}{r} \Big/ \binom{n}{\frac{n}{2}}.
\end{align}

A lower bound for general $d$-row Young diagrams is also useful \cite{Mishchenko1996,Giambruno2015}. 
Suppose the first row $\lambda_1$ and $\lambda_1'$ of $\lambda$ and its conjugate is upper bounded by $\frac{n}{\alpha}$ for some $\alpha > 1$. Then, 
\begin{align}\label{eq:dim-Comparision2}
	\dim S^\lambda \geq \frac{\alpha^n}{n^{d(d+2)/2}}.
\end{align}
We will intensively use these results for specific cases in Section \ref{sec:Charge}, \ref{sec:Covariant} and \ref{sec:QNTK}.

%----------------------------------------------------------------------------------------------------------

\subsection{Schur--Weyl duality and CQA architecture}\label{sec:SchurWeyl}

We now provide a brief review on $S_n$-Convolutional Quantum Alternating Ans{\"a}tze and the group CQA proposed in Ref.~\cite{Zheng2021SpeedingUL}. It motivates the definition of CQA ensembles in the formation of SU$(d)$-symmetric designs \cite{SUd-k-Design2023} and it is also used as the model in our study of QNTK under SU$(d)$ symmetry in Section \ref{sec:QNTK}.

For quantum systems, there is a discrete set of translations corresponding to permuting the qudits as well as a continuous notion of translation corresponding to spatial rotations by elements of $\text{SU}(d)$. To be precise, let $V$ be a $d$-dimensional complex Hilbert space with orthonormal basis $\{e_1,\ldots,e_d\}$. The tensor product space $V^{\otimes n}$ admits two natural representations: the \textit{tensor product representation} $\pi_{\operatorname{SU}(d)}$ of $\text{SU}(d)$, acting as
\begin{align}\label{eq:SUd}
	\pi_{\operatorname{SU}(d)}(g) (e_{i_1} \otimes \cdots \otimes e_{i_n}) \vcentcolon = g \cdot e_{i_1} \otimes \cdots \otimes g \cdot e_{i_n}, 
\end{align}
where $g \cdot e_{i_k}$ is the fundamental representation of $\text{SU}(d)$, and the \textit{permutation representation} $\pi_{S_n}$ of $S_n$ acting as
\begin{align}\label{eq:Sn}
	\pi_{S_n}(\sigma) (e_{i_1} \otimes \cdots \otimes e_{i_n}) \vcentcolon =
	e_{i_{\sigma^{-1}(1)}} \otimes \cdots \otimes e_{i_{\sigma^{-1}(n)}}.
\end{align}
We treat $\mathcal{H} = V^{\otimes n}$ as the Hilbert space of an $n$-qudit system. Schur--Weyl duality states that the action of $\text{SU}(d)$ and $S_n$ on $V^{\otimes n}$ jointly decompose the space into irreducible representations of both groups in the form 
\begin{align}\label{eq:A-SchurWeyl}
	V^{\otimes n} = \bigoplus_\lambda W_\lambda \otimes S^\lambda.
\end{align}
Again, $\lambda$ denotes a Young diagram. In this setting, it corresponds not only to a unique $S_n$ irrep $S^\lambda$, but also an $\text{SU}(d)$ irrep $W_\lambda$ \cite{Goodman2009,Tolli2009}. It should be noted that within an $n$-qudit system, only irreps corresponding to $\lambda$ ranging over {Young diagrams of size $n$ with at most $d$ rows} can be found in the decomposition. 

We denote by $\mathbbm{1}_{m_{\operatorname{SU}(d),\mu}} \cong S^\mu, \mathbbm{1}_{ m_{S_n,\lambda}} \cong W_\lambda$ the multiplicity spaces of $\text{SU}(d)$ and $S_n$ irreps, respectively. Then
\begin{align}
	& \pi_{\operatorname{SU}(d)} \cong \bigoplus_\mu W_\mu \otimes \mathbbm{1}_{m_{\operatorname{SU}(d),\mu}}, \quad  \pi_{S_n} \cong \bigoplus_\lambda \mathbbm{1}_{ m_{S_n,\lambda}} \otimes S^\lambda,
\end{align}
where $m_{\operatorname{SU}(d),\mu}=\dim S^\mu$ and $m_{S_n,\lambda} = \dim W_\lambda$.

An operator $A$ acting on the system being $\text{SU}(d)$-symmetric/invariant means that
\begin{align}
	\pi_{\operatorname{SU}(d)}(g) A = A \pi_{\operatorname{SU}(d)}(g) \text{ or } g^{\otimes n} A = A g^{\otimes n}.
\end{align}
One can check by Eqs.~\eqref{eq:SUd} and \eqref{eq:Sn} that these permutation actions clearly commute with $g^{\otimes n}$. Furthermore, Schur--Weyl duality and the double commutant theorem \cite{Goodman2009,Tolli2009} confirm that $\text{SU}(d)$-symmetric operators are exactly built from permutations in the symmetric group $S_n$. That is, they can be expressed as linear combinations, such as $\sum c_i \sigma_i$, of permutations.

Decomposing the entire space into $\text{SU}(d)$ irreps is a conventional practice in physics. Quantum states living in these subspaces are actually permutation-invariant or $S_n$-symmetric. Since our focus is on quantum circuits with $\text{SU}(d)$ symmetry, we should decompose the entire Hilbert space with respect to $S_n$ irreps (for more details, see Refs.~\cite{Tolli2009,krovi,Zheng2021SpeedingUL}). As a reminder, although the entire Hilbert space is decomposed into smaller subspaces, one should not expect that related problems including computing the ground state energy of $\text{SU}(d)$-symmetric Hamiltonian or constructing a $\text{SU}(d)$-symmetric random quantum circuit, would become easier. There are two reasons in general:
\begin{enumerate}
	\item There are various inequivalent $S_n$ irreps from the decomposition to deal with, and the total number is $p(n,d)$, which scales at most superpolynomially (see \eqref{eq:Ramanujan}) with $n$ and has no closed-form formula for evaluation. 
	
	\item Even for qubits with $d = 2$, using the hook-length formula from Definition \ref{def:YoungDiagram}, we know that (cf.~Eq.~\eqref{eq:two-row-dim})
	\begin{align}
		\dim S^{(m,m)} = \frac{(2m)!}{(m+1)!\, m!} = \frac{2^m}{m+1} \prod_{k=1}^m \frac{2k-1}{k} > \frac{2^m}{m+1} 
	\end{align}
	for the $S_n$ irrep of Young diagram $\lambda = (m,m)$ on a $2m$-qubit system. One can find other examples with exponentially large subspaces respecting the $\text{SU}(d)$ symmetry \cite{Giambruno2015}, which still cause difficulties when approaching the problem.
\end{enumerate}

We now introduce the mathematical definition of the $S_n$-CQA ansatz:

\begin{definition}\label{def:CQA}
	The $S_n$-CQA ansatz is defined as
	\begin{align}\label{eq:CQAansatz}
		\begin{aligned}
			\cdots \exp(-i \sum_{k,l} \beta_{kl} X_k X_l ) \exp(-i\gamma H_S) 
			\exp(-i \sum_{k,l} \beta'_{kl} X_k X_l) \exp(-i\gamma' H_S) \cdots,
		\end{aligned}
	\end{align}
	where $X_k X_l$ are products of YJM elements which are 4-local and still diagonal under the Young basis (see the example given in \eqref{eq:YJMExample}). The Hamiltonian $H_S$ is defined as the summation of adjacent transpositions $\sum_{i = 1}^{n-1} (i,i+1)$. 
\end{definition}

One can also set $k \leq l$ in the above definition because YJM elements are commutative with each other. Moreover, let us define the group generated by alternating exponentials from \eqref{eq:CQAansatz}:
\begin{align}
	\mathrm{CQA} = \Big\langle \exp(-i \sum_{k,l} \beta_{kl} X_k X_l), \quad \exp(-i\gamma H_S) \Big\rangle.
\end{align}
Obviously, CQA is contained in the \emph{group of $\text{SU}(d)$-symmetric unitaries}. To define this group, let $\text{U}(S^\lambda)$ denote the unitary group acting on the representation space $S^\lambda$, i.e., $\text{U}(S^\lambda) \cong \operatorname{U}(\dim S^\lambda)$. A typical element $g$ from the group of $\text{SU}(d)$-symmetric unitaries is then a collection of unitaries:
\begin{align}\label{eq:GroupElements}
	g = \bigoplus_\lambda U_\lambda^{\oplus m_{S_n, \lambda}},
\end{align}
where $U_\lambda \in \operatorname{U}(S^\lambda)$ and $\lambda$ range over all Young diagrams of size $n$ with at most $d$ rows. For simplicity, we omit the multiplicities and denote this group by  
\begin{align}
	\text{either } \bigoplus_\lambda \operatorname{U}(S^\lambda) \ \text{ or } \ \mathcal{U}_\times.
\end{align}

On the other hand, by restricting the phase factors to be 1 on each $S^\lambda$, we have the special unitary group $\text{SU}(S^\lambda)$ as well as $\bigoplus_\lambda \operatorname{SU}(S^\lambda) = S\mathcal{U}_\times$ consisting of $\text{SU}(d)$-symmetric unitaries with unit determinant on each $S_n$ irrep block, i.e., unitaries with trivial relative phase factors with respect to each irrep. We also define $\mathcal{V}_4$ to be the group generated by $\text{SU}(d)$-symmetric 4-local unitaries. It is demonstrated in Ref.~\cite{Zheng2021SpeedingUL} that
\begin{align}
	S\mathcal{U}_\times \subsetneqq \mathrm{CQA} \subsetneqq \mathcal{V}_4 \subsetneqq \mathcal{U}_\times, 
\end{align}
establishing a theoretical guarantee for searching the ground state energy of the frustrated 2D Heisenberg model using the $S_n$-CQA ansatz, because relative phase factors can be ignored when we measure the expectation value in the experiment. 

%----------------------------------------------------------------------------------------------------------

\subsection{Useful properties on quantum entropies}\label{sec:entropy}

We note that the intuitive treatments relating the mutual information, relative entropy, and operator distance measure can be made rigorous. We refer the curious readers to a rather systematic treatment presented in Ch.5 of \cite{watrous_2018}. The key question in the setting presented in Section \ref{sec: backgrounds} is the relation of the mutual information $I(R: \bar{B})$ between the register $R$ and $\bar{B}$, and the (trace) norm distance between $\rho_{R \cup \bar{B}}$ and $\sigma_{R \cup \bar{B}}$. We first review some basic definitions. Assume that the Hilbert space is a composite system $\mathcal{H} = \mathcal{H}_A \otimes \mathcal{H}_{\bar{A}}$ and a pure state $\psi$ defined on the Hilbert space. Its Von Neumann entropy can be uniquely given by
\begin{align}
    S(\rho_A) = - \operatorname{Tr}[\rho_A \log \rho_A],
\end{align}
where $\rho_A \equiv \operatorname{Tr}_A[\psi]$ the partial trace over the subsystem $A$ and  $S(\rho_A) = S(\rho_{\bar{A}})$ as the system is bipartite. It is important to note that although the mathematical notion of Von Neumann entropy is well-defined as long as it is given a positive semi-definite operator acting on $\mathcal{H}$, the physical meaning of entropy of some quantum states might not exist if the system is multipartite or the quantum state is not pure in $\mathcal{H}$, which is still a frontier in quantum information research. Another useful quantity is called the quantum relative entropy given two density matrix $\rho, \sigma$ such that the image of $\rho$ lies inside that of $\sigma$
\begin{align}
    D(\rho \| \sigma) \equiv \operatorname{Tr}[\rho \log \rho - \rho \log \sigma].
\end{align}
Otherwise $D(\rho \| \sigma) = \infty$. In the setting mentioned in Section \ref{sec: backgrounds} with $\rho_{R \cup \bar{B}}$ and $\sigma_{R \cup \bar{B}} \equiv \rho_{R} \otimes \rho_{\bar{B}}$, we have
\begin{align}
    \begin{aligned}
        D(\rho_{R \cup \bar{B}} \| \sigma_{R \cup \bar{B}} ) & =  D(\rho_{R \cup \bar{B}} \| \rho_{R} \otimes \rho_{\bar{B}} ) \\
        & = \operatorname{Tr}\left\{ \rho_{R \cup \bar{B}} \log(\rho_{R \cup \bar{B}}) - \rho_{R \cup \bar{B}} \log(\rho_R \otimes I) - \rho_{R \cup \bar{B}}\log(I \otimes \rho_{\bar{B}}) \right\} \\
        & =\operatorname{Tr}[ \rho_{R \cup \bar{B}} \log(\rho_{R \cup \bar{B}})] - \sum_{ij} \lambda_{ij}\log(r_{i}) \ket{i, j} \bra{i, j} - \lambda_{ij}\log({\bar{b}}_{i}) \ket{i, j} \bra{i, j} \\
        & = S(\rho_{R \cup \bar{B}}) - S(\rho_{R}) - S(\rho_{\bar{B}}) \\
        & = I(R:\bar{B}),
    \end{aligned}
\end{align}
where in the second equality we used the fact that 
\begin{align}
    \operatorname{span}\left\{\operatorname{ker}\left(\rho_R \otimes I\right), \operatorname{ker}\left(I \otimes \rho_{\bar{B}}\right)\right\}=\operatorname{ker}\left(\rho_{R} \otimes \rho_{\bar{B}}\right) \subset \operatorname{ker} \rho_{R \cup \bar{B}}
\end{align}
as one property in defining the relative entropy. The third equality assumes the Schdmit decomposition of $\rho_{R \cup \bar{B}}$:
\begin{align}
    \rho_{R \cup \bar{B}} = \sum^{r}_{n} \sigma_{n} \ket{n}\bra{n}_{R \cup \bar{B}} = \sum_{ij} \lambda_{ij} \ket{i, j} \bra{i, j}.
\end{align}
As a result,
\begin{align}
    \rho_R = \sum_i (\sum_j \lambda_i)\ket{i}\bra{i}_R \equiv \sum_i r_i \ket{i}\bra{i}_R
\end{align}
and similarly for $\rho_{\bar{B}}$. We note the following properties of quantum entropy
\begin{itemize}
    \item $S(\rho \otimes \sigma) = S(\rho) + S(\sigma)$,

    \item $D(\rho_0 \otimes \rho_1 \| \sigma_0 \otimes \sigma_1) = D(\rho_0 \| \sigma_0) + D(\rho_1 \| \sigma_1)$.
\end{itemize}
Using the above properties, we can also rewrite $D(\rho_{R \cup \bar{B}} \| \rho_{R} \otimes \rho_{\bar{B}} ) = I(R : \bar{B}) = |S(\rho_R \otimes \rho_{\bar{B}}) - S(\rho_{R \cup \bar{B}})|$. Mutual information can be thought of as a useful criterion to determine how much information is lost. Alternatively, if we wish to express in terms of fidelity, which is defined in the average case \cite{Swingle2022}, it is shown that the saturation of mutual information $I(R: B \cup \operatorname{MEM})$ is a necessary condition to obtain the optimal recovery. Infidelity measures, such as the ones used in \cite{Tajima2022, kong2022near}, often involve the purified distance $P(\rho, \sigma)$ which is related with the trace norm distance
\begin{align} \label{eq: trace-purified}
    \frac{1}{2}\|\rho-\sigma\|_1 \leq P(\rho, \sigma) \leq \sqrt{2\|\rho-\sigma\|_1}.
\end{align}
It turns out that the smallness of the mutual information $I(R: \bar{B})$ is a necessary condition for the decoupling in terms of the trace norm distance and purified distance. For a pathological review, given two arbitrary density states $\rho, \sigma$, 
\begin{align}
    \begin{aligned}
        D(\rho \| \sigma ) \geq 0, \quad \text{and  } D(\rho \| \sigma )=0 \text{ if and only if } \rho = \sigma.
    \end{aligned}
\end{align}
This inequality is known as the \emph{Klein inequality}. First note that when $\rho_{R \cup \bar{B}}$ is completely decoupled as a product state, it is easy to see that $I(R: \bar{B}) = 0$ and Eq.\eqref{eq: trace-purified} trivially vanishes, while  A sense of continuity exists for the mutual information in this case, known as the \emph{Fannes-Audenaert inequality}. Let
\begin{align}
    \delta = \frac{1}{2} \| \rho_{R \cup \bar{B}} - \rho_{R} \otimes \rho_{\bar{B}} \|_1,
\end{align}
it holds that
\begin{align}
    I(R: \bar{B}) \leq \delta \log(d_{R \cup \bar{B}} -1) + H(\delta, 1 - \delta) \leq \delta \log(d_{R \cup \bar{B}}) + H(\delta, 1 - \delta).
\end{align}
If $\delta$ is assumed to be sufficiently small, the classical Shannon entropy
\begin{align}
    H(\delta, 1-\delta) = - \delta \log(\delta) - (1-\delta) \log(1-\delta) \approx \delta (1 - \delta + \log(\frac{1}{\delta})) \leq \delta\log(\frac{e}{\delta}).
\end{align}
In this regime, we have the bound
\begin{align}
    I(R: \bar{B}) \leq \delta \log(\frac{e d_{R \cup \bar{B}}}{\delta}).
\end{align}
As long as $\frac{1}{\delta} = o(n)$ where $n$ is the total number of physical qudits, the expression $\delta \log(\frac{e d_{R \cup \bar{B}}}{\delta})$ decays asymptotically to $0$. Hence, there must exist $n$ such that, $\forall \epsilon > 0$, $\delta \log(\frac{e d_{R \cup \bar{B}}}{\delta}) < \epsilon$. Hence the continuity relation establishes that the smallness of mutual information is a necessary condition to a good decoupling via trace norm distance or by Eq.\eqref{eq: trace-purified} the purified distance. Conversely, by the quantum Prinsker inequality, we have that
\begin{align}\label{eq: continuity-equation}
    2 \delta^2 \leq I(R: \bar{B}) \leq \delta \log(\frac{e d_{R \cup \bar{B}}}{\delta}),
\end{align}
and thus the (asymptotic) smallness of mutual information to the $1/4$ power can be regarded as a sufficient condition to conclude decoupling in trace distance and purified distance. 

%----------------------------------------------------------------------------------------------------------

\section{Characterizing designs by commutant under group representation}\label{sec:CommutantTheory}

To study whether an ensemble forms a $k$-design, major approaches include computing the frame potential of the ensemble or analyzing the commutant algebra in the representation space. In this appendix, we illustrate these strategies in detail, establish their mathematical relationship, and finally move on to determining the commutant of the group $\mathcal{U}_\times$ of $\text{SU}(d)$-symmetric unitaries (Theorem \ref{Thm:Commutant}) and discussing the limitation of considering frame potentials in the presence of $\text{SU}(d)$ symmetry. These perspectives connect various approaches for characterizing unitary designs such as the tensor product expander \cite{HarrowTEP08,HarrowTEP09,harrow2016local,Haferkamp2021} and the frame potential \cite{RobertsChaos2017,junyu2017chaos,hunter2019unitary,junyu2020chargescrambler,brian2022linear}. Computations involving unitary 2-designs under symmetries in Section \ref{sec:Application} are largely based on theses facts.

%------------------------------------------------------------------------------------------------------------------------------------------------

\subsection{Quantum $k$-fold channel}\label{sec:k-channel}

We first provide the definition of a $k$-fold channel \cite{Dankert2006PRA,harrow2016local,hunter2019unitary} or the $k$-th moment (super-)operator associated with a compact group $G$.

\begin{definition}
	Given a compact group $G$ with Haar measure $\mu$ and a unitary representation $\rho$ on the concerned Hilbert space $\mathcal{H}$. For any operator $M \in \operatorname{End}(\mathcal{H}^{\otimes k})$, the \emph{$k$-fold channel} twirled by the Haar measure $\mu$ over $G$ acting on $M$ is given by 
	\begin{align}\label{eq: tpe}
		T_k^G(M) = \int_G d\mu(g) \rho^{\otimes k}(g) M (\rho^{\otimes k}(g))^\dagger = \int_G dU U^{\otimes k} M U^{\dagger \otimes k},
	\end{align}
	where we denote the matrix representations of group elements simply by $U$ and $V$ on the right-hand side of the above equation. Despite its integral form, as a super-operator, $T_k^G(\cdot)$ is merely a linear map acting on $\operatorname{End}(\mathcal{H}^{\otimes k})$ and can be reformulated as the \emph{$k$-th moment (super)-operator}:
	\begin{align}
		T_k^G = \int_G U^{\otimes k } \otimes \bar{U}^{ \otimes k} dU.
	\end{align}
	Replacing $G$ by an arbitrary ensemble $\mathcal{E}$, $T_k^{\mathcal{E}}$ can be analogously defined, which provides a basis for the study of (approximate) $k$-designs.
\end{definition}

In later contexts, when we write $T_k^G$ for certain a compact group $G$, the integral is automatically understood to be carried out over the Haar measure.
Since the Haar measure is left-invariant, $T_k^G(M)$ commutes with the $k$-fold tensor product representation $\rho^{\otimes k}$ of $G$:
\begin{align}
	V^{\otimes k} T_k^G(M) V^{\dagger \otimes k} = \int_G (VU)^{\otimes k} M (UV)^{\dagger \otimes k} dU = T_k^G(M).
\end{align}
Putting it another way, $T_k^G$ projects $M$ into the \emph{commutant algebra} 
\begin{align}
	\operatorname{Comm}_k(G) \vcentcolon = \{M \in \operatorname{End}(\mathcal{H}^{\otimes k}); U^{\otimes k} M = M U^{\otimes k} \}, 
\end{align}
i.e., the subspace of all operators that commute with the tensor product representation $\rho^{\otimes k}$. The operator $T_k^G$ is surjective since if $M \in \operatorname{Comm}(M)$, by definition we have $T_k^G(M) = M$. Hence it is a \emph{projector} from $\operatorname{End}(\mathcal{H}^{\otimes k})$ onto $\operatorname{Comm}_k(G)$. Then we obtain the following identity either by the invariance of the Haar measure or by the property of projection: 
\begin{align}
	T_k^G(T_k^G(M)) = \int_G dU dV (VU)^{\otimes k} M (UV)^{\dagger \otimes k} = \int_G d(UV) (VU)^{\otimes k} M (UV)^{\dagger \otimes k} = T_k^G(M),
\end{align}
which further implies that $T_k^G$ has eigenvalues either 0 or 1.

For the common case of unitary designs without any symmetry assumptions, $G = \operatorname{U}(d^n) \equiv \operatorname{U}(N)$ with $\rho^{\otimes k}$ is given by the $k$-fold tensor products of fundamental representation of $\text{U}(N)$. Therefore, by the Schur--Weyl duality and the double commutant theorem (cf. Schur--Weyl duality on an $n$-qudit system), the commutant algebra is isomorphic to the representation of the symmetric group algebra $\mathbb{C}[S_k]$ which permutes elements from $\mathcal{H}^{\otimes k}$. In the presence of $\text{SU}(d)$ symmetry, the group of interest is replaced by $\mathcal{U}_\times$ defined in Section \ref{sec:SchurWeyl} and we denote by $T_k^{\mathcal{U}_\times}$ the  corresponding $k$-th moment operator. 

%--------------------------------------------------------------------------------------------------------------------------

\subsection{Approximate generation of unitary $k$-designs }\label{sec:Definitions}

The viewpoint that $T_k^G$ is a projector onto the commutant of $G$ provides a foundation for the characterization of approximate unitary $k$-designs. We adopt the following strong definition \cite{harrow2016local,harrow2023approximate,SUd-k-Design2023} (see also e.g.,~Refs.~\cite{Dankert2006PRA,Harrow2design2009,Liu_2018,hunter2019unitary,Haferkamp2021,Gao2022,vanDam2002,Low2010} for various other definitions as well as comparison of operator norms): 

\begin{definition}\label{def:AppDesign}
	Given a compact group $G$, an ensemble of unitaries $\mathcal{E}$ is called an \emph{$\epsilon$-approximate unitary $k$-design with respect to  $G$} if the following matrix inequality holds in the sense of complete positivity (i.e., $A \leq_{\mathrm{cp}} B$ means $B - A$ is completely positive):
	\begin{align}
		(1 - \epsilon) T_k^G \leq_{\mathrm{cp}} T_k^{\mathcal{E}} \leq_{\mathrm{cp}} (1 + \epsilon) T_k^G. 
	\end{align} 
	We denote by $c_{\mathrm{cp}}(\mathcal{E}, k)$ the smallest constant $\epsilon$ achieving the above bound.
\end{definition}

\begin{remark}
	There are various other conditions for the definition of approximate $k$-designs in the literature, including the following:
	\begin{enumerate}
		\item The induced $2$-norm of the difference of $k$-th moment operators satisfies
		\begin{align}
			\Vert T_k^{\mathcal{E}} - T_k^G \Vert_{2 \to 2} \leq \epsilon.
		\end{align}
		We denote by $g(\mathcal{E}, k)$ the smallest constant $\epsilon$ achieving the above bound. Viewing super-operators $T_k^{\mathcal{E}}, T_k^G$ as ordinary operators, the induced $2$-norm is exactly the infinity norm that we have used in the main text. When the operator is Hermitian and positive semidefinite, $g(\mathcal{E}, k)$ is simply the largest eigenvalue of $T_k^{\mathcal{E}} - T_k^G$.
		
		\item The diamond norm of the difference of $k$-th moment operators satisfies
		\begin{align}
			\Vert T_k^{\mathcal{E}} - T_k^G \Vert_{\diamond} \leq \epsilon.
		\end{align}
		We denote by $c_\diamond(\mathcal{E}, k)$ the smallest constant $\epsilon$ achieving the above bound.
	\end{enumerate}
	These conditions are related by the following inequalities:
	\begin{align}
		& \frac{c_{\mathrm{cp}}(\mathcal{E}, k)}{N^{2k}} \leq c_\diamond(\mathcal{E}, k) \leq 2c_{\mathrm{cp}}(\mathcal{E},k),  \\
		&\frac{g(\mathcal{E},k)}{2N^k} \leq c_{\mathrm{cp}}(\mathcal{E}, k) \leq N^{2k}g(\mathcal{E},k), \label{ineq:cp}
		\\ 
		&\frac{g(\mathcal{E}, k)}{N^k} \leq c_\diamond(\mathcal{E}, k) \leq N^k g(\mathcal{E},k).
	\end{align}
\end{remark}

\begin{lemma}\label{lemma:EnsembleHermitian}
	Recall that $T_k^{\mathcal{E}}$ is merely a linear map on $\operatorname{End}(\mathcal{H}^{\otimes k})$. With further conditions on the measure $\nu$ of $\mathcal{E}$ being specified, $T_k^{\mathcal{E}}$ satisfies the following properties:
	\begin{enumerate}
		\item If $\nu$ is left-invariant, then $T_k^{\mathcal{E}}$ is a projector onto $\operatorname{Comm}_k(\mathcal{E})$.
		
		\item If $\nu$ is invariant under inverse, i.e.,
		\begin{align}
			\int_\mathcal{E} d\nu(g) f(g^{-1}) = \int_\mathcal{E} d\nu(g) f(g)
		\end{align}
		for any function $f$ defined on $\mathcal{E}$, then $T_k^{\mathcal{E}}$ is Hermitian.
	\end{enumerate}
	In particular, when $\mathcal{E}$ is taken as the restricted Haar measure over some compact subgroup of $G$, both of the above properties hold.
\end{lemma}    

We have discussed the first property in Section \ref{sec:k-channel}. The second property is also straightforward by using $U^{-1} = U^\dagger$ for unitary representation:
\begin{align}
	T_k^{\mathcal{E} \dagger} = \int_S V^{\dagger \otimes k} \bar{V}^{\dagger \otimes k} dV = \int_S (V^{-1})^{\otimes k} M (\bar{V}^{-1})^{ \otimes k} dV = T_k^\mathcal{E}. 
\end{align} 
Most ensembles encountered in the literature, such as in Refs.~\cite{Znidaric2008,Oliveira2design2007a,Oliveira2design2007b,Harrow2design2009,Brown_2010,harrow2016local,harrow2023approximate,Haferkamp2021,U(1)Design2023,SUd-k-Design2023}, induce Hermitian $k$-fold channels that can be diagonalized with an operator norm equal to the largest absolute value of their eigenvalues. This fact is consistently used in our study. The notion of the frame potential applies more generally to the non-Hermitian case.

Besides, by applying the bi-invariance of the Haar measure and the Fubini theorem, which holds for well-behaved measures including  restricted Haar measures on compact subgroups, we see that $T_k^{\mathcal{E}}$ commutes with $T_k^G$: 
\begin{align}
	\begin{aligned}
		T^{\mathcal{E}}_k T^G_k(M) & = \int_\mathcal{E} \int_G V^{\otimes k}  U^{\otimes k} M U^{\dagger \otimes k} V^{\dagger \otimes k} dU dV = \int_G U^{\otimes k} M U^{\dagger \otimes k} dU = T^G_k(M) \\
		& = \int_\mathcal{E} \int_G U^{\otimes k} V^{\otimes k} M V^{\dagger \otimes k} U^{\dagger \otimes k} dU dV \\
		& = \int_G  \int_\mathcal{E} U^{\otimes k} V^{\otimes k} M V^{\dagger \otimes k} U^{\dagger \otimes k} dV dU = T^G_k T^\mathcal{E}_k(M). 
	\end{aligned}
\end{align}
By Lemma \ref{lemma:EnsembleHermitian}, $T_k^G$ is always diagonalizable. Assuming that $T_k^\mathcal{E}$ is Hermitian and hence diagonalizable, they can be {simultaneously diagonalized}. For example, 
\begin{align}\label{eq:EigenvaluesExample}
	T^G_k = \begin{pmatrix} 1 & & & & & \\ & 1 & & & & \\ & & & 0 & & \\ & & & & 0 & \\ & & & & & 0 \end{pmatrix}, \quad T^\mathcal{E}_k = \begin{pmatrix} 1 & & & & & \\ & 1 & & & & \\ & & & \lambda & & \\ & & & & \mu & \\ & & & & & \nu \end{pmatrix}.
\end{align} 
Obviously, the eigenspace corresponding to the unit eigenvalue of $T_k^G$ is exactly $\operatorname{Comm}_k(G)$ whose eigenvectors, by definition, also commute with the restricted representation on the ensemble $\mathcal{E}$. Therefore, 
\begin{align}\label{eq:AboveFact}
	\operatorname{Comm}_k(G) \subset \operatorname{Comm}_k(\mathcal{E}).
\end{align}
as instantiated in \eqref{eq:EigenvaluesExample}. It is now clear that only when $0 \leq \vert \lambda \vert, \vert \mu \vert, \vert \nu \vert < 1$, the convolution of  $T^\mathcal{E}_k$ converges to $T_k^G$ and thus forms an approximate $k$-design with respect to $G$ in the sense of Definition \ref{def:AppDesign}. Within this framework, evaluating the upper bound of the \emph{second largest absolute eigenvalue} of  $\lambda,\mu,\nu$ helps  determine the convergence speed of $\mathcal{E}$ to unitary $k$-designs. The case in which $T^\mathcal{E}_k$ is non-Hermitian can be addressed by calculating the frame potential as introduced later.

As a basic application of this method, suppose that $\mathcal{E}$ is taken as a compact subgroup of $G$, e.g, a one-parameter subgroup, equipped with the Haar measure inherited from that of $G$. Then $T^\mathcal{E}_k$ is also a projector and the eigenvalues $\lambda$, $\mu$, and $\nu$ exemplified above is either 0 or 1. Therefore, $T^\mathcal{E}_k = T^G_k$ if and only if $\operatorname{Comm}_k(\mathcal{E}) = \operatorname{Comm}_k(G)$, which further indicates the following simple but important conclusion.

\begin{fact}\label{Thm:GroupDesign}
	If the unitary ensemble $\mathcal{E}$ is given by a compact Lie subgroup of $G$ with restricted Haar measure, then $\mathcal{E}$ either forms an exact unitary $k$-design or it can never generate a unitary $k$-design in the approximate sense, meaning that it cannot generate a unitary $k$-design with arbitrary precision in terms of any measure defined in Definition \ref{def:AppDesign} or converge to a unitary $k$-design.
\end{fact}

%----------------------------------------------------------------------------------------------------------

\subsection{Random walks on compact groups}\label{sec:RandomWalk}

We now review the relationship between unitary $k$-designs and \emph{random walks on compact groups} \cite{Diaconis1988,Applebaum2014,Meckes2019}, which would elaborate on the strategy of computing the second largest eigenvalue with more deep insights from probability and representation theory of groups. To begin with, let us define the convergence of measures over groups:

\begin{definition}
	Suppose that $\{\mu_p\}$ is a sequence of probability densities/measures over a compact group $G$. Then it \emph{converges in the weak star topology},  or simply \emph{converges weakly}, to the Haar measure $\mu$, denoted by $\mu_p \xrightarrow{w} \mu$, if for any continuous (and automatically bounded) function $f$ defined on $G$,
	\begin{align}\label{eq:MeasureConvergence}
		\lim_{p \to \infty} \int_G f(g) d\mu_p(g)  = \int_G f(g) d\mu(g), \ \text{ or equivalently, } \ \lim_{p \to \infty} \mathbb{E}_{\mu_p}(f) = \mathbb{E}_{\mu}(f) .
	\end{align}
	Note that the limit is considered independently for each single function $f$. Requiring \emph{uniform convergence} for all $f$ is somewhat too strong for continuous compact groups \cite{Diaconis1988,Meckes2019}. 
\end{definition}

In order to check whether $\mu_p \xrightarrow{w} \mu$, we can use the following Lévy continuity theorem generalized from classical Euclidean space to compact groups, which translates the convergence of expectations by \emph{Fourier transformation} to the convergence of certain operators acting on irreducible representations of $G$.  

\begin{theorem}[Lévy continuity theorem \cite{Applebaum2014}] 
	Given any irrep $\pi$ of a compact group $G$ and any density function $\nu$, the operator
	\begin{align}
		\hat{\nu}(\pi) \vcentcolon = \int_G \pi(g) d\nu(g)
	\end{align}
	acting on the representation space of $\pi$ is called the Fourier transform or characteristic function of $\nu$. The convergence of $\mu_n \to \mu$ defined above is equivalent to the convergence of matrix entries $\hat{\mu}_n(\pi)_{ij} \to \hat{\mu}(\pi)$ for all inequivalent $G$ irreps.
\end{theorem}

Let us check the Fourier transformation of the Haar measure $\mu$:
\begin{align}
	\hat{\mu}(\pi) = \int_G \pi(g) d\mu(g) = \begin{cases} 1 & \pi \text{ is the trivial representation} \\ 0 & \text{otherwise}
	\end{cases}.
\end{align}
This is due to the so-called \emph{Schur orthogonality} \cite{Goodman2009,Tolli2009}. Since $\hat{\mu}_n(\pi) \equiv 1$ on the trivial representation, the operator norm 
\begin{align}
	\Big\Vert \bigoplus_\lambda \hat{\mu}_n(\pi) - \bigoplus_\lambda \hat{\mu}(\pi) \Big\Vert
\end{align}
evaluated over all inequivalent $G$ irreps or the second largest absolute value of eigenvalues of $\bigoplus_\lambda \hat{\mu}_n(\pi)$ (if it has a discrete spectrum) determines whether and how fast $\mu_n$ converges to $\mu$. 

\begin{remark}
	A \emph{random walk} on the group $G$ is simply a sequence $\{S_n\}$ of \emph{random variables} $S_n = X_1 \cdots X_n$ for which the $X_i$ are independent random variables with values in the group $G$ distributed according to the same density $\nu$. This induces a sequence of densities $\{\nu^{\ast n}\}$, with which one can examine the convergence properties via the previous theorem. 
	
	On the other hand, when studying unitary $k$-designs, we define $k$-th moment operator
	\begin{align}
		T_k^{\mathcal{E}} = \int_\mathcal{E} V^{\otimes k} \otimes \bar{V}^{\otimes k} dV
	\end{align} 
	of an unitary ensemble $\mathcal{E}$ and compare it with $T_k^{G}$. The ensemble is sampled multiple times, imitating a random walk on a quantum circuit. We also note that the tensor product $V^{\otimes k} \otimes \bar{V}^{\otimes k}$ from the integral can  in principle be further decomposed with respect to the irreps of $G$, thus we can interpret $T_k^{\mathcal{E}}$ as a \emph{truncated} Fourier transform of the measure $\nu$ prescribed in $\mathcal{E}$. 
	
	Consequently, the formation of unitary $k$-designs is weaker compared to the convergence of measures. For instance, $\text{SU}(N)$ is an exact $k$-design to $\text{U}(N)$ for arbitrary $k$, but $\text{SU}(N) \neq $ $\text{U}(N)$ so one cannot say that the Haar measure of $\text{SU}(N)$ converges to that of $\text{U}(N)$. Even when comparing the integral with respect to these measures, there are mismatches: let $\det$ denote the determinant function, then
	\begin{align}
		\int_{\operatorname{SU}(N)} \det V dV = 1 \neq  \int_{\operatorname{U}(N)} \det U dU = 0.
	\end{align}
	The formal reason is that there is no guarantee that $V^{\otimes k} \otimes \bar{V}^{\otimes k}$ encompasses all inequivalent irreps even when $k \to \infty$, e.g., the 1-dimensional representation $\det$ given by taking the determinant \cite{Fulton1997,Goodman2009}. However, unitary $k$-designs are more practical and relevant for quantum computation,  where we focus on the conjugate actions of unitaries on density matrices as $U^{\otimes k} \rho U^{\dagger \otimes k}$.
\end{remark}

%----------------------------------------------------------------------------------------------------------

\subsection{Frame potential, spectral form factors, and $k$-invariance}\label{sec:FramePotential}

We now demonstrate that, for the characterization of $k$-design properties, the perspective of defining $k$-fold channels and computing their second largest eigenvalues is closely related to the following two concepts: the \emph{spectral form factor} $R^{\mathcal{E}}_{2k}$ and the \emph{frame potential} $F^{(k)}_{\mathcal{E}}$, which are widely used in recent physics literature \cite{Gross2006,RobertsChaos2017,junyu2017chaos,hunter2019unitary,junyu2020chargescrambler,brian2022linear}. The spectral form factor is defined as
\begin{align}\label{eq: form-factor}
	R^{\mathcal{E}}_{2k} \vcentcolon = \int_{\mathcal{E}} dU |\operatorname{tr}(U)|^{2k}.
\end{align}
Using the facts that traces interchange with integrals and the identity $\operatorname{tr} (U \otimes V) = \operatorname{tr} U \operatorname{tr} V$, we obtain
\begin{align}
	R^{\mathcal{E}}_{2k} = \operatorname{tr} \int_{\mathcal{E}} U^{\otimes k } \otimes \bar{U}^{ \otimes k}  dU = \operatorname{tr} T_k^{\mathcal{E}}. 
\end{align}
In particular, when $\mathcal{E} = G$, $R^G_{2k} = \operatorname{tr} T_k^G$ measures precisely the {dimension} of the commutant $\operatorname{Comm}_k(G)$ explained previously. For the case without symmetry, let $G = U(d^n) \equiv U(N)$. It is well-known by Schur--Weyl duality that $R^{G}_{2k} = k!$ when $k < d^n = N$ \cite{Goodman2009,Tolli2009}. In the most general setting of arbitrarily large $k$, $R^{G}_{2k}$ is proved to be equal to the number of permutations having no \emph{increasing subsequence} of length greater than $d^n$ \cite{Rains1998}, which relates to the so-called \emph{increasing subsequence problem} from combinatorics \cite{Sagan01}.

The frame potential measures the 2-norm distance between a given ensemble and the Haar-random unitary: 
\begin{align}
	F^{(k)}_{\mathcal{E}} = \int_{\mathcal{E}} dU dV \Vert \operatorname{tr}(UV^\dagger) \Vert^{2k}.
\end{align}
Comparing to the spectral form factor, the frame potential is defined for more general choices of ensembles:

\begin{proposition}\label{prop: FramePotential-Commutant}
	Given an arbitrary ensemble $\mathcal{E}$,
	\begin{align}
		F^{(k)}_{\mathcal{E}} = \operatorname{tr} (T_k^{\mathcal{E} \dag} T_k^{\mathcal{E}} ),
	\end{align}
	which is simply the squared 2-norm of $T_k^{\mathcal{E}}$. If $T_k^{\mathcal{E}}$ is Hermitian, $F^{(k)}_{\mathcal{E}} = \operatorname{tr} ( (T_k^{\mathcal{E}})^2 )$. When $\mathcal{E}$ is compact subgroup, 
	\begin{align}
		F^{(k)}_{\mathcal{E}} = R^{\mathcal{E}}_{2k}.
	\end{align}
	In either cases, the frame potential is lower-bounded by $F^{(k)}_G = R^G_{2k}$.
	\begin{proof}
		Similarly to how we derived $R^{\mathcal{E}}_{2k} = \operatorname{tr} T_k^{\mathcal{E}}$  above, 
		\begin{align}
			F_{\mathcal{E}}^{(k)} & = \int_{U,V \in \mathcal{E}} \vert \operatorname{tr}(U^\dagger V) \vert^{2k} dV dU = \int_{U,V \in \mathcal{E}} \operatorname{tr} ((U^\dagger V)^{\otimes k} \otimes (\overline{U^\dagger V})^{\otimes k}) dV dU \notag \\
			& = \operatorname{tr} \int_{U,V \in \mathcal{E}} (U^\dagger V)^{\otimes k} \otimes (\overline{U^\dagger V})^{\otimes k} dV dU = \operatorname{tr} \int_{U \in \mathcal{E}} \int_{V \in \mathcal{E}} \Big( U^{\dagger \otimes k} \otimes \bar{U}^{\dagger \otimes k} \Big)  \Big( V^{\otimes k} \otimes \bar{V}^{\otimes k} \Big) dV dU \\
			& = \operatorname{tr} \big( T_k^{\mathcal{E} \dag} T_k^{\mathcal{E}} \big). \notag
		\end{align}	
		When $\mathcal{E}$ is taken as a compact subgroup of $G$ with the restricted Haar measure, $T_k^{\mathcal{E}} = T_k^{\mathcal{E} \dag}$ becomes a projector by Lemma \ref{lemma:EnsembleHermitian}. Then,
		\begin{align}
			F_{\mathcal{E}}^{(k)} = \operatorname{tr} \big( (T_k^{\mathcal{E}})^2 \big) = \operatorname{tr}T^{\mathcal{E}}_k = R^{\mathcal{E}}_{2k}.
		\end{align}
		Finally, regardless of whether or not $T_k^{\mathcal{E}}$ is Hermitian, $T_k^{\mathcal{E} \dagger} T_k^{\mathcal{E}}$ is always diagonalizable with nonnegative eigenvalues. Let $W_{k,\mathcal{E}}^{\lambda = 1}$ denote its unit eigenspace, for any $M \in \operatorname{Comm}_k(G)$,  
		\begin{align}
			\begin{aligned}
				& U^{\otimes k} M = U^{\otimes k} M, U^{\dag \otimes k} M = U^{\dag \otimes k} M \\
				\implies\quad & T_k^{\mathcal{E} \dagger} T_k^{\mathcal{E}} (M) = M \\
				\implies \quad & F^{(k)}_G = R^G_{2k} = \operatorname{tr}(T_k^G) = \dim \operatorname{Comm}_k(G) \leq \dim W_{k,\mathcal{E}}^{\lambda = 1} \leq F_{\mathcal{E}}^{(k)}.
			\end{aligned}
		\end{align}
		This concludes the proof. 
	\end{proof}
\end{proposition}

Note that the last statement, $F^{(k)}_G \leq F_{\mathcal{E}}^{(k)}$, can be verified directly using the bi-invariance of Haar measure \cite{RobertsChaos2017,junyu2017chaos,hunter2019unitary,junyu2020chargescrambler,brian2022linear} when comparing the real Haar randomness with that assigned by the ensemble $\mathcal{E}$. 

We now introduce the notion of $k$-invariance \cite{junyu2020chargescrambler}, which characterizes how invariant the ensemble is under the Haar-random unitary. For a given ensemble $\mathcal{E}$, its $k$-invariance $I_{\mathcal{E}}^{(k)}$ is defined by
\begin{align}
	I_{\mathcal{E}}^{(k)} = F_{\mathcal{E}}^{(k)} - F_{\tilde{\mathcal{E}}}^{(k)},
\end{align}
where $\tilde{\mathcal{E}}$ is obtained from averaging $\mathcal{E}$ over the Haar measure
\begin{align}
	\tilde{\mathcal{E}} = \left\{\int_{G} d W\left(W U W^{\dagger}\right): U \in \mathcal{E}\right\}.
\end{align}
By employing methods similar to those used above,
\begin{align}\label{eq:k-invariance}
	F^{(k)}_{\tilde{\mathcal{E}}} & = \operatorname{tr}\left( \int_{\tilde{\mathcal{E}}} d \tilde{U} \int_{\tilde{\mathcal{E}}} d \tilde{V} \left( \tilde{U}^{\dagger \otimes k} \otimes \bar{\tilde{U}}^{\dagger \otimes k}\right) \left( V^{\otimes k} \otimes \bar{V}^{\otimes k} \right) \right) \notag \\
	& = \operatorname{tr} \left( \int_{\mathcal{E}} dU \int_{G} dW  \int_{\mathcal{E}} dV \int_{G} dX  \left((WU^\dag W^\dag)^{\otimes k} \otimes  (\overline{WU^\dag W^\dag})^{\otimes k} \right) \Big( (XVX^\dag)^{\otimes k} \otimes (\overline{XVX^\dag})^{\otimes k} \Big) \right) \\
	& = \operatorname{tr}\Big( \big( T^G_{k} T^{\mathcal{E} \dag}_k T^G_{k} \big) \big( T^G_{k} T^{\mathcal{E}}_k T^G_{k} \big) \Big) = \operatorname{tr}\Big( T^G_{k} T^{\mathcal{E} \dag}_k T^{\mathcal{E}}_k T^G_{k} \Big) = \operatorname{tr}( T^G_{k})  \notag.
\end{align}
Obviously, the $k$-invariance $I^{(k)}_{\mathcal{E}} \geq 0$. If $\mathcal{E}$ is an exact $k$-design, $I^{(k)}_{\mathcal{E}} = 0$. We call any ensemble for which $I^{(k)}_{\mathcal{E}} = 0$  $k$-invariant. 

\begin{remark}
	With the introduction of the commutant, the spectral gap of the $k$-fold channel, and the frame potential characterizations of ensembles provided above, we can now discuss their relationship. We assume that $T_k^{\mathcal{E}}$ is Hermitian with nonnegative eigenvalues, i.e., is positive semidefinite. In exotic scenarios in which $T_k^{\mathcal{E}}$ is non-Hermitian, we consider the operator $T_k^{\mathcal{E} \dagger} T_k^{\mathcal{E}}$ instead. 
	
	With this assumption, in the language of commutant theory, we evaluate the \emph{second largest eigenvalue} $\lambda$ of $T_k^{\mathcal{E}}$. It has been shown in Refs.~\cite{HarrowTEP08,HarrowTEP09,harrow2016local} that, for a random walk with $p$ steps (or a random circuit of depth $p$) to achieve an $\epsilon$-approximate $k$-design, i.e.,~achieve $\Vert (T_k^{\mathcal{E}})^p - T_k^G \Vert_{\diamond} \leq \epsilon$, the smallest $p$ needed is 
	\begin{align}
		\tilde{p} = \frac{1}{\log \frac{1}{\lambda}} \log \frac{N^{2k}}{\epsilon}.
	\end{align}
	Since for large $x$ it holds that $x \leq (x+1) \log (x+1)$, we have
	\begin{align}\label{eq:InequalitySecondEigenvalue}
		\frac{1}{\log \frac{1}{\lambda}} = \frac{1}{\log \frac{1 - \lambda}{\lambda} + 1} \leq \frac{\frac{1 - \lambda}{\lambda} + 1}{\frac{1 - \lambda}{\lambda}} = \frac{1}{1- \lambda}\quad\implies\quad  \tilde{p} \leq  \frac{1}{1- \lambda} \log \frac{N^{2k}}{\epsilon}.
	\end{align}
	Therefore, a polynomial \emph{spectral gap} between the first and second largest eigenvalues of $T_k^\mathcal{E}$ guarantees an efficient random circuit scheme (also see Refs.~\cite{vanDam2002,Low2010}). 
	
	On the other hand, suppose that we consider frame potential $F^{(k)}_{\mathcal{E}}(p)$ for each $p$. Then, the inequality
	\begin{align}\label{eq:InequalityFramePotential}
		\Vert (T_k^{\mathcal{E}})^p - T_k^G \Vert_{\diamond}^2 \leq N^{2k} \Big(F^{(k)}_{\mathcal{E}}(p) - F^{(k)}_G \Big)
	\end{align}
	can be applied to bound the difference under the diamond norm \cite{hunter2019unitary,junyu2020chargescrambler,brian2022linear}. By Proposition \ref{prop: FramePotential-Commutant}, the knowledge of both the second largest eigenvalue $\lambda$ and $F^{(k)}_G = \dim \operatorname{Comm}_k(G)$ is sufficient for bounding 
	\begin{align}\label{eq:FramePotential-Eigenvalue}
		F^{(k)}_{\mathcal{E}}(p) = \operatorname{tr}( (T_k^{\mathcal{E}})^{2p}) = \sum_i \lambda_i^{2p},
	\end{align}
	where the $\lambda_i$ denote eigenvalues of $T_k^{\mathcal{E}}$. Conversely, with knowledge of $F^{(k)}_{\mathcal{E}}(p)$ for all $p \in \mathbb{N}$ one can uniquely determine these eigenvalues $\lambda_i$  through the so-called \emph{moment problem} studied in number theory and algebraic geometry \cite{schmudgen2020ten}. Even though it is generally impossible to explicitly solve $\lambda_i$  in Eq.~\eqref{eq:FramePotential-Eigenvalue}, this consideration unifies the concepts of frame potential, commutant, operator traces and eigenvalues in the context of characterizing $k$-design properties.
\end{remark}

%------------------------------------------------------------------------------------------------------------------------------------------------

\subsection{Commutant of the group of $\text{SU}(d)$-symmetric unitaries}\label{sec:SnCommutant}

At the end of this section, we briefly explain the potential difficulty of working with the frame potential in the presence of $\text{SU}(d)$ symmetry, and describe the commutant of $\mathcal{U}_\times$ with an explicit spanning set, which is core to our computation of $T_{k=2}^{\mathrm{CQA}}(M)$ in Section \ref{sec:Application} for the analysis of several applications in this work.

To lay a basis for the proofs, we first consider the commutant algebra $\operatorname{Comm}_k(\operatorname{U}(d^n))$ for a generic $n$-qudit system $H = V^{\otimes n}$ with no symmetry. It is well-known by Schur--Weyl duality and the double commutant theorem \cite{Goodman2009,Tolli2009} that
\begin{align}
	\operatorname{Comm}_k(\operatorname{U}(d^n)) = \operatorname{span}\{ \sigma, \sigma \in S_k \},
\end{align}
where the $\sigma$ should be understood as permutations on the $k$-fold tensor product $\mathcal{H}^{\otimes k}$. To express $\sigma$ explicitly, note that any $M \in \operatorname{Comm}_k(\operatorname{U}(N))$ is just an element from $\operatorname{End}(\mathcal{H}^{\otimes k})$ which has a standard basis given by tensor products of matrix units, i.e., matrices $E_{ij}$ with unit entry in the position $(i,j)$ and zero entries elsewhere. Under the computational basis $\{ \ket{i} \}_{i=1}^N$,
\begin{align}
	E_{ij} = \ket{i} \bra{j}.
\end{align}
Then it is straightforward to check by definition that
\begin{align}\label{eq:Permutation}
	\sum_{i,j} E_{ii} \otimes E_{jj}, \quad \sum_{i,j} E_{ij} \otimes E_{ji}, \text{ and } \sum_{i,j,k,r,s} E_{is} \otimes E_{jk} \otimes E_{kr} \otimes E_{rj} \otimes E_{si}
\end{align}
correspond to the identity matrix, the transposition $(1,2)$ on the first two indices, and the permutation $(15) (234)$, which swaps the first and fifth indices while translating the second, third, and forth indices cyclically, respectively. Note that we define these operators on $\mathcal{H}^{\otimes k}$ for arbitrary $k$, and the cumbersome tensor products with identity matrix $I$ in the above expressions are omitted. A general permutation $\sigma$ of cycle type $\mu = (\mu_1,\ldots,\mu_r) \vdash k$ (Definition \ref{def:CycleType}) can be written out following this procedure: when there is a basis vector label $i$ appearing as a covariant (contravariant) index of some matrix unit from the tensor product, it should be assigned again as a contravariant (covariant) index. Besides using the computational basis, permutations can also be expanded by (generalized) Pauli matrices with nice properties. These expansions are all important for the study of unitary $k$-designs \cite{Oliveira2design2007a,Oliveira2design2007b,Znidaric2008,Harrow2design2009,Brown_2010,harrow2016local,RobertsChaos2017,Gross2021}.

We now discuss the more involved case of $\operatorname{Comm}_k(\mathcal{U}_\times)$. Recall that, as we study $\text{SU}(d)$-symmetric quantum circuits,  by Schur--Weyl duality, the entire Hilbert space $\mathcal{H} = V^{\otimes n}$ of qudits decomposes into irreps $S^\lambda$ of $S_n$ with multiplicities $m_{S_n,\lambda}$ (Section \ref{sec:SchurWeyl}):
\begin{align}
	\mathcal{H} \cong \bigoplus_\lambda \mathbbm{1}_{m_{S_n,\lambda}} \otimes S^\lambda.
\end{align}
In Section \ref{sec:SnTheory}, we have introduced the spanning of these irreps by the Young--Yamanouchi basis. We change from the computational basis $\{\ket{i}\}$ to the Young basis $\{\ket{\alpha_T, m}\}$ with $\alpha_T$ labeling a basis vector and $m$ recording the irrep multiplicity by Schur transform \cite{Tolli2009,krovi,Zheng2021SpeedingUL}, and redefine the matrix unit as 
\begin{align}
	E_{(\alpha_T, m), (\alpha_{T'}, m')} = \ket{\alpha_T, m} \bra{\alpha_{T'}, m'}.
\end{align}
It turns out that the commutant $\operatorname{Comm}_k(\mathcal{U}_\times)$ is spanned by ``permutations'' generalized from 
\eqref{eq:Permutation} using the Young--Yamanouchi basis. As a simple but enlightening example, we have
\begin{align}\label{eq:GeneralPermutation}
	\sum_{T_1,T_2} E_{(\alpha_{T_1}, m_1),(\alpha_{T_1}, m_1')} \otimes E_{(\alpha_{T_2}, m_2),(\alpha_{T_2}, m_2')}
\end{align}
generalized from $\sum_{i,j} E_{ii} \otimes E_{jj}$. However, \eqref{eq:GeneralPermutation} no longer represents the identity matrix because the summation is taken within two $S_n$ irreps labeled by the Young diagrams of $T_1$ and $T_2$, but not over the entire space $\mathcal{H}^{\otimes k}$. Besides, the multiplicity indices can vary arbitrarily as there is {no need} to require $m_1 = m_1'$ or $m_2 = m_2'$. We only write covariant and contravariant basis vector labels in pairs. Moreover, when $k = 2$ or $k=5$,
\begin{align}\label{eq:GeneralPermutation2}
	\sum_{T_1,T_2} & E_{(\alpha_{T_1}, m_1), (\alpha_{T_2}, m_2)} \otimes E_{(\alpha_{T_2}, m_2'),(\alpha_{T_1}, m_1')}, \\
	\sum_{T_1,T_2,T_3,T_4,T_5} & E_{(\alpha_{T_1}, m_1), (\alpha_{T_5}, m_5)} \otimes E_{(\alpha_{T_2}, m_2), (\alpha_{T_3}, m_3)} \\
	& \otimes E_{(\alpha_{T_3}, m_3'), (\alpha_{T_4}, m_4)} \otimes E_{(\alpha_{T_4}, m_4'), (\alpha_{T_2}, m_2')} \otimes E_{(\alpha_{T_5}, m_5'), (\alpha_{T_1}, m_1')}. \notag
\end{align}
(\ref{eq:GeneralPermutation}) and (\ref{eq:GeneralPermutation2}) show how to generalize examples of permutations in \eqref{eq:Permutation}. With all these preparations, we prove the following theorem.

\begin{theorem}\label{Thm:Commutant}
	The commutant $\operatorname{Comm}_k(\mathcal{U}_\times)$ is spanned by the collection of all generalized permutations
	\begin{align}
		\sum E_{(\alpha_{T_1}, m_1), (\alpha_{T_2}, m_2)} \otimes E_{(\alpha_{T_3}, m_3), (\alpha_{T_4}, m_4)} \otimes \cdots \otimes E_{(\alpha_{T_{2k-1}}, m_{2k-1} ), (\alpha_{T_{2k}}, m_{2k} )},
	\end{align}
	where the basis vector labels come in pairs and there are no restrictions on multiplicity indices such as Eqs.~\eqref{eq:GeneralPermutation} and \eqref{eq:GeneralPermutation2}.
\end{theorem}
\begin{proof}
	By definition, matrix representations $U$ of group elements $g \in \mathcal{U}_\times$ are just collections of unitary matrices acting on inequivalent $S_n$ irrep blocks with identical copies on the multiplicity spaces (Eq.~\eqref{eq:GroupElements}). So the conjugation action of $\mathcal{U}_\times$ on any simple tensor product of matrix units is given by 
	\begin{align}
		(U_1 E_{(\alpha_{T_1}, m_1), (\alpha_{T_2}, m_2)} U_2^\dagger) \otimes (U_3 E_{(\alpha_{T_3}, m_3), (\alpha_{T_4}, m_4)} U_4^\dagger) \otimes \cdots \otimes (U_{2k-1} E_{(\alpha_{T_{2k-1}}, m_{2k-1} ), (\alpha_{T_{2k}}, m_{2k} )} U_{2k}^\dagger) 
	\end{align}
	where the $U_i$ are unitaries acting on decomposed $S_n$ irreps spanned by $\{ \ket{\alpha_{T_i}} \}$. A generic element $M \in \operatorname{End}(\mathcal{H}^{\otimes k})$ is a linear combination of simple tensor products of matrix units:
	\begin{align}\label{eq:SimpleTensors}
		M = \sum_{T_i,M_i} c_{T_i,M_i} E_{(\alpha_{T_1}, m_1), (\alpha_{T_2}, m_2)} \otimes E_{(\alpha_{T_3}, m_3), (\alpha_{T_4}, m_4)} \otimes \cdots \otimes E_{(\alpha_{T_{2k-1}}, m_{2k-1}), (\alpha_{T_{2k}}, m_{2k})} ,
	\end{align}
	as they form a standard basis for $\operatorname{End}(\mathcal{H}^{\otimes k})$. Belonging to $\operatorname{Comm}_k(\mathcal{U}_\times)$ means being invariant under the conjugate action of an arbitrary $U = \rho(g)$. 
	
	We first take the $U_i$ as arbitrary diagonal phase matrices. Then the conjugate action accounts for scalar products with phase factors on each simple tensor. Since these simple tensors are linearly independent, being invariant under phase change implies such invariance for individual simple tensors from \eqref{eq:SimpleTensors}, which leads to the requirements on coupling covariant and contravariant basis vector indices. For instance, when $k = 2$, phase-change-invariant simple tensors are of the following forms:
	\begin{align}
		(U_1 E_{(\alpha_{T_1}, m_1), (\alpha_{T_1}, m_1')} U_1^\dagger) \otimes (U_2 E_{(\alpha_{T_2}, m_2), (\alpha_{T_2}, m_2')} U_2^\dagger)
	\end{align}
	or
	\begin{align}
		(U_1 E_{(\alpha_{T_1}, m_1), (\alpha_{T_2}, m_2)} U_2^\dagger) \otimes (U_2 E_{(\alpha_{T_2}, m_2'), (\alpha_{T_1}, m_1')} U_1^\dagger).
	\end{align}
	With this example, let $U_1$ be a matrix that exchanges arbitrary rows and columns as
	\begin{align}
		U_1 = \begin{pmatrix} 1 & 0 & 0 \\ 0 & 0 & 1 \\ 0 & 1 & 0		\end{pmatrix},
	\end{align}
	while  $U_2$ is set to be the identity matrix. Conjugated by these kind of $\mathcal{U}_\times$ group elements, the basis vector label $\alpha_{T_1}$ varies arbitrarily inside the $S_n$ irrep acted on by $U_1$. Therefore, being invariant indicates that we should take summations over $T_1, T_2$, yielding  
	\begin{align}
		\sum_{T_1,T_2} E_{(\alpha_{T_1}, m_1),(\alpha_{T_1}, m_1')} \otimes E_{(\alpha_{T_2}, m_2),(\alpha_{T_2}, m_2')}
	\end{align}
	and
	\begin{align}
		\sum_{T_1,T_2} E_{(\alpha_{T_1}, m_1), (\alpha_{T_2}, m_2)} \otimes E_{(\alpha_{T_2}, m_2'),(\alpha_{T_1}, m_1')}.
	\end{align}
	However, these considerations do not affect the choices of multiplicity labels $m_i$, which is why they are assigned arbitrarily.
	
	We still need to prove that the operators $M$ spanned with the above requirements commute with all other unitaries from $\mathcal{U}_\times$. This is done by considering the Lie algebra $\mathfrak{g} = \mathfrak{L}(\mathcal{U}_\times)$ consisting of 
	\begin{align}
		E \otimes I \otimes \cdots \otimes I + I \otimes E \otimes \cdots \otimes I + \cdots I \otimes \cdots \otimes I \otimes E \in \operatorname{End}(\mathcal{H}^{\otimes k}),
	\end{align}
	where $E$ is an anti-Hermitian matrix respecting the decomposition of $\mathcal{H}^{\otimes k}$ under $\text{SU}(d)$ symmetry. We expand $E$ by matrix units and examine the commutativity. Using the example
	\begin{align}
		M = \sum_{T_1,T_2} E_{(\alpha_{T_1}, m_1), (\alpha_{T_2}, m_2)} \otimes E_{(\alpha_{T_2}, m_2'), (\alpha_{T_1}, m_1')},
	\end{align}
	we have
	\begin{align}
		(E_{(\alpha_T, m), (\alpha_{T'}, m')} \otimes I) M =  \sum_{T_2} E_{(\alpha_T, m), (\alpha_{T_2}, m_2)} \otimes E_{(\alpha_{T_2}, m_2'), (\alpha_{T'}, m')} = M (I \otimes E_{(\alpha_T, m), (\alpha_{T'}, m')})
	\end{align}
	by contracting the same tensor indices. Analogously, $(I \otimes E_{(\alpha_T, m), (\alpha_{T'}, m')}) M = M (E_{(\alpha_T, m), (\alpha_{T'}, m')} \otimes I)$ and hence $M$ commutes with the Lie algebra elements. The general case for arbitrary $k$ can be similarly deduced.
\end{proof}

\begin{remark}
	With this theorem, let us try to compute the frame potential $F^{(k)}_{\mathcal{U}_\times}$ for $G = \mathcal{U}_\times$. The simplest case is $k = 1$, where $\operatorname{Comm}_k( \mathcal{U}_\times)$ is spanned by
	\begin{align}
		\sum_T E_{(\alpha_T, m), (\alpha_T, m')}
	\end{align}
	for each $S_n$ irrep appearing in the decomposition and with arbitrary copies. For $n$-qubit systems, Schur--Weyl duality indicates that there are $p(n,2) = \lfloor \frac{n}{2} \rfloor + 1$ inequivalent $S_n$ irreps corresponding to two-row Young diagrams in the decomposition. Moreover, the multiplicity of the irrep $S^\lambda$ with $\lambda = (n-r,r)$ is $n-2r + 1$. Therefore, for qubits,
	\begin{align}
		F^{(1)}_{\mathcal{U}_\times} = \dim \operatorname{Comm}_{k = 1}\left( \mathcal{U}_\times\right) = \sum_{0 \leq r \neq \leq \lfloor n/2 \rfloor} (n-2r + 1)^2.
	\end{align}
	When $k = 2$, 
	\begin{align}
		\begin{aligned}
			F^{(2)}_{\mathcal{U}_\times} = \dim \operatorname{Comm}_{k = 2}\left( \mathcal{U}_\times\right) = & (n+1)^4 + 2\sum_{1 \leq r \leq \lfloor n/2 \rfloor} (n-2r + 1)^4 \\
			& + 2\sum_{0 \leq r \neq s \leq \lfloor n/2 \rfloor} (n-2r + 1)^2 (n-2s + 1)^2,
		\end{aligned}
	\end{align}
	where the first two terms on the right-hand side appear when the four indices of basis vectors $\alpha_T$ from \eqref{eq:GeneralPermutation} are all selected from an equivalent $S_n$ irrep. As the trivial irrep ($r = 1$) of $S_n$ is 1-dimensional with no freedom for exchanging indices, we separate it from the second term. The third term counts the number of cases in which two pairs of indices are chosen from two inequivalent $S_n$ irreps. Then the general case follows. 
	
	For qudits, the number of inequivalent $S_n$ irreps is $p(n,d)$, which denotes the number of Young diagrams with $n$ boxes and at most $d$ rows (Definition \ref{def:PartitionFunction}). Unfortunately, there is no closed-form formula for $p(n,d)$, except the asymptotic formula due to Ramanujan, Hardy, and Uspensky \cite{Ramanujan1918,Uspensky1920} and  bounds later developed in Refs.~\cite{Maroti2003,Wladimir2009}. Consequently, it may be infeasible to calculate the frame potential $F^{(k)}_{\mathcal{U}_\times}$ for the group of $\text{SU}(d)$-symmetric unitaries, let alone comparing it with $F^{(k)}_{\mathcal{E}}$ in \eqref{eq:InequalityFramePotential}. To circumvent this obstacle to some extent, we will work exclusively with the commutant algebra $\operatorname{Comm}_k(\mathcal{U}_\times)$ when presenting our main results.
\end{remark}

%------------------------------------------------------------------------------------------------------------------------------------------------

\section{Applications of SU$(d)$ Symmetric $2$-design}\label{sec:Application}

We now present the details for analyzing quantum scrambling, quantum error correction and quantum machine learning with the presence of SU$(d)$ symmetry using our previously developed results on SU$(d)$-symmetric $k$-design.

%------------------------------------------------------------------------------------------------------------------------------------------------

\subsection{General setting and explicit expansion for $T_{k=2}^{\operatorname{CQA}}$}\label{sec:Setting}

We first set up the scenario for general quantum information recovery process provide necessary mathematical techniques for computation under continuous symmetry. As described in the main text, the entire $n$-qudit system is divided into two parts $A,\bar{A}$ where $A$ contains $k$ qudits carrying the information entangled with a reference system $R$ depicted as the density matrix $\rho_{R \cup A}$. The left part $\bar{A}$ bears a prescribed mixed state $\rho_{\bar{A}}$, which may be further purified by adding a memory MEM as $\rho_{\bar{A} \cup \text{MEM}}$. After being processed under certain quantum channel, we try to recover or decode the information by $n-t$ qudits from the system, denoted by $B$, even by the memory MEM if it is applicable.

We adopt \emph{abstract index notation} in order to give a clear and precise mathematical description for the process, plus tensor network diagrams illustrating the idea graphically. As a brief introduction, let us rewrite the initial state $\rho_{R \cup A}, \rho_{\bar{A} \cup \text{MEM}}$ by
\begin{align}
	Q^{R A} Q_{R' A'}, \quad  P^{\bar{A} M} P_{\bar{A}' M'},
\end{align}
where $Q^{R A}, P^{\bar{A} M}$ denote the purified states in $A$ and $\bar{A}$, respectively. Superscript indices correspond to vectors (ket) while subscript indices correspond to dual vectors (bra). Then  
\begin{align}
	U\indices{^{B \bar{B}}_{A \bar{A}}} Q^{R A} P^{\bar{A} M}
\end{align}
represents the unitary action $U\indices{^{B \bar{B}}_{A \bar{A}}}$ on initial states. When a pair of the same indices appears, we contract tensors in that place which concretely means that we apply \emph{Einstein summation convention} to sum all components of operators with the same indices in that place. One notable reason to use abstract index notation is that \emph{tensor contractions} turn out to be matrix actions on vectors, matrix products, taking (partial) traces and so forth in different circumstances, which provides a unified framework for our later computations. As a reminder, the initial states may be left mixed in some cases (see Section \ref{sec:Covariant}) and we simply write $P\indices{^{\bar{A}}_{\bar{A}'}}$ for the whole density matrix.

The total density matrix under unitary action is written as
\begin{align}
	D\indices{^{R B \bar{B} M}_{R' B' \bar{B}' M'} } = (U\indices{^{B \bar{B}}_{A \bar{A}}} Q^{R A}  P^{\bar{A} M}) (Q_{R' A'}  P_{\bar{A}' M'} U\indices{_{B' \bar{B}'}^{A' \bar{A}'}} ).
\end{align}
We are going to expand the following terms using abstract indices 
\begin{align}
	\rho_{B \cup \text{MEM}} = D\indices{^{R B \bar{B} M}_{R B' \bar{B} M'} } = D\indices{^{B M}_{B' M'} } 
	\text{ and } \text{Tr}( \rho_{B \cup \text{MEM}}^2 ) & = D\indices{^{B M}_{B' M'} }  D\indices{^{B' M'}_{B M} },
\end{align}
which interestingly yields a form of out-of-time-ordered correlation \cite{roberts2016chaos,RobertsChaos2017} for Pauli operators (Eq.\eqref{eq:S-B-MEM}). The central topic in this section is analytically averaging these quantities or their \emph{variants} under Haar randomness with or without the symmetry. We proceed to show some basic computational techniques and apply them to the concrete questions later. 

%------------------------------------------------------------------------------------------------------------------------------------------------

\subsubsection*{Computation of $\text{Tr}( \rho_{B \cup \text{MEM}}^2 )$}

We first expand $\text{Tr}(\rho_{B \cup \text{MEM}}^2)$ by abstract index notation with
\begin{align}
	\begin{aligned}
		\rho_{B \cup \text{MEM}} = D\indices{^{R B \bar{B} M}_{R B' \bar{B} M'} } = D\indices{^{B M}_{B' M'} } & = (U\indices{^{B \bar{B}}_{A \bar{A}}} Q^{R A} P^{\bar{A} M}) (Q_{R A'} P_{\bar{A}' M'} U\indices{_{B' \bar{B}}^{A' \bar{A}'}} ) \\
		& = U\indices{^{B \bar{B}}_{A \bar{A}}} \mu\indices{^{A}_{A'}} U\indices{_{B' \bar{B}}^{A' \bar{A}'}} P^{\bar{A} M} P_{\bar{A}' M'},
	\end{aligned}
\end{align}
where $Q^{R A} Q_{R A'} = \mu\indices{^{A}_{A'} }$ is the density matrix of the initial state in $A$. Similarly, the matrix product $\rho_{B \cup \text{MEM}}^2$ gives us 
\begin{align}
	P^{\bar{A} M} P_{\bar{A}' M'} P^{\bar{A}'' M'} P_{\bar{A}''' M} = \rho\indices{^{\bar{A}}_{\bar{A}'''} } \rho\indices{^{\bar{A}''}_{\bar{A}'} }
\end{align}
with $\rho$ denote the density matrix of the initial state in $\bar{A}$. Then
\begin{align}\label{eq:Tr-rho-20}
	\text{Tr}( \rho_{B \cup \text{MEM}}^2 ) & = D\indices{^{B M}_{B' M'} }  D\indices{^{B' M'}_{B M} } \notag \\
	& =	\Big( U\indices{^{B \bar{B}}_{A \bar{A}}} \mu\indices{^{A}_{A'}} U\indices{_{B' \bar{B}}^{A' \bar{A}'}} P^{\bar{A} M} P_{\bar{A}' M'} \Big)
	\Big( U\indices{^{B' \bar{B}'}_{A'' \bar{A}''}} \mu\indices{^{A''}_{A'''}} U\indices{_{B \bar{B}'}^{A''' \bar{A}'''}} P^{\bar{A}'' M'} P_{\bar{A}''' M} \Big) \notag \\
	& = (U\indices{^{B \bar{B}}_{A \bar{A}}} 
	\mu\indices{^{A}_{A'}}) 
	(U\indices{_{B' \bar{B}}^{A' \bar{A}'}} \rho\indices{^{\bar{A}''}_{\bar{A}'} } ) 
	(U\indices{^{B' \bar{B}'}_{A'' \bar{A}''}}
	\mu\indices{^{A''}_{A'''}} )
	(U\indices{_{B \bar{B}'}^{A''' \bar{A}'''}}  
	\rho\indices{^{\bar{A}}_{\bar{A}'''} }) \\
	& = \hat{U}\indices{^{B \bar{B}}_{A' \bar{A}}} 
	\tilde{U}\indices{_{B' \bar{B}}^{A' \bar{A}''}}  
	\hat{U}\indices{^{B' \bar{B}'}_{A''' \bar{A}''}}	\tilde{U}\indices{_{B \bar{B}'}^{A''' \bar{A}}} \notag \\
	& = \hat{U}\indices{^{B \bar{B}}_{A' \bar{A}}} \tilde{U}\indices{_{B \bar{B}'}^{A''' \bar{A}}}	\hat{U}\indices{^{B' \bar{B}'}_{A''' \bar{A}''}} 	\tilde{U}\indices{_{B' \bar{B}}^{A' \bar{A}''}}. \notag
\end{align}

Let $T\indices{^{Q_A''' Q_A'}_{A' A'''}}, T\indices{^{Q_B' Q_B}_{B B'}}$ be the transpositions exchanging $A, A'$ and $B, B'$ repetitively. As mentioned in Section \ref{sec:SnCommutant}, it is straightforward to check by definition that transpositions can be written as
\begin{align}\label{eq:SWAP1}
	\sum_{i,j} E_{ij} \otimes E_{ji},
\end{align} 
where $E_{ij}$ is a matrix unit acting on the concerned (sub-)space which is labeled by $A$ or $B$ at present. It is also conventional to adopt another expansion of transpositions given by the \emph{generalized Pauli matrices}. Let $d_A$ denote the dimension of $A$, we define
\begin{align}
	X\ket{j} = \ket{j+1}, \quad  Z\ket{j} = \omega^j \ket{j}
\end{align} 
with $X\ket{d_A} = \ket{1}$ and $\omega = e^{2i\pi/d_A}$. Those are ingredients to build $d_A^2$ generalized Pauli matrices $P = X^i Y^j, i,j = 1,...,d_A$ acting on $d_A$-dimensional spaces. Then by definition,
\begin{align}\label{eq:SWAP2}
	\frac{1}{d_A} \sum_P P \otimes P^\dagger
\end{align}
is exactly the same transposition as Eq.\eqref{eq:SWAP1}. One can restrict to the case of qubits ($d = 2$) and substitute common Pauli matrices and their tensor products to gain more intuition. We expand transpositions on $B, B'$ by similar methods. To distinguish the notation from those for $A,A'$, we denote them by $V$ and $W$ respectively. Then
\begin{align}\label{eq:Tr-rho-2}
	\begin{aligned}
		& \hat{U}\indices{^{B \bar{B}}_{A' \bar{A}}} \tilde{U}\indices{_{B \bar{B}'}^{A''' \bar{A}}}	\hat{U}\indices{^{B' \bar{B}'}_{A''' \bar{A}''}} 	\tilde{U}\indices{_{B' \bar{B}}^{A' \bar{A}''}}  \\
		= & T\indices{^{Q_A''' Q_A'}_{A' A'''}} 	
		T\indices{^{Q_B' Q_B}_{B B'}} 
		\hat{U}\indices{^{B \bar{B}}_{Q_A' \bar{A}}} \tilde{U}\indices{_{Q_B \bar{B}'}^{A''' \bar{A}}} \hat{U}\indices{^{B' \bar{B}'}_{Q_A''' \bar{A}''}} 	\tilde{U}\indices{_{Q_B' \bar{B}}^{A' \bar{A}''}}  \\
		= & \frac{1}{d_A d_B} \sum_{W,V}  W\indices{^{Q_A'''}_{A'}} W\indices{_{A'''}^{Q_A'}} V\indices{^{Q_B'}_{B}} V\indices{_{B'}^{Q_B}}
		\hat{U}\indices{^{B \bar{B}}_{Q_A' \bar{A}}} \tilde{U}\indices{_{Q_B \bar{B}'}^{A''' \bar{A}}} \hat{U}\indices{^{B' \bar{B}'}_{Q_A''' \bar{A}''}} 	\tilde{U}\indices{_{Q_B' \bar{B}}^{A' \bar{A}''}}   \\
		= & \frac{1}{d_A d_B} \sum_{W,V} \Big( \hat{U}\indices{^{B \bar{B}}_{Q_A' \bar{A}}} W\indices{_{A'''}^{Q_A'}} \tilde{U}\indices{_{Q_B \bar{B}'}^{A''' \bar{A}}} \Big) V\indices{_{B'}^{Q_B}}
		\Big( \hat{U}\indices{^{B' \bar{B}'}_{Q_A''' \bar{A}''}} W\indices{^{Q_A'''}_{A'}}	\tilde{U}\indices{_{Q_B' \bar{B}}^{A' \bar{A}''}} \Big) V\indices{^{Q_B'}_{B}} \\
		= & \frac{1}{d_A d_B} \sum_{W,V} \Big( \hat{U}\indices{^{B \bar{B}}_{Q_A' \bar{A}}} W\indices{_{A'''}^{Q_A'}} \tilde{U}\indices{_{Q_B \bar{B}'}^{A''' \bar{A}}} \Big) 
		\Big( V\indices{_{B'}^{Q_B}} \delta\indices{^{\bar{B}'}_{\bar{B}'''}} \Big) 
		\Big( \hat{U}\indices{^{B' \bar{B}'''}_{Q_A''' \bar{A}''}} W\indices{^{Q_A'''}_{A'}}	\tilde{U}\indices{_{Q_B' \bar{B}''}^{A' \bar{A}''}} \Big)
		\Big( V\indices{^{Q_B'}_{B}} \delta\indices{^{\bar{B}''}_{\bar{B}}} \Big)  \\
		= & \frac{1}{d_A d_B} \sum_{W,V} \Big(	U\indices{^{B \bar{B}}_{A \bar{A}}} 
		\big( (\mu\indices{^{A}_{Q_A'}} W\indices{_{A'''}^{Q_A'}}) \rho\indices{^{\bar{A}}_{\bar{A}'''} } \big) U\indices{_{Q_B \bar{B}'}^{A''' \bar{A}'''}} \Big) 
		\Big( V\indices{_{B'}^{Q_B}} \delta\indices{^{\bar{B}'}_{\bar{B}'''}} \Big)  \\
		& \Big( U\indices{^{B' \bar{B}'''}_{A'' \bar{A}''}}
		\big( (\mu\indices{^{A''}_{Q_A'''}} W\indices{^{Q_A'''}_{A'}}) \rho\indices{^{\bar{A}''}_{\bar{A}'} } \big)	U\indices{_{Q_B' \bar{B}''}^{A' \bar{A}'}} \Big)
		\Big( V\indices{^{Q_B'}_{B}} \delta\indices{^{\bar{B}''}_{\bar{B}}} \Big). 
	\end{aligned}
\end{align}
We have the following indices when taking contractions:
\begin{align*}
	(A,A',A'',A'''), \quad (B,B'), \quad (Q_A',Q_A'''), \quad (Q_B,Q_B'), \quad
	(\bar{A},\bar{A}',\bar{A}'',\bar{A}'''), \quad  (\bar{B},\bar{B}',\bar{B}'',\bar{B}''').
\end{align*}

\begin{figure}
	\centering
	\includegraphics[width=7in]{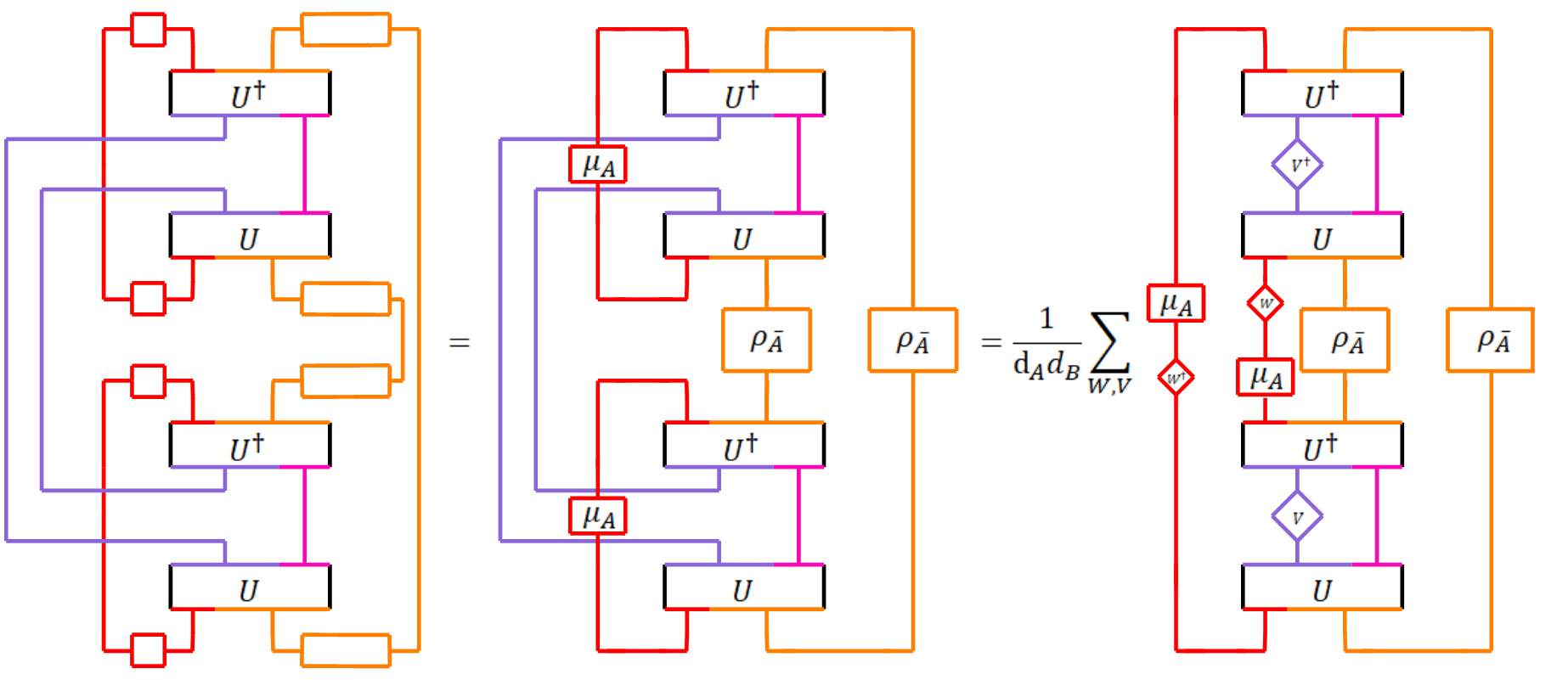}
	\caption{Tensor network diagram illustrating the deduction from Eq.\eqref{eq:Tr-rho-20} to \eqref{eq:Tr-rho-2}.}
	\label{}
\end{figure}

It would be formidable to integrate Eq.\eqref{eq:Tr-rho-2} in its most general form. Instead, assume the initial states are maximally entangled, i.e., $\mu = \frac{1}{d_A}I, \rho = \frac{1}{d_{\bar{A}}}I$. Then
\begin{align}\label{eq:S-B-MEM}
	\begin{aligned}
		\text{Tr}( \rho_{B \cup \text{MEM}}^2 ) = & \frac{1}{d_A d_B d_A^2 d_{\bar{A}}^2 } \sum_{W,V} \Big( U\indices{^{B \bar{B}}_{A \bar{A}}} 
		\big( W\indices{_{A'''}^{A}} \delta\indices{^{\bar{A}}_{\bar{A}'''} } \big) U\indices{_{B \bar{B}'}^{A''' \bar{A}'''}} \Big) 
		\Big( V\indices{_{B'}^{B}}  \delta\indices{^{\bar{B}'}_{\bar{B}'''}} \Big)  \\
		& \Big( U\indices{^{B' \bar{B}'''}_{A'' \bar{A}''}} 
		\big( W\indices{^{A''}_{A'}} \delta\indices{^{\bar{A}''}_{\bar{A}'} } \big)	U\indices{_{B' \bar{B}''}^{A' \bar{A}'}} \Big)
		\Big( V\indices{^{B'}_{B}} \delta\indices{^{\bar{B}''}_{\bar{B}}} \Big) \\
		= & \frac{1}{d^k d^{n-t} d^{2n}} \sum_{W,V} \text{Tr}\Big( (U W^\dagger U^\dagger) V^\dagger (U W U^\dagger) V \Big).
	\end{aligned}
\end{align}

%------------------------------------------------------------------------------------------------------------------

\subsubsection*{2-design Formula}

To average Eq.\eqref{eq:S-B-MEM} with several variants appearing in subsequent subsections, we apply the second moment operator $T_{k=2}^{\text{Haar}}$ to $W^\dagger \otimes W$ and then contract with $V^\dagger \otimes V$. Recall that without symmetry, $\text{Comm}_{k=2}(\operatorname{U}(d^n))$ is spanned by the following two basis elements (see Section \ref{sec:SnCommutant}):
\begin{align}\label{eq:U(N)-basis}
	I = \sum_{i,j=1}^{d^n} E_{ii} \otimes E_{jj}, \quad S = \sum_{i,j=1}^{d^n} E_{ij} \otimes E_{ji},
\end{align}
where the indices $i,j$ label any basis, e.g., the computational basis, of the $d^n$-dimensional entire Hilbert space. We orthogonalize this two basis vectors for the commutant. Then for any $M \in \operatorname{End}(H^{\otimes 2})$,
\begin{align}\label{eq:2-design}
	T_{k=2}^{\text{Haar}}(M) = \int_{\operatorname{U}(d^n)} U^{\otimes 2} M U^{\dagger \otimes 2} dU =  \frac{1}{d^{2n} - 1} \Big( \text{Tr}(M) I + \text{Tr}(SM) S - \frac{1}{d^n} \text{Tr}(M) S - \frac{1}{d^n} \text{Tr}(SM) I \Big).
\end{align} 
Substituting $M = W^\dagger \otimes W$ into the above identity, using the expansions from Eq.\eqref{eq:U(N)-basis} and taking contractions with $V^\dagger \otimes V$, we have ($A$ contains $k$ qudits while $B$ contains $n-t$ qudits)
\begin{align}
	& \sum_W \text{Tr}(W^\dagger \otimes W) = \sum_W  \text{Tr}(W^\dagger) \text{Tr}(W) = d^{2n}, \quad \sum_W \text{Tr}(S (W^\dagger \otimes W)) = d^{2k} d^n, \\
	& \sum_V \text{Tr}(V^\dagger V) = d^{2n-2t} d^n, \quad \sum_{i,j,V} \text{Tr}(E_{ij} V^\dagger E_{ji} V) = \sum_V \text{Tr}(V^\dagger) \text{Tr}(V) = d^{2n},
\end{align}
which assembles to  
\begin{align}
	\int \text{Tr}( \rho_{B \cup \text{MEM}}^2 ) dU = \frac{1}{d^{k+n-t} d^{2n}} \Big( \frac{1}{d^{2n} - 1} ( d^{2n - 2t} d^{3n} + d^{2k}d^{3n} - d^{3n} - d^n d^{2k + 2n - 2t} ) \Big).
\end{align}

This technique is immediately applied to calculate the OTOC for arbitrary operator $W$ as long as we assume $\text{Tr}(W) = 0, \text{Tr}(W^\dagger W) \leq d^n$. For nontrivial local Pauli basis element acting on one qudit ($n - t = 1$),
\begin{align}\label{eq:OTOC-general}
	\frac{1}{d^2 - 1} \frac{1}{d^n} \sum_{V \neq I} \text{Tr}\Big( (U W^\dagger U^\dagger) V^\dagger (U W U^\dagger) V \Big) = \frac{1}{d^2 - 1} \frac{1}{d^n} \Big( \frac{1}{d^{2n} - 1} ( -\frac{1}{d^n} d^n (d^2-1) d^n ) \Big) = -\frac{1}{d^{2n} - 1} \to 0
\end{align}
for large $n$.

The systematic framework to compute $T_k^{\text{Haar}}(M)$ for larger $k$ is called \emph{Weingarten calculus} \cite{Collins2003,RobertsChaos2017,Meckes2019}. For our purpose, we only consider the case when $k = 2$ and reformulate Eq.\eqref{eq:2-design} at the presence of SU$(d)$ symmetry. As illustrated in Section \ref{sec:SnCommutant}, $\text{Comm}_{k=2}(\operatorname{U}(d^n))$ is spanned by basis elements resembling those in \eqref{eq:U(N)-basis}, however we need to replace the computational basis indices by Young basis indices consisting of content vector $\alpha_i$ (see Section \ref{sec:SnTheory}) and multiplicities $m$:
\begin{align}\label{eq:CQA-basis}
	& \sum_{i,j} E_{(\alpha_i^\lambda,m_1), (\alpha_i^\lambda, m_2)} \otimes E_{(\alpha_j^\lambda,m_3), (\alpha_j^\lambda, m_4)}, \quad 
	\sum_{i,j} E_{(\alpha_i^\lambda,m_1), (\alpha_i^\lambda, m_2)} \otimes E_{(\alpha_j^\lambda,m_3), (\alpha_j^\lambda, m_4)}, \\
	& \sum_{i,j} E_{(\alpha_i^\lambda,m_1), (\alpha_i^\lambda, m_2)} \otimes E_{(\alpha_j^\mu,m_1'), (\alpha_j^\mu, m_2')}, \quad 
	\sum_{i,j} E_{(\alpha_i^\lambda,m_1), (\alpha_j^\mu, m_1')} \otimes E_{(\alpha_j^\mu,m_2'), (\alpha_i^\lambda, m_2)}, 
\end{align}
We write $\alpha$ with superscripts $\lambda,\mu$ in order to explicitly show the concerned irreps. The first two terms are selected from one irrep $S^\lambda$ (but maybe from different equivalent copies). The last two terms indicates the exceptions that we also have commutant basis elements formed from inequivalent irreps. Let $d_\lambda = \dim S^\lambda$. We orthogonalize these basis vectors, then
\begin{align}
	T_{k=2}^{\text{Haar}}(M) = T_{k=2}^{\text{CQA}}(M) = \int_{\operatorname{U}(\bigoplus_\lambda(S^\lambda))} U^{\otimes 2} M U^{\dagger \otimes 2} dU
\end{align}
can be expanded in the following two kinds of terms ranging from all $S_n$ irreps with multiplicities by Schur--Weyl duality:
\begin{align}\label{eq:2-design-expansion1}
	\begin{aligned}
		\frac{1}{d_\lambda^2 - 1} \Big( & \text{Tr}\Big( (\sum_{i,j} E_{(\alpha_i^\lambda,m_1), (\alpha_i^\lambda, m_2)} \otimes E_{(\alpha_j^\lambda,m_3), (\alpha_j^\lambda, m_4)} ) M \Big)  (\sum_{i,j} E_{(\alpha_i^\lambda,m_1), (\alpha_i^\lambda, m_2)} \otimes E_{(\alpha_j^\lambda,m_3), (\alpha_j^\lambda, m_4)} ) \\
		+ & \text{Tr}\Big( (\sum_{i,j} E_{(\alpha_i^\lambda,m_1), (\alpha_j^\lambda, m_2)} \otimes E_{(\alpha_j^\lambda,m_3), (\alpha_i^\lambda, m_4)} ) M \Big) (\sum_{i,j} E_{(\alpha_i^\lambda,m_1), (\alpha_j^\lambda, m_2)} \otimes E_{(\alpha_j^\lambda,m_3), (\alpha_i^\lambda, m_4)} ) \\
		- \frac{1}{d_\lambda} & \text{Tr}\Big( (\sum_{i,j} E_{(\alpha_i^\lambda,m_1), (\alpha_i^\lambda, m_2)} \otimes E_{(\alpha_j^\lambda,m_3), (\alpha_j^\lambda, m_4)} ) M \Big) (\sum_{i,j} E_{(\alpha_i^\lambda,m_1), (\alpha_j^\lambda, m_2)} \otimes E_{(\alpha_j^\lambda,m_3), (\alpha_i^\lambda, m_4)} ) \\
		-\frac{1}{d_\lambda} & \text{Tr}\Big( (\sum_{i,j} E_{(\alpha_i^\lambda,m_1), (\alpha_j^\lambda, m_2)} \otimes E_{(\alpha_j^\lambda,m_3), (\alpha_i^\lambda, m_4)} ) M \Big) (\sum_{i,j} E_{(\alpha_i^\lambda,m_1), (\alpha_i^\lambda, m_2)} \otimes E_{(\alpha_j^\lambda,m_3), (\alpha_j^\lambda, m_4)} ).
	\end{aligned}
\end{align} 
and
\begin{align}\label{eq:2-design-expansion2}
	\begin{aligned}
		& \frac{1}{d_\lambda d_\mu} \text{Tr}\Big( (\sum_{i,j} E_{(\alpha_i^\lambda,m_1), (\alpha_i^\lambda, m_2)} \otimes E_{(\alpha_j^\mu,m_1'), (\alpha_j^\mu, m_2')} ) M \Big)  (\sum_{i,j} E_{(\alpha_i^\lambda,m_1), (\alpha_i^\lambda, m_2)} \otimes E_{(\alpha_j^\mu,m_1'), (\alpha_j^\mu, m_2')} ), \\
		& \frac{1}{d_\lambda d_\mu} \text{Tr}\Big( (\sum_{i,j} E_{(\alpha_i^\lambda,m_1), (\alpha_j^\mu, m_1')} \otimes E_{(\alpha_j^\mu,m_2'), (\alpha_i^\lambda, m_2)} ) M \Big)  (\sum_{i,j} E_{(\alpha_i^\lambda,m_1), (\alpha_j^\mu, m_1')} \otimes E_{(\alpha_j^\mu,m_2'), (\alpha_i^\lambda, m_2)} ).
	\end{aligned}
\end{align}
We apply these formulas to study conserved quantity scrambling in the next subsection. 

%-----------------------------------------------------------------------------------------------------------------

\subsection{Spreading of local SU$(d)$ conversed quantities}\label{sec:Charge}

We now apply the expansion of $T_{k=2}^{\text{CQA}}$ to evaluate the out-of-time-ordered correlation (OTOC) $\frac{1}{d^n}\text{Tr}(W^\dagger V^\dagger W V)$ for SU$(d)$ conserved quantities. The local operator basis elements $V$ are first taken from Pauli set by convention and then replaced by SU$(d)$ symmetric local basis defined by permutations. In either case, we verify that for conserved quantities determined by exchanging interaction on qubits ($d = 2$), OTOC is lower bounded by $\Omega(\frac{1}{n^{3/2}})$ (Eq.\eqref{eq:OTOC-SU(2)}) in contrast to exponential decay with no symmetry (Eq.\eqref{eq:OTOC-general}). Moreover, our expansion of $T_{k=2}^{\text{CQA}}$ is easily rephrased for general compact Lie group actions and we showcase an example of OTOC under U$(1)$ symmetry for exchanging interaction and demonstrate a linear decay in Eq.\eqref{eq:OTOC-U(1)}.

%-------------------------------------------------------------------------------------------------

\subsubsection*{OTOC under Local Pauli Basis}

Let $W$ denote a generic SU$(d)$ conserved quantity, i.e., $W$ commutes with $\hat{U}^{\otimes n}$ for any $\hat{U} \in \text{SU}(d)$. The 2-design projection acting on $W^{\otimes 2}$ consists of the following term:
\begin{align}
	\begin{aligned}
		\frac{1}{d_\lambda^2 - 1} \Big( & \text{Tr}\Big( (\sum_{i,j} E_{(\alpha_i^\lambda,m_1), (\alpha_i^\lambda, m_2)} \otimes E_{(\alpha_j^\lambda,m_3), (\alpha_j^\lambda, m_4)} ) (W^\dagger \otimes W) \Big)  (\sum_{i,j} E_{(\alpha_i^\lambda,m_1), (\alpha_i^\lambda, m_2)} \otimes E_{(\alpha_j^\lambda,m_3), (\alpha_j^\lambda, m_4)} ) \\
		+ & \text{Tr}\Big( \sum_{i,j} E_{(\alpha_i^\lambda,m_1), (\alpha_j^\lambda, m_2)} \otimes E_{(\alpha_j^\lambda,m_3), (\alpha_i^\lambda, m_4)} ) (W^\dagger \otimes W) \Big) (\sum_{i,j} E_{(\alpha_i^\lambda,m_1), (\alpha_j^\lambda, m_2)} \otimes E_{(\alpha_j^\lambda,m_3), (\alpha_i^\lambda, m_4)} ) \\
		- \frac{1}{d_\lambda} & \text{Tr}\Big( (\sum_{i,j} E_{(\alpha_i^\lambda,m_1), (\alpha_i^\lambda, m_2)} \otimes E_{(\alpha_j^\lambda,m_3), (\alpha_j^\lambda, m_4)} ) (W^\dagger \otimes W) \Big) (\sum_{i,j} E_{(\alpha_i^\lambda,m_1), (\alpha_j^\lambda, m_2)} \otimes E_{(\alpha_j^\lambda,m_3), (\alpha_i^\lambda, m_4)} ) \\
		-\frac{1}{d_\lambda} & \text{Tr}\Big( (\sum_{i,j} E_{(\alpha_i^\lambda,m_1), (\alpha_j^\lambda, m_2)} \otimes E_{(\alpha_j^\lambda,m_3), (\alpha_i^\lambda, m_4)} ) (W^\dagger \otimes W) \Big) (\sum_{i,j} E_{(\alpha_i^\lambda,m_1), (\alpha_i^\lambda, m_2)} \otimes E_{(\alpha_j^\lambda,m_3), (\alpha_j^\lambda, m_4)} ).
	\end{aligned}
\end{align} 
Since $W$ admits the symmetry,
\begin{align}
	& \text{Tr}\Big( (\sum_{i,j} E_{(\alpha_i^\lambda,m_1), (\alpha_i^\lambda, m_2)} \otimes E_{(\alpha_j^\lambda,m_3), (\alpha_j^\lambda, m_4)} ) (W^\dagger \otimes W) \Big)
	= \sum_{i,j} W^\dagger_{(\alpha_i^\lambda, m_2), (\alpha_i^\lambda, m_1)} W_{(\alpha_j^\lambda, m_4), (\alpha_j^\lambda, m_3)}, \\
	& \text{Tr}\Big( \sum_{i,j} E_{(\alpha_i^\lambda, m_1), (\alpha_j^\lambda, m_2)} \otimes E_{(\alpha_j^\lambda, m_3), (\alpha_i^\lambda, m_4)} ) (W^\dagger \otimes W) \Big)
	= \sum_{i,j} W^\dagger_{(\alpha_j^\lambda, m_2), (\alpha_i^\lambda, m_1)} W_{(\alpha_i^\lambda, m_4), (\alpha_j^\lambda, m_3)} 
\end{align}
vanishes unless $m_1 = m_2 = m_\lambda, m_3 = m_4 = m_\lambda'$, which can be further simplified as
\begin{align}
	& \sum_{i,j} W^\dagger_{(\alpha_i^\lambda, m_2), (\alpha_i^\lambda, m_1)} W_{(\alpha_j^\lambda, m_4), (\alpha_j^\lambda, m_3)}   = \text{Tr}(\Pi_{\lambda,m_\lambda} W^\dagger) \text{Tr}(\Pi_{\lambda,m'_\lambda} W), \\
	& \sum_{i,j} W^\dagger_{(\alpha_j^\lambda, m_2), (\alpha_i^\lambda, m_1)} W_{(\alpha_i^\lambda, m_4), (\alpha_j^\lambda, m_3)} = \sum_{i,j} W^\dagger_{(\alpha_j^\lambda, m_\lambda), (\alpha_i^\lambda, m_\lambda)} W_{(\alpha_i^\lambda, m'_\lambda), (\alpha_j^\lambda, m'_\lambda)} 
\end{align}
where $\Pi_{\lambda,m_\lambda}, \Pi_{\lambda,m'_\lambda}$ restricts the matrix into the irrep $S^\lambda$ corresponding to the multiplicities $m_\lambda,m'_\lambda$ respectively. Obviously, $\Pi_{\lambda,m_\lambda} W = \Pi_{\lambda,m'_\lambda} W$ in terms of their entries, so the last term could be understood as the trace of the matrix product of $W$ with its conjugate in $S^\lambda$. Symbolically, let $\Pi_{\lambda,m_\lambda,m'_\lambda} = \sum_i E_{(\alpha_i,m_\lambda), (\alpha_i, m'_\lambda)}$,
\begin{align}\label{eq:TraceOfProduct}
	\sum_{i,j} W^\dag_{(\alpha_j, m_\lambda), (\alpha_i, m_\lambda)} W_{(\alpha_i, m'_\lambda), (\alpha_j, m'_\lambda)} = \text{Tr}( \Pi_{\lambda,m_\lambda} W^\dag \Pi_{\lambda,m_\lambda,m'_\lambda} W ).
\end{align}

The expansion still contains another two kinds of terms for a pair of inequivalent irreps $S^\lambda \ncong S^\mu$,
\begin{align}
	& \frac{1}{d_\lambda d_\mu} \text{Tr}\Big( (\sum_{i,j} E_{(\alpha_i^\lambda,m_1), (\alpha_i^\lambda, m_2)} \otimes E_{(\alpha_j^\mu,m_1'), (\alpha_j^\mu, m_2')} ) (W^\dagger \otimes W) \Big)  (\sum_{i,j} E_{(\alpha_i^\lambda,m_1), (\alpha_i^\lambda, m_2)} \otimes E_{(\alpha_j^\mu,m_1'), (\alpha_j^\mu, m_2')} ), \\
	& \frac{1}{d_\lambda d_\mu} \text{Tr}\Big( (\sum_{i,j} E_{(\alpha_i^\lambda,m_1), (\alpha_j^\mu, m_1')} \otimes E_{(\alpha_j^\mu,m_2'), (\alpha_i^\lambda, m_2)} ) (W^\dagger \otimes W) \Big)  (\sum_{i,j} E_{(\alpha_i^\lambda,m_1), (\alpha_j^\mu, m_1')} \otimes E_{(\alpha_j^\mu,m_2'), (\alpha_i^\lambda, m_2)} ).
\end{align}
The first term remains when $m_1 = m_2 = m_\lambda, m_1' = m_2' = m_\mu$ but the second term always vanishes since $W$ is assumed to be $\text{SU}(d)$-symmetric.

To compute
\begin{align}
	\frac{1}{d^n} \int_{\mathcal{U}_\times} \text{Tr}\Big( (U W^\dagger U^\dagger) V^\dagger (U W U^\dagger) V \Big),
\end{align}
we contract the above terms with a general basis operator $V$ that may not respect the symmetry. It gives
\begin{align}\label{eq:Charge1}
	\begin{aligned}
		\frac{1}{d_\lambda^2 - 1} \Big( & \text{Tr}(\Pi_{\lambda,m_\lambda} W^\dagger) \text{Tr}(\Pi_{\lambda,m'_\lambda} W) \text{Tr}( \Pi_{\lambda,m_\lambda} V^\dagger \Pi_{\lambda,m'_\lambda} V) \\
		+  & (\sum_{i,j} W^\dagger_{(\alpha_j^\lambda, m_\lambda), (\alpha_i^\lambda, m_\lambda)} W_{(\alpha_i^\lambda, m'_\lambda), (\alpha_j^\lambda, m'_\lambda)} ) (\sum_{i,j} V^\dagger_{(\alpha_j^\lambda, m_\lambda),(\alpha_j^\lambda,m'_\lambda)} V_{ (\alpha_i^\lambda, m'_\lambda) (\alpha_i^\lambda,m_\lambda)} ) \\
		- & \frac{1}{d_\lambda} \text{Tr}(\Pi_{\lambda,m_\lambda} W^\dagger) \text{Tr}(\Pi_{\lambda,m'_\lambda} W) (\sum_{i,j} V^\dagger_{(\alpha_j^\lambda, m_\lambda),(\alpha_j^\lambda,m'_\lambda)} V_{ (\alpha_i^\lambda, m'_\lambda) (\alpha_i^\lambda,m_\lambda)} ) \\
		- & \frac{1}{d_\lambda} (\sum_{i,j} W^\dagger_{(\alpha_j^\lambda, m_\lambda), (\alpha_i^\lambda, m_\lambda)} W_{(\alpha_i^\lambda, m'_\lambda), (\alpha_j^\lambda, m'_\lambda)} )  \text{Tr}( \Pi_{\lambda,m_\lambda} V^\dagger \Pi_{\lambda,m'_\lambda} V),\\
		= \frac{1}{d_\lambda^2 - 1} \Big( & \text{Tr}(\Pi_{\lambda,m_\lambda} W^\dagger) \text{Tr}(\Pi_{\lambda,m'_\lambda} W) \text{Tr}( \Pi_{\lambda,m_\lambda} V^\dagger \Pi_{\lambda,m'_\lambda} V) \\
		+  & \text{Tr}( \Pi_{\lambda,m_\lambda} W^\dagger \Pi_{\lambda,m_\lambda,m'_\lambda} W) \text{Tr}( \Pi_{\lambda,m'_\lambda,m_\lambda} V^\dagger ) \text{Tr}( \Pi_{\lambda,m_\lambda,m'_\lambda} V) \\
		- & \frac{1}{d_\lambda} \text{Tr}(\Pi_{\lambda,m_\lambda} W^\dagger) \text{Tr}(\Pi_{\lambda,m'_\lambda} W)
		\text{Tr}( \Pi_{\lambda,m'_\lambda,m_\lambda} V^\dagger ) \text{Tr}( \Pi_{\lambda,m_\lambda,m'_\lambda} V)  \\
		- & \frac{1}{d_\lambda} \text{Tr}( \Pi_{\lambda,m_\lambda} W^\dagger \Pi_{\lambda,m_\lambda,m'_\lambda} W) \text{Tr}( \Pi_{\lambda,m_\lambda} V^\dagger \Pi_{\lambda,m'_\lambda} V),
	\end{aligned}
\end{align}
with
\begin{align}\label{eq:Charge2}
	\frac{1}{d_\lambda d_\mu} \text{Tr}(\Pi_{\lambda,m_\lambda} W^\dagger) \text{Tr}(\Pi_{\mu,m_\mu} W) \text{Tr}( \Pi_{\lambda,m_\lambda} V^\dagger \Pi_{\mu,m_\mu} V).
\end{align}
As a reminder, if $S^\lambda$ is the 1-dimensional trivial representation (we omit the 1-dimensional sign representation appears whose occurrence demands $d \geq n$), Eq.\eqref{eq:Charge1} degenerates to
\begin{align}
	\text{Tr}(\Pi_{\lambda} W^\dagger) \text{Tr}(\Pi_{\lambda} W) \text{Tr}( \Pi_{\lambda} V^\dagger \Pi_{\lambda} V) = \text{Tr}( \Pi_{\lambda} V^\dagger \Pi_{\lambda} V).
\end{align}

Let us comment on each term before proceeding: 
\begin{enumerate}
	\item Matrix products with restricted projections like $\Pi_{\lambda,m_\lambda}$ have straightforward meaning.
	
	\item As mentioned earlier, Eq.\eqref{eq:TraceOfProduct} can be seen as the \emph{trace of the matrix product} of copies of $W^\dagger$ and $W$ in isomorphic $S_n$ irrep $S^\lambda$, which equals $\text{Tr}(\Pi_{\lambda,m_\lambda} W^\dagger W)$ quantitatively.
	
	\item Resembling \emph{product of traces}, 
	\begin{align}\label{eq:ProductOfTrace}
		\sum_{i,j} V^\dagger_{(\alpha_j^\lambda, m_\lambda),(\alpha_j^\lambda,m'_\lambda)} V_{ (\alpha_i^\lambda, m'_\lambda) (\alpha_i^\lambda,m_\lambda)} = \text{Tr}( \Pi_{\lambda,m'_\lambda,m_\lambda} V^\dagger ) \text{Tr}( \Pi_{\lambda,m_\lambda,m'_\lambda} V) 
	\end{align}
	counts off-diagonal entries of $V$ if $m_\lambda \neq m'_\lambda$ represents different isomorphic copies of $S^\lambda$, which will vanish if $V$ is SU$(d)$ symmetric.
\end{enumerate}

%----------------------------------------------------------------------------------------------------

We still need summing over all inequivalent irreps with different multiplicities. We now take the convention using $V = V_r$ as a Pauli basis operator located at the $r$th site of the system with
\begin{align}
	W = (i,j) - \frac{1}{d}I = \frac{1}{d} \sum_{P \neq I} P_i \otimes P^\dagger_j,
\end{align}  
being the conserved quantity determined by SWAPs, or \emph{exchanging interactions} for $d = 2$:
\begin{align}
	W = (i,j) - \frac{1}{2}I = \frac{1}{2} \sum_{P \neq I} P_i \otimes P^\dagger_j = 2\hat{S}_i \cdot \hat{S}_j,
\end{align}  
where $\hat{S}_i = \frac{1}{2} (X_i + Y_i + Z_i)$ is the familiar spin operator \cite{Tasaki2020}. In order to make an appropriate comparison with the case when no symmetry is considered, $\text{Tr}(W) = 0$ is prescribed and we may further multiply $W$ with the factor $\frac{1}{\sqrt{1 - 1/d^2}}$ with which $\text{Tr}(W^\dagger W) = d^n$.

As a common property of $S_n$ group characters \cite{Ingram1950,Roichman1996,Lassalle2008},
\begin{align}
	\Big\vert \text{Tr}(\Pi_{\lambda,m_\lambda} \sigma) \Big\vert = \Big\vert \chi_\lambda(\sigma) \Big\vert \leq d_\lambda \implies  \Big\vert \text{Tr}(\Pi_{\lambda,m_\lambda} W) \Big\vert \leq (1 - \frac{1}{d}) d_\lambda.
\end{align}
Moreover, since $V$ is unitary,
\begin{align}
	& 0 \leq \text{Tr}( \Pi_{\lambda,m_\lambda} V^\dagger \Pi_{\lambda,m_\lambda'} V) \leq d_\lambda, \quad 0 \leq \text{Tr}( \Pi_{\lambda,m_\lambda} V^\dagger \Pi_{\lambda,m_\mu} V) \leq \min\{d_\lambda,d_\mu\}.
\end{align}
Let $\#(S^\lambda)$ denote the number of multiplicities of $S^\lambda$, which equals the dimension of the dual SU$(d)$ irrep under Schur--Weyl duality. Since there are at most super-polynomially many inequivalent $S_n$ irreps (see Eq.\eqref{eq:Ramanujan}), the sum of absolute value of the last term in Eq.\eqref{eq:Charge1}
\begin{align}\label{eq:LastTerm}
	\frac{1}{d^n} \sum_\lambda \sum_{m_\lambda,m'_\lambda} \frac{1}{d_\lambda^2 - 1} \frac{1}{d_\lambda} \Big\vert \text{Tr}( \Pi_{\lambda,m_\lambda} W^\dagger \Pi_{\lambda,m_\lambda,m'_\lambda} W) \Big\vert \text{Tr}( \Pi_{\lambda,m_\lambda} V^\dagger \Pi_{\lambda,m'_\lambda} V)  \leq \frac{1}{d^n} \sum_\lambda \frac{\#^2(S^\lambda) }{d_\lambda}
\end{align}
decays faster than any inverse polynomial of $n$. It would be omitted later after we study the scaling behavior of other terms from Eq.\eqref{eq:Charge1} and \eqref{eq:Charge2}.

%----------------------------------------------------------------------------------------------------

As a basic result in $S_n$ representation theory, except the trivial representation (and sign representation), the second lowest dimension of $S_n$ irreps is $n - 1$ \cite{Rasala1977,Sagan01,Sellke2020}. Thus $d_\lambda^2 - 1 \approx d_\lambda^2$ for nontrivial case and we sum the first term in Eq.\eqref{eq:Charge1} with Eq.\eqref{eq:Charge2} as
\begin{align}\label{eq:Charge3}
	\begin{aligned}
		\frac{1}{d^n} \Big( & \sum_\lambda \sum_{m_\lambda, m'_\lambda} \frac{1}{d_\lambda^2} \text{Tr}(\Pi_{\lambda,m_\lambda} W^\dagger) \text{Tr}(\Pi_{\lambda,m'_\lambda} W) \text{Tr}( \Pi_{\lambda,m_\lambda} V_r^\dagger \Pi_{\lambda,m'_\lambda} V_r) \\
		& + \sum_{\lambda \neq \mu} \sum_{m_\lambda, m_\mu} \frac{1}{d_\lambda d_\mu} \text{Tr}(\Pi_{\lambda,m_\lambda} W^\dagger) \text{Tr}(\Pi_{\mu,m_\mu} W) \text{Tr}( \Pi_{\lambda,m_\lambda} V_r^\dagger \Pi_{\mu,m_\mu} V_r) \Big) \\
		= \frac{1}{d^n} & \sum_{\lambda, \mu} \sum_{m_\lambda, m_\mu} \frac{1}{d_\lambda d_\mu} \text{Tr}(\Pi_{\lambda,m_\lambda} W^\dagger) \text{Tr}(\Pi_{\mu,m_\mu} W) \text{Tr}( \Pi_{\lambda,m_\lambda} V_r^\dagger \Pi_{\mu,m_\mu} V_r) \\
		= \frac{1}{d^n} & \sum_{\lambda, \mu} \sum_{m_\lambda, m_\mu} \Big( \frac{\chi_\lambda(i,j)}{d_\lambda} - \frac{1}{d}\Big) \Big(\frac{\chi_\mu(i,j)}{d_\mu} - \frac{1}{d} \Big)  \text{Tr}( \Pi_{\lambda,m_\lambda} V_r^\dagger \Pi_{\mu,m_\mu} V_r). 
	\end{aligned}
\end{align}
We are still left with
\begin{align}\label{eq:Charge4}
	\begin{aligned}
		\frac{1}{d^n} \sum_\lambda \sum_{m_\lambda, m'_\lambda} \frac{1}{d_\lambda^2 - 1} \Big( & \text{Tr}( \Pi_{\lambda,m_\lambda} W^\dagger \Pi_{\lambda,m_\lambda,m'_\lambda} W) \text{Tr}( \Pi_{\lambda,m'_\lambda,m_\lambda} V_r^\dagger ) \text{Tr}( \Pi_{\lambda,m_\lambda,m'_\lambda} V_r) \\
		- & \frac{1}{d_\lambda} \text{Tr}(\Pi_{\lambda,m_\lambda} W^\dagger) \text{Tr}(\Pi_{\lambda,m'_\lambda} W)
		\text{Tr}( \Pi_{\lambda,m'_\lambda,m_\lambda} V_r^\dagger ) \text{Tr}( \Pi_{\lambda,m_\lambda,m'_\lambda} V_r) \Big) \\
		= \frac{1}{d^n} \sum_\lambda \sum_{m_\lambda, m'_\lambda} \frac{1}{d_\lambda^2 - 1} \Big( & d_\lambda - (1 - \frac{2}{d}) \chi_\lambda(i,j) - \frac{1}{d_\lambda} \chi_\lambda(i,j)^2 \Big) \text{Tr}( \Pi_{\lambda,m'_\lambda,m_\lambda} V_r^\dagger ) \text{Tr}( \Pi_{\lambda,m_\lambda,m'_\lambda} V_r) \Big) 
	\end{aligned}
\end{align}
which is nonnegative and we are going to estimate them in turns. One more trick may further simplify the computation is that Pauli matrices are similar under unitary transformation: for any two Pauli matrices $P,Q$ there exists $\hat{U} \in \operatorname{SU}(d)$ such that 
\begin{align}\label{eq:PauliSymmetry}
	\hat{U} P \hat{U}^\dagger = Q \implies \text{Tr}(\tilde{W} Q_r \tilde{W} Q_r ) = \text{Tr}(\tilde{W} \hat{U}^{\otimes n} P_r  \hat{U}^{\dagger \otimes n} \tilde{W} \hat{U}^{\otimes n} P_r  \hat{U}^{\dagger \otimes n}) = \text{Tr}(\tilde{W} P_r \tilde{W} P_r ).
\end{align}
because $W$ as well as its time evolution is assumed to be SU$(d)$ symmetric. This fact only holds for non-symmetric case after taking Haar average.

%----------------------------------------------------------------------------------------------------

\subsubsection*{Explicit Computation under SU$(2)$ Symmetry}

To make further computation, we have to restrict ourselves to systems of qubits ($d = 2$) because we need calculating terms like $\text{Tr}( \Pi_{\lambda,m'_\lambda,m_\lambda} V^\dagger )$ under explicit Young basis (total spin basis) constructing by sequential coupling (addition of SU$(2)$ quantum angular momenta) \cite{Harrow06,sergii1,Zheng2021SpeedingUL} and evaluating $\chi_\lambda(i,j)$ in precise. To begin with, as we introduce at the end of Section \ref{sec:SnTheory}, Lemma \ref{lemma: total-ordering2} confirms that for two-row Young diagrams $\lambda = (n-r,r)$,
\begin{align}\label{eq:2RowCharacter}
	\frac{n - 4}{2n - 2} = \chi_{(n/2,n/2)}(i,j) \leq \chi_\lambda(i,j) = \frac{n^2 - (2r+1)n + 2r^2 - 2r}{n^2 -n} \leq \chi_{(n)}(i,j) = 1.
\end{align}
If $n$ is odd, the lower bound is similar and can be calculated through Eq.\eqref{eq:CharacterValue} or \eqref{eq:P_l2}. Thus Eq.\eqref{eq:Charge3} has non-positive terms due to 
\begin{align}
	\Big( \frac{\chi_\lambda(i,j)}{d_\lambda} - \frac{1}{2} \Big) \Big(\frac{\chi_\mu(i,j)}{d_\mu} - \frac{1}{2} \Big)
\end{align}
when one of the characters is larger than $\frac{1}{2}$ and the other is lower. The critical point for $\frac{\chi_\lambda(i,j)}{d_\lambda} - \frac{1}{2} = 0$ happens when $\lambda = (n - r_0,r_0)$ with
\begin{align}
	r_0 = \frac{n}{2} + \frac{1}{2} - \sqrt{\frac{3 n}{4} + \frac{1}{4}}.
\end{align}

One the other hand, let $\lambda = (n-r,r)$ we notice that (we set $V = Z$ by Eq.\eqref{eq:PauliSymmetry})
\begin{align}
	\text{Tr}( \Pi_{\lambda,m_\lambda} Z_r^\dagger \Pi_{\mu,m_\mu} Z_r) = 0 \text{ unless } \mu = (n-r-1,r+1) \text{ or } (n-r+1,r-1).
\end{align}
That is, $\lambda,\mu$ are consecutive two-row Young diagrams with respect to the dominance relation (Definition \ref{def:Dominance}). To see the reason,
\begin{align}
	\Pi_{\lambda,m_\lambda} \Big( \sum_{i=1}^n Z_i \Big) \Pi_{\mu,m_\mu}  
\end{align}
is a zero matrix for any $\lambda \neq \mu$ because the sum $\sum_{i=1}^n Z_i$ of spin-$z$ operators is decomposed according to SU$(2)$ irreps also labeled by $\lambda,\mu$ by Schur--Weyl duality. Hence
\begin{align}
	\Pi_{\lambda,m_\lambda} Z_n \Pi_{\mu,m_\mu} = -\Pi_{\lambda,m_\lambda} \Big( \sum_{i=1}^{n-1} Z_i \Big) \Pi_{\mu,m_\mu}  
\end{align}
is given by the spin-$z$ operator $\sum_{i=1}^{n-1} Z_i$ on the first $n-1$ qubits. From the perspective of either SU$(2)$ sequential coupling (Clebsch-Gordan decomposition) or $S_n$ branching rule \cite{Sagan01,Goodman2009,Tolli2009,Zheng2021SpeedingUL}, it may have nonzero entries only when $\lambda,\mu$ are consecutive because this is only way when they share equivalent SU$(2)$ or $S_{n-1}$ irreps on $n-1$ qubits (this argument fails when $d \geq 3$ with even intricate branching behavior of Young diagrams). Otherwise, after taking trace,
\begin{align}
	\text{Tr}( \Pi_{\lambda,m_\lambda} Z_n^\dagger \Pi_{\mu,m_\mu} Z_n) = 0 \implies \text{Tr}( \Pi_{\lambda,m_\lambda} Z_r^\dagger \Pi_{\mu,m_\mu} Z_r) = 0
\end{align}
as $\Pi_{\lambda,m_\lambda},\Pi_{\mu,m_\mu}$ are invariant under $S_n$ permutations. Together with the previous analysis on $S_n$ characters, non-positive terms in Eq.\eqref{eq:Charge3} only appear for the following pair of Young diagrams
\begin{align}\label{eq:Critical}
	\lambda_0 = (n - \lfloor r_0 \rfloor, \lfloor r_0 \rfloor), \quad \lambda_1 = (n - \lceil r_0 \rceil, \lceil r_0 \rceil).
\end{align}
Let $\epsilon = r_0 - \lfloor r_0 \rfloor \in [0,1)$, 
\begin{align}
	\Big\vert \Big( \frac{\chi_{\lambda_0}(i,j)}{d_{\lambda_0}} - \frac{1}{2} \Big) \Big(\frac{\chi_{\lambda_1}(i,j)}{d_{\lambda_1}} - \frac{1}{2} \Big) \Big\vert = \frac{4\epsilon(1-\epsilon) (\epsilon + \sqrt{1+3n}) (-1 + \epsilon + \sqrt{1+3n})}{n^4 - n^2} = O(\frac{1}{n^3}).
\end{align}
Making use of dimension comparison of $S_n$ irreps of two-row Young diagram (Eq.\eqref{eq:dim-Comparision}), we find that 
\begin{align}
	\frac{1}{2^n} d_{\lambda_0} \approx \frac{1}{2^n} d_{\lambda_1} \approx \frac{1}{2^n} \frac{n - 2r_0 +1}{n - r_0 + 1}\binom{n}{r_0} = \Theta(\frac{1}{n}).
\end{align}
Counting their multiplicity gives $(n - 2r_0)^2 = \Theta(n)$ approximately and in summary, non-positive terms have the following scaling behavior:
\begin{align}
	\Big\vert \frac{2}{2^n} \sum_{m_{\lambda_0}, m_{\lambda_1}} \Big( \frac{\chi_{\lambda_0}(i,j)}{d_{\lambda_0}} - \frac{1}{2} \Big) \Big(\frac{\chi_{\lambda_1}(i,j)}{d_{\lambda_1}} - \frac{1}{2} \Big)  \text{Tr}( \Pi_{\lambda_0,m_{\lambda_0}} V_r^\dagger \Pi_{\lambda_1,m_{\lambda_1}} V_r) \Big\vert = O(\frac{1}{n^3}). 
\end{align}

\vspace{2mm}
%----------------------------------------------------------------------------------------------------

We now show that the above non-positive term is dominated by positive terms with $\lambda = \lambda_0$ in Eq.\eqref{eq:Charge4} which equals
\begin{align}\label{eq:Charge4'}
	 \frac{1}{2^n} \sum_{m_{\lambda_0}} \frac{1}{d_{\lambda_0}^2 - 1} \Big( & d_{\lambda_0} - \frac{1}{d_{\lambda_0}} \chi_{\lambda_0}(i,j)^2 \Big) \text{Tr}( \Pi_{\lambda,m_{\lambda_0}} Z_r ) \text{Tr}( \Pi_{\lambda,m_{\lambda_1}} Z_r) 
	 \approx \frac{1}{2^n} \sum_{m_{\lambda_0}} \frac{1}{d_{\lambda_0}^2 - 1} \frac{3}{4}d_{\lambda_0} \text{Tr}( \Pi_{\lambda,m_{\lambda_0}} Z_r ) \text{Tr}( \Pi_{\lambda,m_{\lambda_0}} Z_r) 
\end{align}
for now. We perform a similar but simpler trick, which only depending on the permutation invariance of $\Pi_{\lambda,m'_\lambda m_\lambda}$ as before:
\begin{align}
	\text{Tr}( \Pi_{\lambda,m'_\lambda m_\lambda} Z_r) = \frac{1}{n} \text{Tr}( \Pi_{\lambda,m'_\lambda m_\lambda} \sum_{i=1}^n Z_i) = \begin{cases} \frac{1}{n} m d_\lambda, & m'_\lambda = m_\lambda, \\ 0, & m'_\lambda \neq m_\lambda,
	\end{cases}
\end{align}
where $m$ is the spin-$z$ component of basis states determined by the multiplicities $m_\lambda$. Suppose $\lambda = (n-r,r)$, it ranges among $n-2r,n-2r-2,...,-n+2r+2,-n+2r$. Let $r = r_0$. Substituting the above identity into Eq.\eqref{eq:Charge4'}, plus the previous dimension comparison, we obtain
\begin{align}\label{eq:OTOC-SU(2)}
	\frac{1}{2^n} \sum_{m_{\lambda_0}} \frac{1}{d_{\lambda_0}^2 - 1} \frac{3}{4}d_{\lambda_0} \text{Tr}( \Pi_{\lambda,m_{\lambda_0}} Z_r ) \text{Tr}( \Pi_{\lambda,m_{\lambda_0}} Z_r)
	\approx \frac{3}{2^{n+2}} \sum_{m = -n+2r_0}^{n-2r_0} \frac{d_\lambda^3}{d_{\lambda_0}^2 - 1} \frac{1}{n^2} m^2 = \Omega(\frac{1}{n^{3/2}}).
\end{align}
Therefore, the desired OTOC of $W$ describing exchanging interaction, which is estimated by the sum of Eq.\eqref{eq:Charge3} and\eqref{eq:Charge4}, is positive and at least $\Omega(\frac{1}{n^{3/2}})$. As a comparison, for general random circuits, as long as $\text{Tr}(W) = 0, \text{Tr}(W^\dagger W) \leq d^n$, OTOC always converges to zero exponentially fast with respect to the system size $n$ (see Eq.\eqref{eq:OTOC-general}). We understand that the current estimation is merely a rough lower bound because there are Young diagrams $\lambda,\mu \neq \lambda_0, \lambda_1$ remained for computation, which demands more advanced techniques to deal with the above summations and we leave it to future work. 

%----------------------------------------------------------------------------------------------------

\subsubsection*{OTOC under SU$(d)$ Symmetric Local Operator Basis}

Even being able to span any operator acting on the $d^n$-dimensional space, generalized Pauli strings does not commute with the SU$(d)$ tensor product action on qudits. Hence we also discuss conserved quantities spreading evaluated under \emph{SU$(d)$ symmetric basis}, which is defined to be an orthogonal basis capable of spanning all SU$(d)$ symmetric operators. By double commutant theorem \cite{Goodman2009,Tolli2009}
\begin{align}
	\text{span}\{\sigma \in S_n\} = \{M \in \operatorname{End}(H), M \hat{U}^{\otimes n} = \hat{U}^{\otimes n} M \},
\end{align} 
thus simply conducting Gram-Schmidt orthogonalization to these permutations synthesizes such a basis. Since we only focus on SU$(d)$ symmetric local operators, especially those 2-local ones, we focus on the following basis elements: 
\begin{align}
	\{I, \frac{1}{\sqrt{1 - 1/d^2}} (\tau - \frac{1}{d}I); \tau = (i,j) \in S_n \}.
\end{align} 
Coincidentally, a single basis element $\frac{1}{\sqrt{1 - 1/d^2}} (\tau - \frac{1}{d}I)$ is also taken as a conserved quantity before.

Applying Eq.\eqref{eq:Charge1} with the fact that $\chi_\lambda$ is invariant for permutations of the same cycle type (Definition \ref{def:CycleType}), we abbreviate $\chi_\lambda(i,j) = \chi_\lambda, \bar{\chi}_\lambda = \frac{\chi_\lambda}{d_\lambda}$ and obtain the following expansion:
\begin{align}\label{eq:ChargeSymmetry}
	\begin{aligned}
		\text{OTOC}_{\text{sym}} = \frac{1}{d^n} \sum_\lambda \frac{\#(S^\lambda)}{d_\lambda^2 - 1} & \frac{1}{1 - 1/d^2} \Big( \Big( \chi_\lambda - \frac{d_\lambda}{d} \Big)^2 \Big( (1 + \frac{1}{d^2} )d_\lambda - \frac{2}{d} \chi_\lambda \Big) 
		+ \Big( (1 + \frac{1}{d^2} )d_\lambda - \frac{2}{d} \chi_\lambda \Big) \Big( \chi_\lambda - \frac{d_\lambda}{d} \Big)^2 \\
		& - \frac{1}{d_\lambda} \Big( \chi_\lambda - \frac{d_\lambda}{d} \Big)^4
		- \frac{1}{d_\lambda} \Big( (1 + \frac{1}{d^2} )d_\lambda - \frac{2}{d} \chi_\lambda \Big)^2 \Big) \\
		= \frac{1}{d^n} \sum_\lambda \frac{\#(S^\lambda) d_\lambda^3}{d_\lambda^2 - 1} & \frac{d^2}{d^2 - 1} \Big( \Big( \bar{\chi}_\lambda - \frac{1}{d} \Big)^2 \Big( 2 + \frac{1}{d^2} - \frac{2}{d} \bar{\chi}_\lambda - \bar{\chi}_\lambda^2 \Big) 
		- \frac{1}{d_\lambda^2} \Big( 1 + \frac{1}{d^2} - \frac{2}{d} \bar{\chi}_\lambda \Big)^2 \Big) \\
		\approx \frac{1}{d^n} \sum_\lambda \#(S^\lambda) d_\lambda & \frac{d^2}{d^2 - 1} \Big( \bar{\chi}_\lambda - \frac{1}{d} \Big)^2 \Big( 2 + \frac{1}{d^2} - \frac{2}{d} \bar{\chi}_\lambda - \bar{\chi}_\lambda^2 \Big) 
	\end{aligned}
\end{align}
where we omit the last term as analogy to what we did in Eq.\eqref{eq:LastTerm}. As mentioned before, if $S^\lambda$ is the 1-dimensional trivial representation, $\bar{\chi}_\lambda = 1$ and we write 
\begin{align}
	\text{Tr}(\Pi_{\lambda} W^\dagger) \text{Tr}(\Pi_{\lambda} W) \text{Tr}( \Pi_{\lambda} V^\dagger \Pi_{\lambda} V) = \Big(1 - \frac{1}{d} \Big)^4
\end{align}
which is already incorporated into the above approximation. Note that each term is positive now. To  lower bound the OTOC under local symmetric basis, we take a closer look at the irrep $S^\lambda$ for which
\begin{align}
	\lambda = (\frac{n-k}{d} + k, \frac{n-k}{d},...,\frac{n-k}{d})
\end{align}
where $k$ is taken from $\{0,...,d-1\}$ such that $n-k$ can be divided by $d$. By Eq.\eqref{eq:CharacterValue},
\begin{align}\label{eq:CharacterDifference}
	\bar{\chi}_\lambda = -\frac{ (d-1) ((d-1)n - kd - k^2)}{n(n-1)d} \implies \Big( \bar{\chi}_\lambda - \frac{1}{2} \Big)^2 = \Theta(\frac{1}{n^2}).
\end{align}
Let $\lambda_0 = (\frac{n-k}{d}, \frac{n-k}{d},...,\frac{n-k}{d})$ obtained from $\lambda$ by discarding $k$ boxes from its first row. Assume $d \leq \frac{n-k}{d}$, the $S_{n-k}$ irrep $S^{\lambda_0}$ is obviously contained in $S^\lambda$. Then Eq.\eqref{eq:dim-Comparision2} indicates that
\begin{align}
	d_\lambda > d_{\lambda_0} \geq \frac{d^{n-k}}{(n-k)^{d(d+2)/2}} > \frac{d^n}{n^{d(d+4)/2}}. 
\end{align}
On the other hand, 
\begin{align}
	\min_{\lambda} \Big( 2 + \frac{1}{d^2} - \frac{2}{d} \bar{\chi}_\lambda - \bar{\chi}_\lambda^2 \Big) = \Big(1 - \frac{1}{d} \Big)^2,
\end{align}
where the minimum is attained on the trivial irrep. Therefore,
\begin{align}
	\text{OTOC}_{\text{sym}} \geq \frac{1}{d^n} d_\lambda & \frac{d^2}{d^2 - 1} \Big( \bar{\chi}_\lambda - \frac{1}{d} \Big)^2 \Big(1 - \frac{1}{d} \Big)^2 = \Omega( \frac{1}{n^{(d+2)^2/2}}).
\end{align}

We wish to note that the calculation is not tight. In the case of qubit, it is still possible to further refine the above lower bound. We first carefully check the scaling behavior of characters of $\lambda = \lambda_0$ or $\lambda_1$ defined in \eqref{eq:Critical}:
\begin{align}
	\bar{\chi}_{\lambda_0} - \frac{1}{2} = \frac{2\epsilon (\sqrt{3n+1} + \epsilon)}{n^2 - n}, \quad \bar{\chi}_{\lambda_1} - \frac{1}{2} = -\frac{2(1 - \epsilon) (\sqrt{3n+1} - 1 + \epsilon)}{n^2 - n},
\end{align}
where we recall that $\epsilon = r_0 - \lfloor r_0 \rfloor \in [0,1)$. We claim that at least one of these two $S_n$ irreps, together with all other sectors corresponding to two-row Young diagrams from $n$-qubit system satisfies
\begin{align}\label{eq:minCharacter}
	\Big( \bar{\chi}_\lambda - \frac{1}{2} \Big)^2 = \Omega(\frac{1}{n^{3}}).
\end{align}
Plugging into Eq.\eqref{eq:ChargeSymmetry}, we conclude that
\begin{align}
	\text{OTOC}_{\text{sym}} = \Big( \frac{1}{2^n} \sum_{\lambda \neq \lambda_0} \#(S^\lambda) d_\lambda \Big) \Omega(\frac{1}{n^{3}}) = \Omega(\frac{1}{n^{3}}) 
\end{align}
where $S^{\lambda_0}$ is omitted because $\Big( \bar{\chi}_{\lambda_0} - \frac{1}{2} \Big)^2$ is assumed to be even smaller than the given scaling. Then the sum is solved using $\sum_\lambda \#(S^\lambda) d_\lambda = 2^n$ by definition and
\begin{align}
	\frac{1}{2^n}d_{\lambda_0} = \Theta(\frac{1}{n}), \quad \#(S^{\lambda_0}) = \Theta(\sqrt{n})
\end{align}
clarified in the previous subsection.

To demonstrate Eq.\eqref{eq:minCharacter}, we find by \eqref{eq:2RowCharacter} that given any $\epsilon \in [0,1)$ the squares of
\begin{align}
	\bar{\chi}_{\lambda_2} - \frac{1}{2} = \frac{2(1+\epsilon) (\sqrt{3n+1} + 1 + \epsilon)}{n^2 - n}, \quad \bar{\chi}_{\lambda_3} - \frac{1}{2} = -\frac{2(2 - \epsilon) (\sqrt{3n+1} - 2 + \epsilon)}{n^2 - n}
\end{align}
are $\Omega(\frac{1}{n^{3}})$ for
\begin{align}
	\lambda_2 \vcentcolon = (n - \lfloor r_0 \rfloor + 1, \lfloor r_0 \rfloor - 1), \quad \lambda_3 \vcentcolon = (n - \lceil r_0 \rceil - 1, \lceil r_0 \rceil + 1).
\end{align}
Other cases follows immediately by the frequently used fact that two-row Young diagrams, as well as the corresponding $S_n$ characters, are totally ordered (Lemma \ref{lemma: total-ordering2}).   

%-------------------------------------------------------------------------------

\subsection{Spreading of local U$(1)$ conversed quantities}\label{sec:U(1)charge}

The expansion of $T_{k=2}^{\text{CQA}}$ (Eq.\eqref{eq:2-design-expansion1} and \eqref{eq:2-design-expansion2}) as well as its restriction to conserved quantity (Eq.\eqref{eq:Charge1} and \eqref{eq:Charge2}) formally works for arbitrary compact Lie group $G$ acting on an $n$-qudit system with the following generic irrep decomposition 
\begin{align}
	H = V^{\otimes n} \cong \bigoplus_\lambda I_{m_\lambda} \otimes S^\lambda
\end{align}
where $S^\lambda$ here denotes one irrep of $G$ and $I_{m_\lambda}$ denotes its multiplicities space. As a concrete example, let us compute the OTOC of $W = (i,j) - \frac{1}{2} I$ under U$(1)$ symmetry, where the concerned group $G$ consists of all unitaries $U$ for which
\begin{align}
	[U, \sum_i \frac{Z_i+I}{2}] = 0 \Leftrightarrow [U, \sum_i Z_i] = 0.
\end{align}
Accordingly, U$(1)$ charge sectors are subspaces of quantum states with the same spin-$z$ components or magnetic quantum numbers. Under computational basis,
\begin{align}
	H = V^{\otimes n} \cong \bigoplus_{k = 0}^n  S^k, \quad S^k = \text{span}\{ \ket{i_1,...,i_n}; k \text{ many of them equal } 0 \}.
\end{align}
and there is obviously no nontrivial multiplicities space. Rewriting Eq.\eqref{eq:Charge1} and \eqref{eq:Charge2} for U$(1)$ symmetry, we have 
\begin{align}\label{eq:U1Charge1}
	\begin{aligned}
		\frac{1}{d_k^2 - 1} \Big( & \text{Tr}(\Pi_k W^\dagger) \text{Tr}(\Pi_k W) \text{Tr}( \Pi_k V^\dagger \Pi_k V) 
		+ \text{Tr}( \Pi_k W^\dagger \Pi_k W) \text{Tr}( \Pi_k V^\dagger) \text{Tr}( \Pi_k V) \\
		- & \frac{1}{d_k} \text{Tr}(\Pi_k W^\dagger) \text{Tr}(\Pi_k W)
		\text{Tr}( \Pi_k V^\dagger ) \text{Tr}( \Pi_k V) 
		- \frac{1}{d_k} \text{Tr}( \Pi_k W^\dagger \Pi_k W) \text{Tr}( \Pi_k V^\dagger \Pi_k V),
	\end{aligned}
\end{align}
with
\begin{align}\label{eq:U1Charge2}
	\frac{1}{d_k d_l} \text{Tr}(\Pi_k W^\dagger) \text{Tr}(\Pi_l W) \text{Tr}( \Pi_k V^\dagger \Pi_l V).
\end{align}
With no necessity to care about multiplicities, these expansions looks more like that for the second moment over U$(d^n)$ (see Eq.\eqref{eq:2-design}). As a caveat, the argument using unitary similar transformation Eq.\eqref{eq:PauliSymmetry} does not hold here even after we average over U$(1)$ symmetric Haar randomness, so we meed to calculate for $V = X,Y,Z$ independently (but the position of one-qubit Pauli operators is still insignificant).

%-------------------------------------------------------------------------------

When $V = Z$, $\Pi_k V \Pi_l V = 0$ unless $k = l$, so we simply calculate Eq.\eqref{eq:U1Charge1}:
\begin{align}
	\text{Tr}( \Pi_k Z \Pi_k Z) = \text{Tr}( \Pi_k Z^2) = d_k = \binom{n}{k}, \quad \text{Tr}( \Pi_k Z) = \binom{n-1}{k-1} - \binom{n-1}{k} = \frac{2k-n}{n} \binom{n}{k}.
\end{align}
and
\begin{align}
	& \text{Tr}(\Pi_k W^\dagger) = \text{Tr}\Big(\Pi_k ((i,j) - \frac{1}{2} I) \Big) = \binom{n-2}{k-2} + \binom{n-2}{k} - \frac{1}{2} \binom{n}{k} = \frac{4k^2 - 4kn + n(n-1)}{2n(n-1)} \binom{n}{k}, \\
	& \text{Tr}(\Pi_k W^\dagger \Pi_k W) = \text{Tr}\Big(\Pi_k ((i,j) - \frac{1}{2} I)^2 \Big) = \frac{5}{4} \binom{n}{k} - \binom{n-2}{k-2} - \binom{n-2}{k} = \frac{-8k^2 + 8kn + n(n-1)}{4n(n-1)} \binom{n}{k}.
\end{align}
Before taking total summation, we check the last term in Eq.\eqref{eq:U1Charge1}:
\begin{align}
	\begin{aligned}
		\sum_{k=1}^{n-1} \frac{1}{d_k^2 - 1} \frac{1}{d_k} \text{Tr}( \Pi_k W^\dagger \Pi_k W) \text{Tr}( \Pi_k V^\dagger \Pi_k V) & = \sum \frac{1}{d_k^2 - 1} \frac{1}{d_k} \frac{-8k^2 + 8kn + n(n-1)}{4n(n-1)} \binom{n}{k}^2 \\
		& \approx \sum_{k=1}^{n-1} \frac{1}{\binom{n}{k}} \frac{-8k^2 + 8kn + n(n-1)}{4n(n-1)}
	\end{aligned}
\end{align}
which decays faster than any inverse polynomial after being normalized by the factor $1/2^n$. Other terms are approximately gathered to 
\begin{align}
	& \sum_{k = 0}^n  \frac{1}{d_k^2} \Big( \Big( \frac{4k^2 - 4kn + n(n-1)}{2n(n-1)} \binom{n}{k} \Big)^2 \binom{n}{k}
	+ \Big( \frac{-8k^2 + 8kn + n(n-1)}{4n(n-1)} \binom{n}{k} \Big) \Big( \frac{2k-n}{n} \binom{n}{k} \Big)^2 \notag \\
	& - \frac{1}{d_k} \Big( \frac{4k^2 - 4kn + n(n-1)}{2n(n-1)} \binom{n}{k} \Big)^2 \Big( \frac{2k-n}{n} \binom{n}{k} \Big)^2 \Big) \notag \\
	= & \sum_{k = 0}^n \Big(  \Big( \frac{4k^2 - 4kn + n(n-1)}{2n(n-1)} \Big)^2
	+ \Big( \frac{-8k^2 + 8kn + n(n-1)}{4n(n-1)} \Big) \Big( \frac{n-2k}{n} \Big)^2 
	- \Big( \frac{4k^2 - 4kn + n(n-1)}{2n(n-1)} \Big)^2 \Big( \frac{n-2k}{n} \Big)^2 \Big) \binom{n}{k} \notag \\
	= & \frac{2^{n-1}}{(n-1) n} + \frac{2^{n-2} (3 n-4)}{n^2} - \frac{2^{n-1} (5 n-8)}{(n-1) n^3} 
	= \frac{3n^3 - 5n^2 - 6n + 16 }{n^3 (n-1)} 2^{n-2}. 
\end{align}
This summation is solved by noticing that for fixed $n$, each term from above is merely a polynomial of $k$ with degree no more than $4$ product with $\binom{n}{k}$. To solve them, we use
\begin{align}
	\sum_{k = 0}^n \binom{n}{k}e^{kx} = (1+e^x)^n, \quad \frac{d}{dx^r}\Big\vert_{x = 0} \sum_{k = 0}^n \binom{n}{k}e^{kx} = \sum_{k = 0}^n k^r \binom{n}{k}.  
\end{align}

%-------------------------------------------------------------------------------

When $V = X,Y$, $\Pi_k V = 0$ by definition, so we simply calculate Eq.\eqref{eq:U1Charge2}:
\begin{align}
	\text{Tr}( \Pi_k X \Pi_l X) = \text{Tr}( \Pi_k Y \Pi_l Y) = \begin{cases} \binom{n-1}{k-1} = \frac{k}{n}\binom{n}{k}, & l = k - 1,  \\ \binom{n-1}{k} = \frac{n-k}{n}\binom{n}{k},  & l = k + 1, \\ 0, & \text{otherwise}.	\end{cases}
\end{align}
with $\text{Tr}(\Pi_k W^\dagger)$ being given above. For either $X$ or $Y$, Eq.\eqref{eq:U1Charge2} equals 
\begin{align}
	\begin{aligned}
		& \sum_{k = 0}^{n-1} \frac{1}{\binom{n}{k} \binom{n}{k+1}} \Big( \frac{4k^2 - 4kn + n(n-1)}{2n(n-1)} \binom{n}{k} \Big) \Big( \frac{4(k+1)^2 - 4(k+1)n + n(n-1)}{2n(n-1)} \binom{n}{k+1} \Big) \Big( \frac{n-k}{n}\binom{n}{k} \Big) \\
		& + \sum_{k = 0}^{n-1} \frac{1}{\binom{n}{k+1} \binom{n}{k}} \Big( \frac{4(k+1)^2 - 4(k+1)n + n(n-1)}{2n(n-1)} \binom{n}{k+1} \Big) \Big( \frac{4k^2 - 4kn + n(n-1)}{2n(n-1)} \binom{n}{k} \Big) \Big( \frac{k}{n}\binom{n}{k} \Big) \\
		= & \sum_{k = 0}^{n-1}  \Big( \frac{4k^2 - 4kn + n(n-1)}{2n(n-1)} \Big) \Big( \frac{4(k+1)^2 - 4(k+1)n + n(n-1)}{2n(n-1)} \Big) \binom{n}{k} \\
		= & \frac{1}{n(n-1)} 2^{n-1} - \frac{n^2 + 3n + 4}{4 n(n-1)}
	\end{aligned}
\end{align}

In conclusion, the OTOC of $W = (i,j) - \frac{1}{2} I$ under U$(1)$ symmetry approximates
\begin{align}\label{eq:OTOC-U(1)}
	\frac{1}{3 \cdot 2^n} \Big(\frac{3n^3 - 3n^2 - 6n + 16)}{n^3(n-1)} 2^{n-2} - \frac{n^2 + 3n + 4}{4 n(n-1)} \Big) = \frac{3n^3 - 3n^2 - 6n + 16)}{12n^3 (n-1)} - \frac{n^2 + 3n + 4}{3n(n-1) 2^n} = \Theta(\frac{1}{n})
\end{align}
for large $n$. 

%-------------------------------------------------------------------------------

\subsection{SU$(d)$ covariant error correction code}\label{sec:Covariant}

We estimate the Choi error of SU$(d)$ covariant quantum error correction codes using computational techniques developed before. To be specific, we follow \cite{kong2022near} and consider a system $S$ of $n$ qudits where we wish to encode the logical information from a logical space of $k$ qudits and then decode it. The procedure is mathematically described similarly as in Section \ref{sec:Setting} where we have logical states $\rho_{R \cup A}$ entangled with the reference $R$ and another initial state $\rho_{\bar{A}}$ with no memory and commutes with $\hat{U}^{\dagger \otimes n - k})$ for any $\hat{U} \in \text{SU}(d)$. Then given a global SU$(d)$ symmetric unitary $U$ acting on the system (with identity map acting on the reference), the following SU$(d)$ \emph{equivariance} or \emph{covariance} is satisfied:
\begin{align}\label{eq:SU(d)-covariance}
	\begin{aligned}
		\hat{U}^{\otimes n} U \rho_{R \cup A} \otimes \rho_{\bar{A}} U^\dagger \hat{U}^{\dagger \otimes n} = U (\hat{U}^{\otimes n} \rho_{R \cup A} \otimes \rho_{\bar{A}} \hat{U}^{\dagger \otimes n}) U^\dagger & = U (\hat{U}^{\otimes k} \rho_{R \cup A} \hat{U}^{\dagger \otimes k}) \otimes ( \hat{U}^{\otimes n-k} \rho_{\bar{A}} \hat{U}^{\dagger \otimes n - k}) U^\dagger \\
		& = U (\hat{U}^{\otimes k} \rho_{R \cup A} \hat{U}^{\dagger \otimes k}) \otimes \rho_{\bar{A}} U^\dagger.
	\end{aligned}
\end{align}
As an immediate result from Schur--Weyl duality and double commutant theorem (see Section \ref{sec:SchurWeyl} or \cite{Goodman2009,Tolli2009} for more details), $\rho_{\bar{A}}$ is generally expanded as
\begin{align}\label{eq:n-kState}
	\rho_{\bar{A}} = \sum_i p_i \sigma_{\bar{A},i},
\end{align}
where the combination coefficients satisfy $\sum_i p_i \text{Tr}(\sigma_{\bar{A},i}) = 1$ with other conditions making $\rho_{\bar{A}}$ a valid density matrix. For instance, it can be defined by SWAPs:
\begin{align}\label{eq:n-k-SWAP-State}
	\rho_{\bar{A}} = \frac{1}{d^{n - k - 1} + d^{n - k}} ( (i,j) + I_{\bar{A}}).
\end{align}
or more generally,
\begin{align}
	\rho_{\bar{A}} = \frac{1}{2 (d^{\# \text{cycles}(\sigma_{\bar{A}})} + d^{n - k}) } (\sigma_{\bar{A}} + I_{\bar{A}})(\sigma_{\bar{A}}^{-1} + I_{\bar{A}}) = \frac{1}{2 (d^{\# \text{cycles}(\sigma_{\bar{A}})} + d^{n - k}) } (\sigma_{\bar{A}} + \sigma_{\bar{A}}^{-1} + 2I_{\bar{A}})
\end{align}
where $\# \text{cycles}(\sigma_{\bar{A}})$ counts the number of cycles (including trivial 1-cycles) of $\sigma$ on $n-k$ qudits. With a more explicit application of representation theory, we can define $\rho_{\bar{A}}$ as
\begin{align}
	\rho_{\bar{A}} = \frac{1}{\#(S^\lambda_0)} I_{W_{\lambda_0}} \otimes \ket{\psi_{\lambda_0}} \bra{\psi_{\lambda_0}} 
\end{align}
with $\bra{\psi_\lambda}$ being a pure state (e.g., a Young basis element) restricted to a certain $S_{n-k}$ irrep of Young diagram $\lambda_0$ and $I_{W_{\lambda_0}}$ being the identity matrix recording its multiplicities according to Schur--Weyl duality Eq.\eqref{eq:A-SchurWeyl}. This perspective is used later when computing Eq.\eqref{eq:PartialTraceOfPermutation}.

Then we implement a decoding map $\mathcal{D}$ acting on $n-t$ non-erased qudits, denoted by $B$, of the system. Our objective is appraising the error correction performance by calculating the Choi error
\begin{align}\label{eq:ChoiError1}
	& F_{\text{Choi}} = \max_{\mathcal{D}} F((\text{id}_R \otimes \mathcal{D}) \text{Tr}_{\bar{B}} (U \rho_{R \cup A} \otimes \mu_{\bar{A} } U^\dagger), \rho_{R \cup A}), \\
	& \epsilon_{\text{Choi}} = \sqrt{1 - F_{\text{Choi}}^2} = \min_{\mathcal{D}} P((\text{id}_R \otimes \mathcal{D}) \text{Tr}_{\bar{B}} (U \rho_{R \cup A} \otimes \mu_{\bar{A} } U^\dagger), \rho_{R \cup A}),
\end{align}
where $\rho_{R \cup A}$ is assumed to be maximally entangled and $F(\rho,\sigma) = \text{Tr}\sqrt{\sqrt{\rho} \sigma \sqrt{\rho}}$ is the \emph{state fidelity} with \emph{purified distance} $P(\rho,\sigma) = \sqrt{1 - F(\rho,\sigma)^2}$.

Using the so-called \emph{complementary channel formalism} \cite{Beny2010,kong2022near}, we have
\begin{align}\label{eq:ChoiError2}
	\min_{\mathcal{D}} P((\text{id}_R \otimes \mathcal{D}) \text{Tr}_{\bar{B}} (U \rho_{R \cup A} \otimes \mu_{\bar{A} } U^\dagger), \rho_{R \cup A}) 
	= \min_{\zeta} P( \text{Tr}_B (U \rho_{R \cup A} \otimes \mu_{\bar{A} } U^\dagger), \text{Tr}_A ( \rho_{R \cup A}) \otimes \zeta ),
\end{align}
where $\zeta$ is a variable density matrix to minimize the concerned purified distance. For maximally entangled state, $\text{Tr}_A ( \rho_{R \cup A}) = \frac{1}{d_A} I_R$. Let us denote $\text{Tr}_B (U \rho_{R \cup A} \otimes \mu_{\bar{A} } U^\dagger)$ by $\rho_{R \cup \bar{B}}$. By triangular inequality
\begin{align}
	\begin{aligned}
		\epsilon_{\text{Choi}} = \min_{\zeta} P( \rho_{R \cup \bar{B}}, \frac{1}{d_A} I_R \otimes \zeta ) & \leq P( \rho_{R \cup \bar{B}}, \rho_{R \cup \bar{B}, \text{avg}} ) + \min_{\zeta} P(\rho_{R \cup \bar{B}, \text{avg}}, \frac{1}{d_A} I_R \otimes \zeta) \\
		& \leq \sqrt{2\Vert \rho_{R \cup \bar{B}} - \rho_{R \cup \bar{B}, \text{avg}} \Vert_1} + \min_{\zeta} P(\rho_{R \cup \bar{B}, \text{avg}}, \frac{1}{d_A} I_R \otimes \zeta),
	\end{aligned}
\end{align}
with $\rho_{R \cup\bar{B}, \text{avg}} = \int \rho_{R \cup \bar{B} } dU$ being averaged over the group $\operatorname{U}(\bigoplus_\lambda S^\lambda)$ of SU$(d)$ symmetric unitaries. In the following context, we calculate the second term on the RHS of the above inequality, and then average the first term to obtain the expectation of $\epsilon_{\text{Choi}}$.

%------------------------------------------------------------------------------------------------------------------------------------------------

\subsubsection*{Computation of $\rho_{R \cup \bar{B}, \text{avg}}^2$}

To illustrate the idea, we assume that the erasure part $\bar{B}$ contains one qudit ($t = 1$) while $A$ is allowed to encode arbitrary $k$ qudits (later we shall assume that $k = o(n)$). By comparing commutant algebras (see Section \ref{sec:CommutantTheory}), the first Haar projector $T_{k=1}^{\text{Haar}} = T_{k=1}^{\text{CQA}}$ can be easily replaced by that using $S_n$ group elements (however this is not true for higher order moments), so
\begin{align}
	\rho_{R \cup B \cup \bar{B}, \text{avg}} = \frac{1}{n!}\sum_{\sigma \in S_n} \sigma (\rho_{R \cup A} \otimes \rho_{\bar{A}}) \sigma^{-1} = \frac{1}{n!} \sum_i p_i \sigma ( \rho_{R \cup A} \otimes \sigma_{\bar{A},i} ) \sigma^{-1}.
\end{align} 
Let $\rho_{R \cup \{i_1,...,i_k\}}$ denotes the maximally entangled state of $R$ and sites $i_1,...,i_k$ permuted from $A$. The above expansion can be divided into two cases depending on whether $\{i_1,...,i_k\} \cap \bar{B}$ is $\emptyset$ or not. For nonempty case, taking partial trace over $B$ yields
\begin{align}
	\frac{1}{n} \sum_{a=1}^k \rho_{R \cup \bar{B},a}
\end{align}
where $\rho_{R \cup \bar{B},a}$ is defined as when the $a$-th qudit in $A$ being permuted to $\bar{B}$ with all left ones getting traced out. We instantiate the case when $a = 1$:
\begin{align}
	\begin{aligned}
		\rho_{R \cup \bar{B},1} = & \text{Tr}_{2,...,k} \Big( \frac{1}{d^k} \sum_{i_1,...i_k, j_1,...,j_k} \ket{i_1^R i_2^R...i_k^R, i_1^A i_2^A ... i_k^A} \bra{j_1^R j_2^R...j_k^R, j_1^A j_2^A ... j_k^A} \Big) \\
		= & \frac{1}{d^{k-1}} \sum_{i_1,...i_k, j_1,...,j_k}   \ket{i_1^R i_2^R...i_k^R, i_1^A} \bra{j_1^R i_2^R...i_k^R, j_1^A} \\
		= & \Big( \frac{1}{d} \sum_{i,j} \ket{i_1^R,i_1^A} \bra{j_1^R, j_1^A} \Big) \frac{1}{d^{k-1}} \sum_{i_2,...,i_k} \ket{i_2^R...i_k^R} \bra{i_2^A...i_k^A}
	\end{aligned}
\end{align}
which equals the maximal entangled state between $a$th register in $R$ with $\bar{B}$ coupled with the maximally mixed state for remaining qudits in $R$.

When $\{i_1,...,i_k\} \cap \bar{B} = \emptyset$, we expand $\rho_{\bar{A}}$ by Eq.\eqref{eq:n-kState} and analyze partial trace for arbitrary permutations $\sigma_{\bar{A},i}$:
\begin{align}\label{eq:PartialTraceOfPermutation}
	\begin{aligned}
		& \text{Tr}_B \Big(\sigma ( \rho_{R \cup A} \otimes \sigma_{\bar{A},i}) \sigma^{-1} \Big) = \frac{1}{d^k} I_R \otimes d^{\# \text{cycles}(\sigma_{\bar{A},i}) - 1} I_{\bar{B}}  = d^{\# \text{cycles}(\sigma_{\bar{A},i})} \frac{1}{d^k} I_R \otimes \frac{1}{d}I_{\bar{B}} \\
		\implies & \text{Tr}_B \Big(\sigma ( \rho_{R \cup A} \otimes \rho_{\bar{A}}) \sigma^{-1} \Big) = \sum_i p_i d^{\# \text{cycles}(\sigma_{\bar{A},i})} \frac{1}{d^k} I_R \otimes \frac{1}{d}I_{\bar{B}} = \frac{1}{d^k} I_R \otimes \frac{1}{d}I_{\bar{B}}.
	\end{aligned}
\end{align} 
Therefore,
\begin{align}
	\rho_{R \cup \bar{B}, \text{avg}} = \frac{1}{n} \sum_{a=1}^k \rho_{R \cup \bar{B},a} + \frac{n-k}{n} \frac{1}{d^k} I_R \otimes \frac{1}{d}I_{\bar{B}} = \frac{1}{n} \sum_{a=1}^k \frac{1}{d^{k-1}} I_{R - a} \otimes \ket{\Psi_{a,\bar{B}}} \bra{\Psi_{a,\bar{B}}} + \frac{n-k}{n} \frac{1}{d^k} I_R \otimes \frac{1}{d}I_{\bar{B}}
\end{align}
where $\ket{\Psi_{a,\bar{B}}}$ denotes the maximal entangled state. Let $k = 1$, the above result is already obtained in \cite{kong2022near} with further computation demonstrating that
\begin{align}
	\mathbb{E}_{\text{Choi}} \leq \frac{\sqrt{d^2 - 1}}{2n} + O(n^{-2}).
\end{align}    

Let $\zeta = \frac{1}{d}I_{\bar{B}}$, by Eq.\eqref{eq:ChoiError1} and \eqref{eq:ChoiError2},
\begin{align}
	\begin{aligned}
		& F(\rho_{R \cup \bar{B}, \operatorname{avg}}, \frac{1}{d_A} I_R \otimes \zeta) = \operatorname{Tr} \sqrt{ \sqrt{\frac{1}{d_A} I_R \otimes \frac{1}{d}I_{\bar{B}}}  \rho_{R \cup \bar{B}, \operatorname{avg}} \sqrt{\frac{1}{d_A} I_R \otimes \frac{1}{d}I_{\bar{B}}} } = \operatorname{Tr} \Big( \sqrt{\frac{1}{d^{k+1}} \rho_{R \cup \bar{B}, \operatorname{avg}} } \Big) \\
		\implies & P(\rho_{R \cup \bar{B}, \operatorname{avg}}, \frac{1}{d_A} I_R \otimes \zeta) = \sqrt{1 - \frac{1}{d^{k+1}} \text{Tr}\big(\sqrt{\rho_{R \cup \bar{B}, \text{avg}}} \big)^2 }.
	\end{aligned}
\end{align}
To estimate the trace of the square root of $\rho_{R \cup \bar{B}, \text{avg}}$, we take the following strategy using a kind of matrix \emph{arithmetical mean - quadratic mean inequality}. Let $A_1,A_2$ be positive semidefinite matrices. Then
\begin{align}
	& A_1^2 + A_2^2 - \frac{1}{2} (A_1 + A_2)^2 = \frac{1}{2} (A_1^2 + A_2^2 - A_1 A_2 - A_2 A_1) = \frac{1}{2} (A_1 - A_2 )^2 \geq 0 \\
	\implies & A_1^2 + A_2^2 \geq \frac{1}{2} (A_1 + A_2)^2,
\end{align}
where $A \geq B$ means $A - B$ is positive semidefinite. Inductively, we assume
\begin{align}
	\sum_{i=1}^k A_i^2 \geq \frac{1}{k} (\sum_{i=1}^k A_i)^2.
\end{align}
The inequality for $k+1$ follows directly by simplifying
\begin{align}
	\begin{aligned}
		& k(A_1^2 + A_2^2 + \cdots + A_{k-1}^2 + A_k^2) + (A_1^2 + \cdots + A_{k-1}^2 + A_{k+1}^2) + \cdots + (A_2^2 +\cdots + A_{k-1}^2 + A_k^2) \\
		\geq & \Big( (A_1 + \cdot + A_{k-1} + A_k)^2 + (A_1 + \cdot + A_{k-1} + A_{k+1})^2 + \cdots + (A_2 + \cdot + A_{k-1} + A_k)^2 \Big).
	\end{aligned}
\end{align} 
To cope with our problem, we rewrite
\begin{align}
	\begin{aligned}
		\rho_{R \cup \bar{B}, \text{avg}} & = \frac{1}{n} \sum_{a=1}^k \frac{1}{d^{k-1}} I_{R - a} \otimes \ket{\Psi_{a,\bar{B}}} \bra{\Psi_{a,\bar{B}}} + \frac{n-k}{n} \frac{1}{d^k} I_R \otimes \frac{1}{d}I_{\bar{B}} \\
		& = \frac{1}{k} \sum_{a=1}^k \Big( \frac{k}{n} \frac{1}{d^{k-1}} I_{R - a} \otimes \ket{\Psi_{a,\bar{B}}} \bra{\Psi_{a,\bar{B}}} + \frac{n-k}{n} \frac{1}{d^k} I_R \otimes \frac{1}{d}I_{\bar{B}} \Big) \\
		& = \sum_{i = 1}^k P_i.
	\end{aligned}
\end{align}
Substituting $A_i$ by $\sqrt{P_i}$, we have
\begin{align}
	\sum_{i=1}^k P_i \geq \frac{1}{k} (\sum_{i=1}^k \sqrt{P_i})^2.
\end{align}
Since $A \geq B$ indicates that $\lambda_i(A) \geq \lambda_i(B)$ when the eigenvalues are ordered in the same way, taking square root on both sides from above, we can confirm that
\begin{align}
	\text{Tr}(\sqrt{\rho_{R \cup \bar{B}, \text{avg}}}) = \text{Tr} \Big( \sqrt{\sum_{i=1}^k P_i} \Big) \geq \frac{1}{\sqrt{k}} \text{Tr} (\sum_{i=1}^k \sqrt{P_i}) = \frac{1}{\sqrt{k}} \sum_{i=1}^k \text{Tr} (\sqrt{P_i}).
\end{align}
It is straightforward to evaluate, by definition, that
\begin{align}
	\text{Tr} (\sqrt{P_i}) = \frac{1}{\sqrt{k}} \Big( \sqrt{ \frac{k}{n} \frac{1}{d^{k-1}} + \frac{n-k}{n} \frac{1}{d^{k+1}} } d^{k-1} + \sqrt{ \frac{n-k}{n} \frac{1}{d^{k+1}}} (d^{k+1} -  d^{k-1}) \Big)
\end{align}
for any $i=1,...,k$. As a result, taking large $n$ limit, we obtain
\begin{align}
	\begin{aligned}
		\sqrt{1 - \frac{1}{d^{k+1}} \text{Tr}\big(\sqrt{\rho_{R \cup \bar{B}, \text{avg}}} \big)^2 } & \leq \sqrt{1 - \frac{1}{d^{k+1}} \Big( \sqrt{ \frac{k}{n} \frac{1}{d^{k-1}} + \frac{n-k}{n} \frac{1}{d^{k+1}} } d^{k-1} + \sqrt{ \frac{n-k}{n} \frac{1}{d^{k+1}}} (d^{k+1} -  d^{k-1}) \Big)^2 } \\
		& = \sqrt{1 - \frac{1}{d^2} \Big( \sqrt{ \frac{k}{n} + \frac{n-k}{n} \frac{1}{d^2} } + \sqrt{ \frac{n-k}{n} \frac{1}{d^2}} (d^2 - 1) \Big)^2 } \approx \frac{k\sqrt{d^2 - 1}}{2n}.
	\end{aligned}
\end{align}

\begin{remark}
	One potential issue to generalize $\bar{B}$ to arbitrary $t$ qudits occurs when computing Eq.\eqref{eq:PartialTraceOfPermutation}: if $t > 1$, the form of $\text{Tr}_B \Big(\sigma ( \rho_{R \cup A} \otimes \sigma_{\bar{A},i}) \sigma^{-1} \Big)$ becomes depending on the intersection of $\bar{B}$ and the \emph{nontrivial support} of $\sigma \sigma_{\bar{A},i} \sigma^{-1}$. If the interaction contains at most one qudits, the result is the same as before. Otherwise, after taking partial trace, $\sigma \sigma_{\bar{A},i} \sigma^{-1}$ is partially contracted and degenerate to lower order, but nontrivial, cycles. One may consider initializing a state like Eq.\eqref{eq:n-k-SWAP-State}, but that conflicts our approach to bound $\Vert \rho_{R \cup \bar{B}} - \rho_{R \cup \bar{B}, \text{avg}} \Vert_1$ in the following and we leave problem to future research work.
\end{remark}

%-------------------------------------------------------------------------------

\subsubsection*{Computation Using Abstract Indices}

By matrix Hölder inequality \cite{Manjegani2007,Baumgartner2011},
\begin{align}
	\Vert \rho_{R \cup \bar{B}} - \rho_{R \cup \bar{B}, \text{avg}} \Vert_1 \leq (d_R d_{\bar{B}})^{\frac{1}{2}} \Vert \rho_{R \cup \bar{B}} - \rho_{R \cup \bar{B}, \text{avg}} \Vert_2^{\frac{1}{2}}.
\end{align}
Moreover,
\begin{align}
	\mathbb{E} \Big[ \Vert \rho_{R \cup \bar{B}} - \rho_{R \cup \bar{B}, \text{avg}} \Vert_2 \Big] = \mathbb{E} \Big[ \text{Tr} ((\rho_{R \cup \bar{B}} - \rho_{R \cup \bar{B}, \text{avg}})^2)^{\frac{1}{2}} \Big]. 
\end{align}
Then, we use Jensen's inequality to get
\begin{align}\label{eq:Jensen-Covariant}
	\mathbb{E} \Big[ \Vert \rho_{R \cup \bar{B}} - \rho_{R \cup \bar{B}, \text{avg}} \Vert_2 \Big]  \leq  \mathbb{E} \Big[ \text{Tr} ((\rho_{R \cup \bar{B}} - \rho_{R \cup \bar{B}, \text{avg}})^2) \Big]^{\frac{1}{2}} = \Big( \mathbb{E} \Big[ \text{Tr} (\rho_{R \cup \bar{B}}^2) \Big] - \text{Tr}(\rho_{R \cup \bar{B}, \text{avg}}^2) \Big)^{\frac{1}{2}}. 
\end{align}
As a result,
\begin{align}
	\mathbb{E} [\sqrt{\Vert \rho_{R \cup \bar{B}} - \rho_{R \cup \bar{B}, \text{avg}} \Vert_1} ] \leq \sqrt{\mathbb{E} [ \Vert \rho_{R \cup \bar{B}} - \rho_{R \cup \bar{B}, \text{avg}} ] \Vert_1} \leq 
	\sqrt{ (d_R d_{\bar{B}})^{1/2} \big(\mathbb{E} [ \text{Tr} (\rho_{R \cup \bar{B}}^2)] - \text{Tr}(\rho_{R \cup \bar{B}, \text{avg}}^2) \big)^{1/2} }.
\end{align}
A precise calculation of $\mathbb{E}[\text{Tr} (\rho_{R \cup \bar{B}}^2)] - \text{Tr}(\rho_{R \cup \bar{B}, \text{avg}}^2)$ not only enables us to estimate $\mathbb{E}[ \epsilon_{\text{Choi}}]$ but also bounds the probability for large Choi error by Markov inequality. Ref.\cite{kong2022near} tackles the problem by the \emph{partial decoupling theorem} established in \cite{Wakakuwa2021Decoupling}. We elucidate our attempts based on results of SU$(d)$-symmetric 2-design developed before. 

Using abstract indices, the initial state under action is
\begin{align}
	D\indices{^{R B \bar{B}}_{R' B' \bar{B}'} } =  U\indices{^{B \bar{B}}_{A \bar{A}}} Q^{R A} P\indices{^{\bar{A}}_{\bar{A}'}} Q_{R' A'} U\indices{_{B' \bar{B}'}^{A' \bar{A}'}}.
\end{align}
It should be careful that $Q^{R A}$ is a state purified by the reference $R$ but $P\indices{^{\bar{A}}_{\bar{A}'}}$ is a general (mixed) state with no entanglement to other qudits for a convenient definition of group equivariance (see Eq.\eqref{eq:SU(d)-covariance}).

By definition
\begin{align}
	\text{Tr}(\rho_{R \cup \bar{B}, \text{avg}}^2 ) = \Big( \int D\indices{^{R B \bar{B}}_{R' B \bar{B}'} } dU \Big) \Big( \int D\indices{^{R' B' \bar{B}'}_{R B' \bar{B}} } dV \Big) = \Big( \int D\indices{^{R \bar{B}}_{R' \bar{B}'} } dU \Big) \Big( \int D\indices{^{R' \bar{B}'}_{R \bar{B}} } dV \Big).
\end{align}
We denote $P\indices{^{\bar{A}}_{\bar{A}'}} = \rho\indices{^{\bar{A}}_{\bar{A}'}},  Q^{R A} Q_{R A'''} = \mu \indices{^{A}_{A'''}}$ as in Section \ref{sec:Setting}, then the integrand is expanded as
\begin{align}\label{eq:rho_avg^2_1}
	\begin{aligned}
		& D(U)\indices{^{R B \bar{B}}_{R' B \bar{B}'} }  D(V)\indices{^{R' B' \bar{B}'}_{R B' \bar{B}} } \\
		= & (U\indices{^{B \bar{B}}_{A \bar{A}}} Q^{R A} P\indices{^{\bar{A}}_{\bar{A}'}}  Q_{R' A'} U\indices{_{B \bar{B}'}^{A' \bar{A}'}} )  
		(V\indices{^{B' \bar{B}'}_{A'' \bar{A}''}} Q^{R' A''} P\indices{^{\bar{A}''}_{\bar{A}'''}} Q_{R A'''} V\indices{_{B' \bar{B}}^{A''' \bar{A}'''}} ) \\
		= & \mu\indices{^{A}_{A'''}} \mu\indices{^{A''}_{A'}} \rho\indices{^{\bar{A}}_{\bar{A}'}} \rho\indices{^{\bar{A}''}_{\bar{A}'''}}  
		U\indices{^{B \bar{B}}_{A \bar{A}}}  U\indices{_{B \bar{B}'}^{A' \bar{A}'}} 
		V\indices{^{B' \bar{B}'}_{A'' \bar{A}''}} V\indices{_{B' \bar{B}}^{A''' \bar{A}'''}} \\
		= & (U\indices{^{B \bar{B}}_{A \bar{A}}} \mu\indices{^{A}_{A'''}}  \rho\indices{^{\bar{A}}_{\bar{A}'}})
		U\indices{_{B \bar{B}'}^{A' \bar{A}'}}  
		(V\indices{^{B' \bar{B}'}_{A'' \bar{A}''}} \mu\indices{^{A''}_{A'}} \rho\indices{^{\bar{A}''}_{\bar{A}'''}} )
		V\indices{_{B' \bar{B}}^{A''' \bar{A}'''}} \\
		= & \check{U}\indices{^{B \bar{B}}_{A''' \bar{A}'}} U\indices{_{B \bar{B}'}^{A' \bar{A}'}}   
		\check{V}\indices{^{B' \bar{B}'}_{A' \bar{A}'''}} V\indices{_{B' \bar{B}}^{A''' \bar{A}'''}}. 
	\end{aligned}
\end{align}
It would be further cleaned up by taking transpositions $T\indices{^{Q_A''' Q_A'}_{A' A'''}}, T\indices{^{Q_B' Q_B}_{B B'}}$ (Pauli matrices are denoted by $P,Q$ for now):
\begin{align}\label{eq:rho_avg^2_2}
	& T\indices{^{Q_A''' Q_A'}_{A' A'''}} T\indices{^{Q_B' Q_B}_{B B'}} \check{U}\indices{^{B \bar{B}}_{Q_A''' \bar{A}'}} U\indices{_{Q_B \bar{B}'}^{A' \bar{A}'}}   
	\check{V}\indices{^{B' \bar{B}'}_{Q_A' \bar{A}'''}} V\indices{_{Q_B' \bar{B}}^{A''' \bar{A}'''}}  \notag \\
	= & \frac{1}{d_A d_B} \sum_{P,Q} P\indices{^{Q_A'''}_{A'}} P\indices{_{A'''}^{Q_A'}} Q\indices{^{Q_B}_{B'}} Q\indices{_{B}^{Q_B'}} \check{U}\indices{^{B \bar{B}}_{A''' \bar{A}}}
	(U\indices{^{B \bar{B}}_{A \bar{A}}} \mu\indices{^{A}_{Q_A'''}}  \rho\indices{^{\bar{A}}_{\bar{A}'}})
	U\indices{_{Q_B \bar{B}'}^{A' \bar{A}'}}  
	(V\indices{^{B' \bar{B}'}_{A'' \bar{A}''}} \mu\indices{^{A''}_{Q_A'}} \rho\indices{^{\bar{A}''}_{\bar{A}'''}} )
	V\indices{_{Q_B' \bar{B}}^{A''' \bar{A}'''}} \notag \\
	= & \frac{1}{d_A d_B d_A^2} \sum_{P,Q}   
	(U\indices{^{B \bar{B}}_{A \bar{A}}}
	P\indices{^{A}_{A'}} \rho\indices{^{\bar{A}}_{\bar{A}'}}) 
	U\indices{_{Q_B \bar{B}'}^{A' \bar{A}'}}  
	Q\indices{^{Q_B}_{B'}} 
	(V\indices{^{B' \bar{B}'}_{A'' \bar{A}''}} P\indices{_{A'''}^{A''}}  \rho\indices{^{\bar{A}''}_{\bar{A}'''}} )
	V\indices{_{Q_B' \bar{B}}^{A''' \bar{A}'''}} 	Q\indices{_{B}^{Q_B'}}
\end{align}
Rewritten in a concise form we get
\begin{align}\label{eq:rho_avg^2_3}
	\text{Tr}( \rho_{R \cup \bar{B}, \text{avg}}^2 ) = \frac{1}{d_A^3 d_B} \sum_{P_A, P_B} & \int \text{Tr}\Big( U (P_A \otimes \rho_{\bar{A}}) U^\dagger P_B V (P_A^\dagger \otimes \rho_{\bar{A}}) V^\dagger P_B^\dagger \Big) dU dV.
\end{align}
By exactly the same method, we immediately have
\begin{align}
	\mathbb{E}[ \text{Tr}( \rho_{R \cup \bar{B}}^2 ) ] = \frac{1}{d_A^3 d_B} \sum_{P_A, P_B} & \int \text{Tr}\Big( U (P_A \otimes \rho_{\bar{A}}) U^\dagger P_B U (P_A^\dagger \otimes \rho_{\bar{A}}) U^\dagger P_B^\dagger \Big) dU.
\end{align}

\begin{figure}
	\centering
	\includegraphics[width=6in]{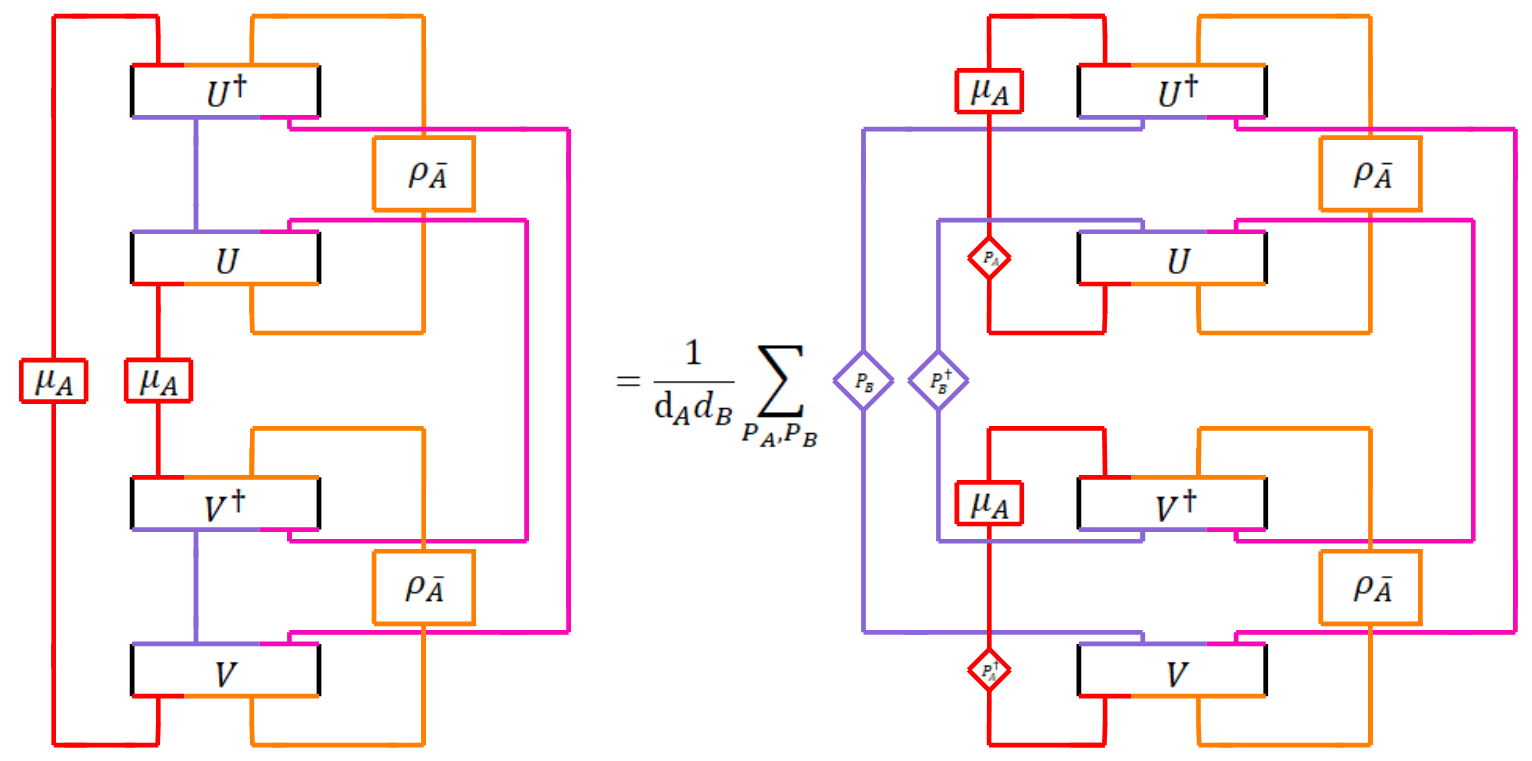}
	\caption{Tensor network diagram illustrating the deduction from Eq.\eqref{eq:rho_avg^2_1} to \eqref{eq:rho_avg^2_3}.}
	\label{}
\end{figure}

We hope to perform similar techniques to evaluate the above integrals as in the last subsection, but the matrix $P_A \otimes \rho_{\bar{A}}$ does not admit SU$(d)$ symmetry on the entire system of $n$ qudits which leads to intractable terms when expanding by $T_{k=2}^{\text{CQA}}$. One possible strategy to remedy the problem is selecting $\rho_{\bar{A}}$ from any trivial SU$(d)$ irrep from the decomposition of the Hilbert space of $n - k$ qudits, which demands $n-k$ to be divided by the integer $d$ according to Littlewood-Richardson rule \cite{Fulton1997,Goodman2009}. The advantage is that $\rho_{\bar{A}}$, as a 1-dimensional density matrix in some SU$(d)$ trivial representation, is obviously pure and we explain in the following that this choice helps to overcome the mathematical difficulty when averaging the $\text{Tr}( \rho_{R \cup \bar{B}}^2 )$ over the group $\operatorname{U}(\bigoplus_\lambda S^\lambda)$ of SU$(d)$ symmetric unitaries. For instance, when $d = 2$ or $3$, we initialize $\rho_{\bar{A}} = \ket{\psi_{\bar{A}}} \bra{\psi_{\bar{A}}}$ as tensor products of SU$(2)$ or SU$(3)$ singlets:
\begin{align}
	\Big( \frac{1}{\sqrt{2}} ( \ket{01} - \ket{10} ) \Big)^{\otimes (n-k)/2}, \quad \Big( \frac{1}{\sqrt{6}} ( \ket{uds} - \ket{usd} + \ket{dsu} - \ket{dus} + \ket{sud} - \ket{sdu} ) \Big)^{\otimes (n-k)/3}.
\end{align}

With this assumption, it is now legal to decompose $P\indices{^{\bar{A}}_{\bar{A}'}}$ as $P^{\bar{A}} P_{\bar{A}'}$. Then we rewrite Eq.\eqref{eq:rho_avg^2_1}:
\begin{align}
	\begin{aligned}
		& D(U)\indices{^{R B \bar{B}}_{R' B \bar{B}'}}  D(V)\indices{^{R' B' \bar{B}'}_{R B' \bar{B}}} \\
		= & (U\indices{^{B \bar{B}}_{A \bar{A}}} Q^{R A} P\indices{^{\bar{A}}_{\bar{A}'}}  Q_{R' A'} U\indices{_{B \bar{B}'}^{A' \bar{A}'}} )  
		(V\indices{^{B' \bar{B}'}_{A'' \bar{A}''}} Q^{R' A''} P\indices{^{\bar{A}''}_{\bar{A}'''}} Q_{R A'''} V\indices{_{B' \bar{B}}^{A''' \bar{A}'''}} ) \\
		= & \mu\indices{^{A}_{A'''}} \mu\indices{^{A''}_{A'}} \rho\indices{^{\bar{A}}_{\bar{A}'''}} \rho\indices{^{\bar{A}''}_{\bar{A}'}} 
		U\indices{^{B \bar{B}}_{A \bar{A}}}  U\indices{_{B \bar{B}'}^{A' \bar{A}'}} 
		V\indices{^{B' \bar{B}'}_{A'' \bar{A}''}} V\indices{_{B' \bar{B}}^{A''' \bar{A}'''}} \\
		= & ( U\indices{^{B \bar{B}}_{A \bar{A}}} \mu\indices{^{A}_{A'''}} \rho\indices{^{\bar{A}}_{\bar{A}'''}} )
		( U\indices{_{B \bar{B}'}^{A' \bar{A}'}}  \mu\indices{^{A''}_{A'}}  \rho\indices{^{\bar{A}''}_{\bar{A}'}}   )
		V\indices{^{B' \bar{B}'}_{A'' \bar{A}''}} V\indices{_{B' \bar{B}}^{A''' \bar{A}'''}} \\
		= & \check{U}\indices{^{B \bar{B}}_{A''' \bar{A}'''}} \check{U}\indices{_{B \bar{B}'}^{A'' \bar{A}''}} 
		V\indices{^{B' \bar{B}'}_{A'' \bar{A}''}} V\indices{_{B' \bar{B}}^{A''' \bar{A}'''}} 
	\end{aligned}
\end{align}
The difference happens when we write $\rho\indices{^{\bar{A}}_{\bar{A}'''}} \rho\indices{^{\bar{A}''}_{\bar{A}'}}$ instead of $\rho\indices{^{\bar{A}}_{\bar{A}'}} \rho\indices{^{\bar{A}''}_{\bar{A}'''}}$ in Eq.\eqref{eq:rho_avg^2_1}. This simple modification allows us to swap only a pair of indices now e.g., $\bar{B}$ and $\bar{B}'$:
\begin{align}
	\begin{aligned}
		& T\indices{^{\bar{Q}_B' \bar{Q}_B}_{\bar{B} \bar{B}'}} 
		\check{U}\indices{^{B \bar{B}}_{A''' \bar{A}'''}} \check{U}\indices{_{B \bar{Q}_B'}^{A'' \bar{A}''}} 
		V\indices{^{B' \bar{B}'}_{A'' \bar{A}''}} V\indices{_{B' \bar{Q}_B}^{A''' \bar{A}'''}} \\
		= & \frac{1}{d_{\bar{B}} d_A^2} \sum_P P\indices{^{\bar{Q}_B'}_{\bar{B}}} P\indices{_{\bar{B}'}^{\bar{Q}_B}} 
		( U\indices{^{B \bar{B}}_{A''' \bar{A}}}  \rho\indices{^{\bar{A}}_{\bar{A}'''}} )
		( U\indices{_{B \bar{Q}_B'}^{A'' \bar{A}'}}   \rho\indices{^{\bar{A}''}_{\bar{A}'}}  )
		V\indices{^{B' \bar{B}'}_{A'' \bar{A}''}} V\indices{_{B' \bar{Q}_B}^{A''' \bar{A}'''}} \\
		= & \frac{1}{d_{\bar{B}} d_A^2} \sum_P 
		( U\indices{^{B \bar{B}}_{A''' \bar{A}}}  \rho\indices{^{\bar{A}}_{\bar{A}'''}} )
		V\indices{_{B' \bar{Q}_B}^{A''' \bar{A}'''}} 	
		P\indices{_{\bar{B}'}^{\bar{Q}_B}}
		V\indices{^{B' \bar{B}'}_{A'' \bar{A}''}} 
		( U\indices{_{B \bar{Q}_B'}^{A'' \bar{A}'}}   \rho\indices{^{\bar{A}''}_{\bar{A}'}}  )
		P\indices{^{\bar{Q}_B'}_{\bar{B}}}. 	
	\end{aligned}
\end{align}
Therefore,
\begin{align}\label{eq:Code1}
	\text{Tr}( \rho_{R \cup \bar{B},\text{avg}}^2 ) = \frac{1}{d_A^2 d_{\bar{B}}} \sum_{P_{\bar{B}}} \int \text{Tr}\Big( U (I_A \otimes \rho_{\bar{A}}) V^\dagger P_{\bar{B}} V (I_A \otimes \rho_{\bar{A}}) U^\dagger P_{\bar{B}}^\dagger \Big) dU dV
\end{align}
with
\begin{align}\label{eq:Code2}
	\mathbb{E}[ \text{Tr}( \rho_{R \cup \bar{B}}^2 ) ] = \frac{1}{d_A^2 d_{\bar{B}}} \sum_{P_{\bar{B}}} \int \text{Tr}\Big( U (I_A \otimes \rho_{\bar{A}}) U^\dagger P_{\bar{B}} U (I_A \otimes \rho_{\bar{A}}) U^\dagger P_{\bar{B}}^\dagger \Big) dU.
\end{align}

\begin{figure}
	\centering
	\includegraphics[width=6in]{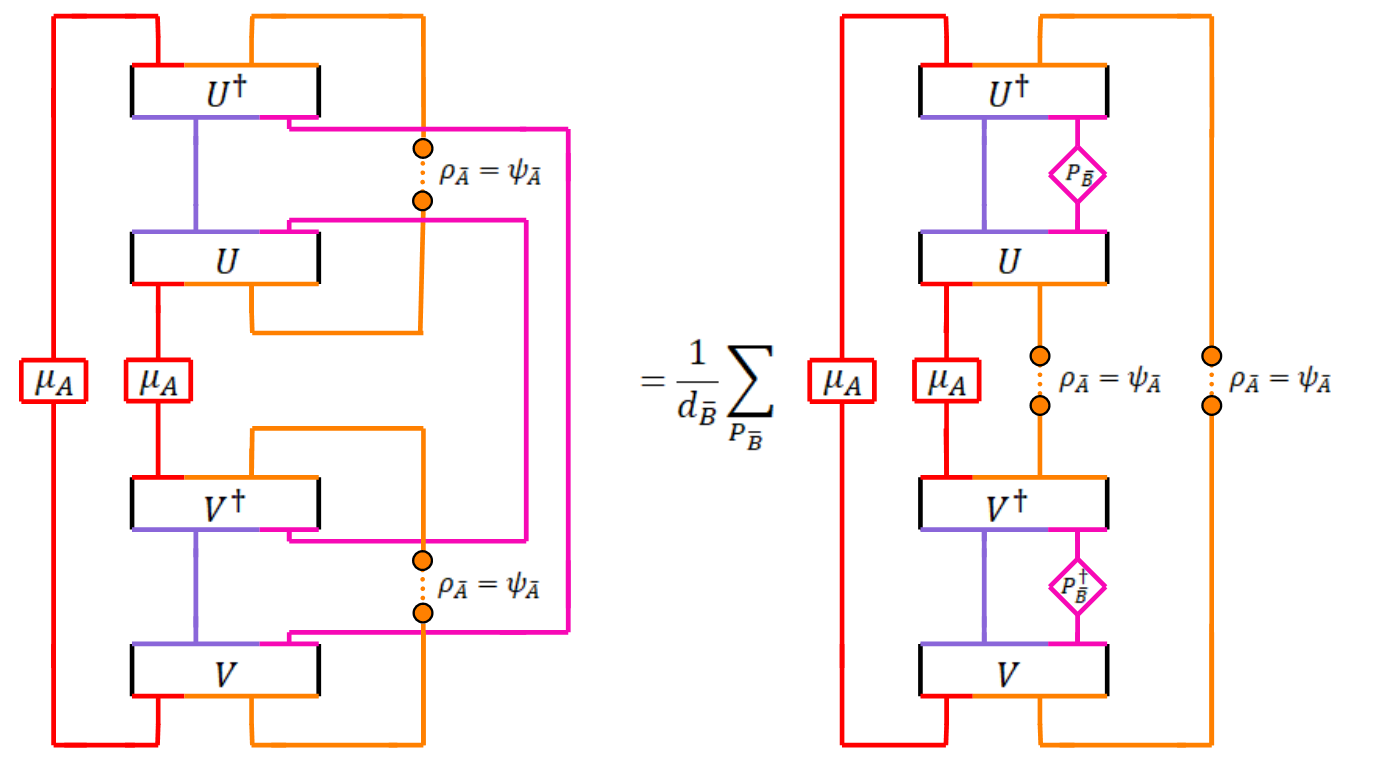}
	\caption{Tensor network diagram illustrating how to obtain Eq.\eqref{eq:Code1} when $\rho_{\bar{A}}$ is initialized as a pure state represented as dots connected by dashed line in the diagram.}
	\label{}
\end{figure}

%------------------------------------------------------------------------------------------------------------------------------------------------

We first integrate Eq.\eqref{eq:Code2} by the expansion of $T_{k=2}^{\operatorname{CQA}}$ in Eq.\eqref{eq:2-design-expansion1} and \eqref{eq:2-design-expansion2}, which has been decorated for SU$(d)$ symmetric conserved quantities in Eq.\eqref{eq:Charge1} and \eqref{eq:Charge2} and can be directly reused here:  
\begin{align}\label{eq:Code3}
	\begin{aligned}
		\frac{1}{d_\lambda^2 - 1} \Big( & \text{Tr}(\Pi_{\lambda,m_\lambda} (I_A \otimes \rho_{\bar{A}}) ) \text{Tr}(\Pi_{\lambda,m'_\lambda} (I_A \otimes \rho_{\bar{A}}) ) \text{Tr}( \Pi_{\lambda,m_\lambda} P_{\bar{B}} \Pi_{\lambda,m'_\lambda} P_{\bar{B}}^\dagger ) \\
		+ & \text{Tr}( \Pi_{\lambda,m_\lambda} (I_A \otimes \rho_{\bar{A}}) \Pi_{\lambda,m_\lambda,m'_\lambda} (I_A \otimes \rho_{\bar{A}}) ) \text{Tr}( \Pi_{\lambda,m'_\lambda,m_\lambda} P_{\bar{B}} ) \text{Tr}( \Pi_{\lambda,m_\lambda,m'_\lambda} P_{\bar{B}}^\dagger ) \\
		- & \frac{1}{d_\lambda} \text{Tr}(\Pi_{\lambda,m_\lambda} W^\dagger) \text{Tr}(\Pi_{\lambda,m'_\lambda}  (I_A \otimes \rho_{\bar{A}}) )
		\text{Tr}( \Pi_{\lambda,m'_\lambda,m_\lambda} P_{\bar{B}} ) \text{Tr}( \Pi_{\lambda,m_\lambda,m'_\lambda} P_{\bar{B}}^\dagger ) \\
		- & \frac{1}{d_\lambda} \text{Tr}( \Pi_{\lambda,m_\lambda} (I_A \otimes \rho_{\bar{A}}) ) \Pi_{\lambda,m_\lambda,m'_\lambda} (I_A \otimes \rho_{\bar{A}}) ) \text{Tr}( \Pi_{\lambda,m_\lambda} P_{\bar{B}} \Pi_{\lambda,m'_\lambda} P_{\bar{B}}^\dagger ),
	\end{aligned}
\end{align}
with
\begin{align}\label{eq:Code4}
	\frac{1}{d_\lambda d_\mu} \text{Tr}(\Pi_{\lambda,m_\lambda} (I_A \otimes \rho_{\bar{A}}) ) \text{Tr}(\Pi_{\mu,m_\mu} (I_A \otimes \rho_{\bar{A}}) ) \text{Tr}( \Pi_{\lambda,m_\lambda} P_{\bar{B}} \Pi_{\mu,m_\mu} P_{\bar{B}}^\dagger ).
\end{align}

On the other hand, the integrand of $\text{Tr}( \rho_{R \cup \bar{B},\text{avg}}^2 )$ embraces product of $T_{k=1}^{CQA}(P_{\bar{B}}), T_{k=1}^{CQA}(P_{\bar{B}}^\dagger)$ (1-design) which is straightforward to be expanded as: 
\begin{align}
	& \frac{1}{d_\lambda} \text{Tr}\Big( (\sum_i E_{(\alpha^\lambda_i, m_1), (\alpha^\lambda_i, m_2)}) P_{\bar{B}} \Big) (\sum_i E_{(\alpha^\lambda_i, m_1), (\alpha^\lambda_i, m_2)}), \quad 
	\frac{1}{d_\mu}  \text{Tr}\Big( (\sum_j E_{(\alpha^\mu_j, m_1'), (\alpha^\mu_j, m_2')}) P_{\bar{B}}^\dagger \Big) (\sum_j E_{(\alpha^\mu_j, m_1'), (\alpha^\mu_j, m_2')})
\end{align}
with contractions with $(I_A \otimes \rho_{\bar{A}})$ and yielding
\begin{align}\label{eq:Code5}
	\begin{aligned}
		\frac{1}{d_\lambda d_\mu} & \text{Tr}\Big( (\sum_i E_{(\alpha^\lambda_i, m_1), (\alpha^\lambda_i, m_2)}) P_{\bar{B}} \Big) \text{Tr}\Big( (\sum_j E_{(\alpha^\mu_j, m_1'), (\alpha^\mu_j, m_2')}) P_{\bar{B}}^\dagger \Big) \cdot \\
		& \text{Tr}\Big(  (\sum_i E_{(\alpha^\lambda_i, m_1), (\alpha^\lambda_i, m_2)}) (I_A \otimes \rho_{\bar{A}})  (\sum_j E_{(\alpha^\mu_j, m_1'), (\alpha^\mu_j, m_2')}) (I_A \otimes \rho_{\bar{A}}) \Big).
	\end{aligned}
\end{align}
Since $I_A \otimes \rho_{\bar{A}}$ is SU$(d)$ symmetric, we need $\lambda = \mu$ with $m_2 = m_1' = m_\lambda, m_1 = m_2' = m'_\lambda$ to get nonzero trace from the RHS from above, which equals
\begin{align}
	\begin{aligned}
		& \text{Tr}\Big( (\sum_i E_{(\alpha^\lambda_i, m_\lambda'), (\alpha^\lambda_i, m_\lambda)}) (I_A \otimes \rho_{\bar{A}})  (\sum_j E_{(\alpha^\lambda_j, m_\lambda), (\alpha^\lambda_j, m_\lambda')}) (I_A \otimes \rho_{\bar{A}}) \Big) \\
		= & \sum_{i,j} (I_A \otimes \rho_{\bar{A}})_{(\alpha^\lambda_i, m_\lambda), (\alpha^\lambda_j, m_\lambda)}) (I \otimes \rho_{\bar{A}})_{(\alpha^\lambda_j, m_\lambda'), (\alpha^\lambda_i, m_\lambda')}.
	\end{aligned}
\end{align}
As a result, Eq.\eqref{eq:Code5} can be rewritten as
\begin{align}\label{eq:Code5'}
	\frac{1}{d_\lambda^2} \text{Tr}( \Pi_{\lambda,m'_\lambda,m_\lambda} P_{\bar{B}} ) \text{Tr}( \Pi_{\lambda,m_\lambda,m'_\lambda} P_{\bar{B}}^\dagger) \text{Tr}( \Pi_{\lambda,m_\lambda} (I_A \otimes \rho_{\bar{A}}) \Pi_{\lambda,m_\lambda,m'_\lambda} (I_A \otimes \rho_{\bar{A}}) ),
\end{align}
which is used to cancel the second term in Eq.\eqref{eq:Code4} because $d_\lambda^2 - 1 \approx d_\lambda^2$ for nontrivial $S_n$ irrep. For the trivial $S_n$ irrep, we reformulate the 2-design expansion in Eq.\eqref{eq:Code4} and find that these terms are still identical. Finally,
\begin{align}
	\begin{aligned}
		& \mathbb{E}[ \text{Tr}( \rho_{R \cup \bar{B}}^2 ) ] - \text{Tr}( \rho_{R \cup \bar{B}, \text{avg} }^2 ) \\
		\approx & \frac{1}{d_A^2 d_{\bar{B}}} \sum_{P_{\bar{B}}} \Big( \sum_{\lambda,\mu, m_\lambda,m_\mu} \frac{1}{d_\lambda d_\mu} \text{Tr}(\Pi_{\lambda,m_\lambda} (I_A \otimes \rho_{\bar{A}}) ) \text{Tr}(\Pi_{\mu,m_\mu} (I_A \otimes \rho_{\bar{A}}) ) \text{Tr}( \Pi_{\lambda,m_\lambda} P_{\bar{B}} \Pi_{\mu,m_\mu} P_{\bar{B}}^\dagger ) \\
		& - \sum_{\lambda, m_\lambda,m'_\lambda} \frac{1}{(d_\lambda^2 - 1)d_\lambda} \text{Tr}(\Pi_{\lambda,m_\lambda} (I_A \otimes \rho_{\bar{A}}) ) \text{Tr}(\Pi_{\lambda,m'_\lambda} (I_A \otimes \rho_{\bar{A}}) )
		\text{Tr}( \Pi_{\lambda,m'_\lambda,m_\lambda} P_{\bar{B}} ) \text{Tr}( \Pi_{\lambda,m_\lambda,m'_\lambda} P_{\bar{B}}^\dagger) \\
		& - \frac{1}{(d_\lambda^2 - 1)d_\lambda} \text{Tr}( \Pi_{\lambda,m_\lambda} (I_A \otimes \rho_{\bar{A}})  \Pi_{\lambda,m_\lambda,m'_\lambda} (I_A \otimes \rho_{\bar{A}})  ) \text{Tr}( \Pi_{\lambda,m_\lambda} P_{\bar{B}} \Pi_{\lambda,m'_\lambda} P_{\bar{B}}^\dagger ) \Big)
	\end{aligned}
\end{align}
We analyze the scaling behavior for each term in the above identity as we did in Section \ref{sec:Charge}: 
\begin{enumerate}
	\item Obviously,  
	\begin{align}
		0 \leq \text{Tr}(\Pi_{\lambda,m_\lambda} (I_A \otimes \rho_{\bar{A}}) ) \leq \text{Tr}(I_A \otimes \rho_{\bar{A}}) = d_A, \quad 0 \leq \text{Tr}( \Pi_{\lambda,m_\lambda} P_{\bar{B}} \Pi_{\mu,m_\mu} P_{\bar{B}}^\dagger ) \leq \min\{d_\mu, d_\lambda\}.
	\end{align}
	
	\item Since the initial state $\rho_{\bar{A}}$ is selected from the SU$(d)$ trivial irrep, or by Schur--Weyl duality, the $S_n$ irrep to the Young diagram $\lambda_0 = (\frac{n-k}{d},...,\frac{n-k}{d})$. Using $S_n$ branching rule \cite{Sagan01,Goodman2009}, the matrix $I_A \otimes \rho_{\bar{A}}$ is exactly decomposed into $S_n$ irreps of Young diagrams $\lambda$ built by adding $k$ more boxes to $\lambda_0$, which satisfy $\lambda_1, \lambda_1' \approx \frac{n}{d}$ as long as $k = o(n)$. Their dimension is lower bounded by (see the end of Section \ref{sec:SnTheory})
	\begin{align}
		\dim S^\lambda = d_\lambda \geq \frac{d^n}{n^{d(d+2)/2}}.
	\end{align} 
	
	\item The total number of $\lambda$ stemmed from $\lambda_0$ is $p(k,d)$ (Definition \ref{def:PartitionFunction}), which is upper bounded by Eq.\eqref{eq:Ramanujan} with $n$ being replaced by $k$. The multiplicity for each of these irreps is also polynomially bounded in $n$.  
\end{enumerate}
In summary, given $k = o(n)$,
\begin{align}\label{eq:CodeError2}
	\mathbb{E}[ \text{Tr}( \rho_{R \cup \bar{B}}^2 ) ] - \text{Tr}( \rho_{R \cup \bar{B}, \text{avg} }^2 ) \leq 
	d_{\bar{B}} \sum_{\lambda,\mu, m_\lambda,m_\mu} \frac{1}{d_\lambda d_\mu} \min\{d_\mu, d_\lambda\} = \frac{1}{e^{\Omega(n)}}.
\end{align}

%------------------------------------------------------------------------------------------------------------------------------------------------ 

\subsection{Overparametrization regime with SU$(d)$ symmetric local quantities}\label{sec:QNTK}

We now discuss the application of the unitary $k$-design to the trainability of the variational ansatz. Let us consider the CQA ansatz $U(\vec{\theta})$ that respects the symmetry of SU($d$):
\begin{align}
	\begin{aligned}
		U(\vec{\theta}) = \exp(-i H_{\text{YJM}}(\beta) ) \exp(-i H_S(\gamma) ) & =
		\exp(-i \sum_{k,l} \beta_{kl} X_k X_l) \exp(-i\gamma \sum_{j=1}^{n-1} (j,j+1) ) \\
		& = \Big( \prod_{k,l} \exp(-i \beta_{kl} X_k X_l) \Big)  \exp(-i\gamma \sum_{j=1}^{n-1} (j,j+1) ),
	\end{aligned}
\end{align}
where $X_k X_l$ are product of YJM-elements (Definition \ref{def:YJM}) and $(j,j+1)$ are adjacent SWAPs with parameters $\vec{\theta} = (\beta_{kl}, \gamma)$. Since YJM-elements commutes with each other, we have $\exp(-i \sum_{k,l} \beta_{kl} X_k X_l) = \prod_{k,l} \exp(-i \beta_{kl} X_k X_l)$. Decomposing the second exponential of adjacent SWAPs however introduces Trotter errors. Then we consider some state initialization $\ket{\psi^\lambda} = \ket{\alpha_T^\lambda,m}$ in one $S_n$ irrep $S^\lambda$. We would also consider the statistical ensemble $\rho_\lambda$ of these states later.

For a given observable $\hat{O}$ that is SU($d$) symmetric, e.g., the Heisenberg Hamiltonian where we could write down: 
\begin{align}
	\hat{O} = \sum_{i=1}^N O_i,
\end{align}
where each $O_i$ contains a single term. In the case of quantum many-body with locality assumption, we typically assume the number $N$ of $O_i$ scales linearly or polynomially with the number of qubits. Then we define the loss: 
\begin{align}
	\frac{1}{2} \Big( \text{Tr} (U(\theta) \rho_\lambda U^\dag(\theta)  \hat{O}) - E_0 \Big)^2 = \Big( \text{Tr} ( \rho_\lambda U^\dag(\theta) \hat{O} U(\theta) ) - E_0 \Big)^2 \equiv \frac{1}{2} \varepsilon^2.
\end{align}
It is proved in \cite{SUd-k-Design2023} that this ansatz approximate SU($d$)-symmetric $k$-designs where $k = \Omega(n^2)$ the square of the number of qudits. Note that for the particular form of the ansatz, we can write: 
\begin{align}
	U(\theta)_{\operatorname{CQA}} = U_{-, l} U_{+, l}, 
\end{align}
where $l = 1, \cdots, L$ is the index for the variational angles. Note that our ansatz can be further Trotterized into products of time evolution of unitaries and each of which is only parameterized by one variational angle. Then the differentiation with chain rule easily reads:
\begin{align}
	\frac{\partial U(\theta)}{\partial \theta_l} = U_{-, l} (-i H_{\text{CQA},l} ) U_{+,l}, \quad \frac{\partial U^\dag (\theta)}{\partial \theta_l } = U^\dag_{+, l} (i H_{\text{CQA},l} ) U^\dag_{-, l},
\end{align}
where $H_{\text{CQA},l}$ denotes the generator (YJM-elements or SWAPs) of the CQA ansatz driven by the parameter $\theta_l$. The QNTK is given by: 
\begin{align}
	\begin{aligned}
		K = \sum_l \frac{\partial \bar{\varepsilon}}{\partial \theta_l}\frac{\partial \varepsilon}{\partial \theta_l}  & = \sum_l \Big\vert \bra{\psi^\lambda} U^\dag_{+, l} \left[ H_{\text{CQA},l}, U^\dag_{-, l} \hat{O} U_{-, l}  \right]  U_{+, l} \ket{\psi^\lambda} \Big\vert^2 \\
		& = - \Big( \bra{\psi^\lambda} U^\dag_{+, l} \left[ H_{\text{CQA},l}, U^\dag_{-, l} \hat{O} U_{-, l}  \right]  U_{+, l} \ket{\psi^\lambda} \Big)^2,\\
		&= - \Big( \sum_{i=1}^N \bra{\psi^\lambda} U^\dag_{+, l} \left[ H_{\text{CQA},l}, U^\dag_{-, l} O_i U_{-, l}  \right]  U_{+, l} \ket{\psi^\lambda} \Big)^2,
	\end{aligned}
\end{align}
since the trace of products of Hermitian and anti-Hermitian matrices is pure imaginary. Recall that the observable $\hat{O}$ is SU$(d)$ symmetric and $\ket{\psi^\lambda}$ (or $\rho_\lambda$) is taken from one $S_n$ irrep, we exclusively concern the unitary $U^\lambda$ restricted to that irrep block and
\begin{align}
	\begin{aligned}
		K & = - \sum_l \sum_{i,j=1}^N \text{Tr}( \rho_\lambda U^{\lambda \dag}_{+, l} \left[  H_{\text{CQA},l}^{\lambda}, O^{\lambda,I} _i \right]  U^\lambda_{+, l} \rho_\lambda U^{\lambda \dag}_{+, l} \left[ H_{\text{CQA},l}^{\lambda}, O^{\lambda,I}_j  \right]  U^\lambda_{+, l} ),
	\end{aligned}
\end{align}
where we define the interaction picture observable $\hat{O}^I = U^{\lambda \dag}_{-, l} \hat{O} U^\lambda_{-, l}$ (and similarly for $O^{\lambda, I}_i$) with general statistical ensemble $\rho_\lambda$. The assumption to respecting SU$(d)$ symmetry is practical in the experiment which also provides an accessible way to average the above identity by Haar randomness of SU$(d)$ symmetric unitaries. As explained in the previous context, twirling a general operator $M$ with no assumption on symmetry leads to most intricate properties in this topic (see Section \ref{sec:SnCommutant} and \ref{sec:Setting}). 

%--------------------------------------------------------------------------------------------------------

We now compute the average $\bar{K}$ when $U_+,U_-$ are independent SU$(d)$ symmetric unitaries and match the Haar measure up to the second moment:  
\begin{align}
	\bar{K} & = - \sum_l \int \Big( \int \text{Tr}( \rho_\lambda U^{\lambda \dag}_{+, l} \left[  H_{\text{CQA},l}^{\lambda}, O^{\lambda,I}  \right]  U^\lambda_{+, l} \rho_\lambda U^{\lambda \dag}_{+, l} \left[  H_{\text{CQA},l}^{\lambda}, O^{\lambda,I}  \right]  U^\lambda_{+, l} ) dU_{+,l} \Big) dU_{-,l}.
\end{align}
The integral involving $U_{+,l}$ is evaluated by first projecting $\left[ H_{\text{CQA},l}^\lambda, O^{\lambda,I}  \right] \otimes \left[  H_{\text{CQA},l}^{\lambda}, O^{\lambda,I}  \right]$ by the $2$th-fold channel $T_{k=2}^{\operatorname{CQA}}$ and then we contract with $\rho_\lambda \otimes \rho_\lambda$. Then we integrate $U_{-,l}$ similarly. Since we assume the symmetry and restrict to one $S_n$ irrep, $T_{k=2}^{\text{CQA}}(M)$ takes a quite simple form resembling the Haar projection with no symmetry (cf.Eq.\eqref{eq:2-design}): 
\begin{align}
	T_{k=2}^{\text{CQA}}(M) = \frac{1}{d_\lambda^2 - 1} ( \text{Tr}(M) I_\lambda^{\otimes 2} + \text{Tr}(S_\lambda M) S_\lambda - \frac{1}{d_\lambda} \text{Tr}(M) S_\lambda - \frac{1}{d_\lambda} \text{Tr}(S_\lambda M) I_\lambda^{\otimes 2}),
\end{align}
where $I_\lambda$ is the identity matrix restrict to the irrep $S^\lambda$ and $S_\lambda$ exchanges tensor products from $S^\lambda \otimes S^\lambda$ (cf.Eq.\eqref{eq:U(N)-basis}). In our cases,
\begin{align}
	\begin{aligned}
		T_{k=2}^{\text{CQA}}( \left[ H_{\text{CQA},l}^{\lambda}, O^{\lambda,I}  \right]^{\otimes 2} ) = \frac{1}{d_\lambda^2 - 1} ( & \text{Tr}(\left[ H_{\text{CQA},l}^{\lambda}, O^{\lambda,I}  \right])^2 I_\lambda^{\otimes 2} + \text{Tr}(\left[ H_{\text{CQA},l}^{\lambda}, O^{\lambda,I}  \right]^2) S_\lambda \\
		& - \frac{1}{d_\lambda} \text{Tr}( \left[  H_{\text{CQA},l}^{\lambda}, O^{\lambda,I}  \right] )^2 S_\lambda - \frac{1}{d_\lambda} \text{Tr}(\left[  H_{\text{CQA},l}^{\lambda}, O^{\lambda,I}  \right]^2 ) I_\lambda^{\otimes 2})
	\end{aligned}
\end{align}
Since the trace of any Lie bracket vanishes, contracting with $\rho_\lambda^{\otimes 2}$, we obtain
\begin{align}
	\begin{aligned}
		& \int \text{Tr}( \rho_\lambda U^{\lambda \dag}_{+, l} [  H_{\text{CQA},l}^{\lambda}, O^{\lambda,I}  ]  U^\lambda_{+, l} \rho_\lambda U^{\lambda \dag}_{+, l} [  H_{\text{CQA},l}^{\lambda}, O^{\lambda,I} ]  U^\lambda_{+, l} ) dU_{+,l} \\
		= & \frac{1}{d_\lambda^2 - 1} \Big( \text{Tr}([ H_{\text{CQA},l}^{\lambda}, O^{\lambda,I} ]^2) \text{Tr}(\rho_\lambda)^2 - \frac{1}{d_\lambda} \text{Tr}([  H_{\text{CQA},l}^{\lambda}, O^{\lambda,I}  ]^2 ) \text{Tr}(\rho_\lambda^2) \Big).
	\end{aligned}
\end{align}
To integrate $U_{-,l}$, we note that $\text{Tr}([  H_{\text{CQA},l}^{\lambda}, O^{\lambda,I}  ]^2 = 2\text{Tr}(H_{\text{CQA},l}^{\lambda} O^{\lambda,I} H_{\text{CQA},l}^{\lambda}  O^{\lambda,I} ) - 2\text{Tr}( (H_{\text{CQA},l}^{\lambda})^2 (O^{\lambda,I})^2 )$ and apply the above method again:
\begin{align}
	\begin{aligned}
		& \int \text{Tr}(H_{\text{CQA},l}^{\lambda} O^{\lambda,I} dU_{-,l} \\
		= & \frac{1}{d_\lambda^2 - 1} \Big( \text{Tr}(O^\lambda)^2 \text{Tr}(H_{\text{CQA},l}^{\lambda \ 2}) + \text{Tr}( O^{\lambda \ 2}) \text{Tr}(H_{\text{CQA},l}^\lambda)^2 -  \frac{1}{d_\lambda} \text{Tr}(O^\lambda)^2 \text{Tr}(H_{\text{CQA},l}^{\lambda})^2 - \frac{1}{d_\lambda} \text{Tr}(O^{\lambda \ 2}) \text{Tr}(H_{\text{CQA},l}^{\lambda \ 2}) \Big).
	\end{aligned}
\end{align}
On the other hand $\text{Tr}( (H_{\text{CQA},l}^{\lambda})^2 (O^{\lambda,I})^2 ) = \text{Tr}( (H_{\text{CQA},l}^{\lambda})^2 U_{-,l}^\dagger O^{\lambda \ 2} U_{-,l}$ is evaluated by the 1-design projection $T_{k=1}^{\text{CQA}}$:
\begin{align}
	\int \text{Tr}( (H_{\text{CQA},l}^{\lambda})^2 (O^{\lambda,I})^2 ) dU_{-,l} = \frac{1}{d_\lambda} \text{Tr}(O^{\lambda \ 2}) \text{Tr}( (H_{\text{CQA},l}^{\lambda})^2).
\end{align}

When $\rho_\lambda$ is pure, the results are gathered as
\begin{align}
	\begin{aligned}
		& \frac{2}{d_\lambda^2 + d_\lambda} \sum_l \Big( \frac{d_\lambda \text{Tr}(O^{\lambda \ 2}) - \text{Tr}(O^\lambda)^2 }{d_\lambda^2 - 1} \text{Tr}(H_{\text{CQA},l}^{\lambda \ 2}) 
		- \frac{d_\lambda \text{Tr}(O^{\lambda \ 2}) - \text{Tr}(O^\lambda)^2 }{d_\lambda^3 - d_\lambda} \text{Tr}(H_{\text{CQA},l}^{\lambda})^2 \Big) \\ 
		= & \frac{2}{(d_\lambda + 1)(d_\lambda^2 - 1)} \Big( \text{Tr}(O^{\lambda \ 2}) - \frac{\text{Tr}(O^\lambda)^2 }{d_\lambda} \Big) \sum_l \Big( \text{Tr}(H_{\text{CQA},l}^{\lambda \ 2}) - \frac{\text{Tr}(H_{\text{CQA},l}^{\lambda})^2 }{d_\lambda} \Big).
	\end{aligned}
\end{align}
By power mean inequality
\begin{align}
	\Big(\frac{1}{n}\sum_i \lambda_i^2 \Big)^{\frac{1}{2}} \geq \frac{1}{n} \sum_i \vert \lambda_i \vert,
\end{align}
the above term is always nonnegative. As a easy way to bound $\text{Tr}(H_{\text{CQA},l}^{\lambda \ 2}) - \frac{\text{Tr}(H_{\text{CQA},l}^{\lambda})^2 }{d_\lambda}$, we may assume the ansatz is Trotterized consisting of time evolution of 2-, 3-, $(2,2)$-cycles due to YJM-elements. Then 
\begin{align}
	& \text{Tr}( (i,j)^2) - \frac{\text{Tr}( (i,j))^2 }{d_\lambda} = (1 - \bar{\chi}_\lambda (i,j)) d_\lambda, \\
	& \text{Tr}( ((i,j,k) + (k,j,i) )^{\lambda \ 2}) - \frac{\text{Tr}( (i,j,k) + (k,j,i) )^2 }{d_\lambda} = 2(1 - \bar{\chi}_\lambda (i,j,k)) d_\lambda, \\
	& \text{Tr}( ((i,j)(k,l))^2) - \frac{\text{Tr}( (i,j)(k,l) )^2 }{d_\lambda} = (1 - \bar{\chi}_\lambda (i,j)(k,l)) d_\lambda, 
\end{align}
where $(i,j,k) + (k,j,i)$ is defined to make the operator Hermitian. For qubits ($d = 2$) and two-row Young diagrams $\lambda = (n-r,r)$ \cite{Ingram1950,Roichman1996,Lassalle2008},
\begin{align}
	& \bar{\chi}_\lambda (i,j,k) = \frac{n^2 - (3r + 1)n + 3r^2 - 3r}{n(n-1)}, \\
	& \bar{\chi}_\lambda (i,j)(k,l) = \frac{n^4 - 2(2r + 3) n^3 + (8r^2 + 12r + 11) n^2 - 2(4 r^3+4 r^2+4 r+3)n + 4r^4 - 8r^3 + 28r^2 - 24r }{n(n-1)(n-2)(n-3)}.
\end{align}
For $S_n$ irreps with large dimension ($r \to \frac{n}{2}$), these character values tend to $c \cdot d_\lambda$ as we saw in Section \ref{sec:Charge} where $\bar{\chi}_\lambda(i,j)$ is nearly $\frac{1}{2} d_\lambda$  Therefore,
\begin{align}
	\bar{K} \approx \sum^N_{ij} \frac{L}{d_\lambda^2} \Big( \text{Tr}(O^{\lambda }_i O^\lambda_j) - \frac{\text{Tr}(O^\lambda_i) \text{Tr}(O^\lambda_j)}{d_\lambda} \Big) \approx \frac{N^2 L }{d_\lambda},
\end{align}
where we assume that the observable is a sum of SWAPs. Then we use again the fact that the character of these local operators scales as $c \cdot d_\lambda$ for large $d_\lambda$ given by
\begin{align}
	d_\lambda = \dim S^{(n-r,r)} = \binom{n}{r} - \binom{n}{r-1} = \frac{n - 2r + 1}{n - r + 1} \binom{n}{r}
\end{align}
in Section \ref{sec:SnTheory}.

\end{document}